\documentclass[fleqn,usenatbib,useAMS]{mnras}

\usepackage{graphicx}	% Including figure files
\usepackage{amsmath}	% Advanced maths commands
\usepackage{amssymb}	% Extra maths symbols
\usepackage{multicol}        % Multi-column entries in tables
\usepackage{bm}		% Bold maths symbols, including upright Greek
\usepackage{pdflscape}	% Landscape pages
\usepackage{hyperref}
\usepackage{xcolor}
\usepackage{multirow}

\usepackage{float}
\usepackage{listings}

\usepackage{caption}
\usepackage{subcaption}
\usepackage[T1]{fontenc}
\usepackage{ae,aecompl}

\usepackage{newtxtext,newtxmath}

\makeatletter
\newcommand\footnoteref[1]{\protected@xdef\@thefnmark{\ref{#1}}\@footnotemark}
\makeatother

\title[Exoplanet spectroscopy with JWST NIRISS]{Exoplanet Spectroscopy with JWST NIRISS: Diagnostics and Case Studies}

\author[Holmberg \& Madhusudhan]{M\aa ns Holmberg$^{1}$\thanks{E-mail: \href{mailto:mlh58@ast.cam.ac.uk}{mlh58@ast.cam.ac.uk}}, 
Nikku Madhusudhan$^{1}$\thanks{E-mail: \href{mailto:nmadhu@ast.cam.ac.uk}{nmadhu@ast.cam.ac.uk}}
\\
$^{1}$Institute of Astronomy, University of Cambridge, Madingley Road, Cambridge, CB3 0HA, UK}

\date{Accepted May 19, 2023. Received in original form Dec 21, 2022}

\pubyear{2023}

\begin{document}
\label{firstpage}
\pagerange{\pageref{firstpage}--\pageref{lastpage}}
\maketitle

% Abstract of the paper
\begin{abstract}
The James Webb Space Telescope (JWST) is ushering in a new era in remote sensing of exoplanetary atmospheres. Atmospheric retrievals of exoplanets can be highly sensitive to high-precision JWST data. It is, therefore, imperative to characterise the instruments and noise sources using early observations to enable robust characterisation of exoplanetary atmospheres using JWST-quality spectra. The present work is a step in that direction, focusing on the NIRISS SOSS instrument mode, with a wavelength coverage of 0.6 - 2.8 $\mu$m and $R \sim 700$. Using a custom-built pipeline, JExoRes, we investigate key diagnostics of NIRISS SOSS with transit spectroscopy of two giant exoplanets, WASP-39~b and WASP-96~b, as case studies. We conduct a detailed evaluation of the different aspects of the data reduction and analysis, including sources of contamination, 1/f noise, and system properties such as limb darkening. The slitless nature of NIRISS SOSS makes it susceptible to contamination due to background sources. We present a method to model and correct for dispersed field stars which can significantly improve the accuracy of the observed spectra. In doing so, we also report an empirically determined throughput function for the instrument. We find significant correlated noise in the derived spectra, which may be attributed to 1/f noise, and discuss its implications for spectral binning. We quantify the covariance matrix which would enable the consideration of correlated noise in atmospheric retrievals. Finally, we conduct a comparative assessment of NIRISS SOSS spectra of WASP-39~b reported using different pipelines and highlight important lessons for exoplanet spectroscopy with JWST NIRISS. 

\end{abstract}


\section{Introduction}

The study of exoplanetary atmospheres has evolved rapidly since the first theoretical studies \citep[e.g.,][]{seager_theoretical_2000, burrows_theory_2001, brown_transmission_2001} and detections of atmospheric signatures nearly two decades ago with the Hubble Space Telescope \citep[HST;][]{charbonneau_detection_2002,vidal-madjar_extended_2003}. Spectroscopic observations have since allowed the atmospheric characterisation of dozens of exoplanets, leading to constraints on chemical compositions, temperature structures, and other atmospheric properties of these distant worlds \citep[see e.g. reviews by][]{seager2010,madhusudhan_exoplanetary_2014,crossfield2015, heng_atmospheric_2015, madhusudhan_exoplanetary_2019,fortney2021}. In particular, transmission spectroscopy of transiting exoplanets has been one of the most successful methods in this pursuit, unveiling a diverse set of atmospheric properties. The method relies on measuring wavelength-dependent variations in the fraction of starlight blocked by the day-night terminator of a planet's atmosphere during transit \citep[e.g.,][]{seager_theoretical_2000, brown_transmission_2001}. The resulting spectrum can then be inverted using atmospheric retrievals to derive the underlying atmospheric properties \citep{madhusudhan_temperature_2009, madhu2018}. Transmission spectroscopy in the near-infrared is of particular  interest due to strong molecular absorption features in this wavelength range which can be tracers of important physical and chemical processes. The HST Wide Field Camera 3 (WFC3) spectrograph, a near-infrared (1.1 - 1.7 $\mu$m) instrument, pioneered high-precision detections of H$_2$O in numerous exoplanet atmospheres using transit spectroscopy \citep[e.g.][]{deming_infrared_2013, mccullough_water_2014,kreidberg_precise_2014,sing_continuum_2016}. Today, precise transmission spectroscopy is routinely conducted using both space-based and ground-based observatories \citep[see e.g. reviews in][]{kreidberg2018,madhusudhan_exoplanetary_2019}.

We are entering a new era of atmospheric characterisation of exoplanets thanks to the successful launch and commissioning of JWST \citep{gardner_james_2006, beichman_observations_2014, kalirai_scientific_2018, greene_characterizing_2016, rigby_science_2023}. Using JWST transmission spectroscopy observations of the hot Jupiter WASP-39 b, obtained as part of the JWST Early Release Science Program \citep[ERS;][]{batalha_transiting_2017, stevenson_transiting_2016, bean_transiting_2018}, recent studies have already reported detections of several chemical species, including H$_2$O, CO$_2$, SO$_2$, CO, Na, and K in the atmosphere of WASP-39~b \citep{jwst_transiting_exoplanet_community_early_release_science_team_identification_2022, ahrer_early_2023, alderson_early_2023, rustamkulov_early_2023, feinstein_early_2023}. JWST's large collecting area (6.5 m), wide wavelength coverage (0.6 - 28 $\mu$m), and high spectral resolution (up to $R \sim 3000$) all come together to greatly outperform what is currently possible with other facilities -- promising to revolutionise our understanding of exoplanet atmospheres. 

One of the prime observing modes for exoplanet spectroscopy with JWST is the Single Object Slitless Spectroscopy (SOSS) mode of the Near Infrared and Slitless Spectrograph \citep[NIRISS;][]{doyon_jwst_2012}, optimised for high precision and spectrophotometric stability of bright targets  ($ 7 < J < 12$) during time-series observations (TSOs). The observing mode covers a wide wavelength region in the near-infrared (0.6 - 2.8 $\mu$m) with a resolving power of $R \sim 700$ at $\sim 1.4$ $\mu$m; and will thus play an important role in constraining key chemical abundances, via transit spectroscopy, such as H$_2$O, CO and K in hot Jupiters \citep{feinstein_early_2023}, and H$_2$O, CH$_4$, NH$_3$ in Warm Neptunes and beyond \citep{greene_characterizing_2016, constantinou_characterizing_2022}. NIRISS SOSS is the only observing mode with a sensitivity reaching into optical wavelengths (at $R > 100$), meaning that it is in principle sensitive to the scattering slope resulting from clouds and hazes \citep{wakeford2015,pinhas2017}, as well as to spectral features from neutral alkali metals such as K and (the wings of) Na \citep{allard_kh_2016,allard_new_2019, welbanks_massmetallicity_2019}.

NIRISS SOSS uses the GR700XD grism to produce three cross-dispersed spectral orders. The first order covers the red end, between 0.9 - 2.8 $\mu$m at $R \sim 700$, while the blue end is covered by the second order, spanning 0.6 - 1.4 $\mu$m at twice the resolution. Order 3 (0.6 - 1.0 $\mu$m) has low throughput and is therefore unlikely to be usable. The GR700XD grism includes a weak cylindrical lens to defocus the incoming light in the spatial direction, spreading it over $\sim 25$ pixels, to minimise pixel-to-pixel sensitivity variations, pointing jitter, and to allow for the observation of brighter targets without saturating the detector. Depending on readout mode, the bright-star limit for NIRISS SOSS is $J \sim 7 - 8$ mag. NIRISS employs a single $2048 \times 2048$ pixel Teledyne HgCdTe HAWAII-2RG detector \citep[e.g.,][]{gardner_james_2006, rauscher_commentary_2012}, also used by the Near Infrared Camera (NIRCam) and the Near Infrared Spectrograph (NIRSpec), providing low readout noise and high quantum efficiency; based on the same successful technology implemented by HST WFC3 and other near-infrared instruments. Combined with no slit losses, these design choices are expected to make NIRISS SOSS especially suited for transmission spectroscopy.

In this work, we present early insights into the performance of NIRISS SOSS for transmission spectroscopy and perform a detailed assessment of the data reduction given the novelty of the instrument. For this purpose, we use transit observations of two hot Jupiters, WASP-39 b \citep{faedi_WASP-39b_2011} and WASP-96 b \citep{hellier_transiting_2014}, obtained as part of the JWST Transiting Exoplanet Community ERS program and the Early Release Observations (ERO), respectively, of exoplanet transit spectroscopy. To analyse this data, we have developed an end-to-end pipeline, JExoRES, for NIRISS SOSS transmission spectroscopy observations, performing everything from the spectrum extraction to the light curve fitting. Motivated by the presence of correlated noise, likely due to 1/f noise from the HgCdTe detector \citep{schlawin_jwst_2020}, we go on to estimate the full covariance matrix of the transmission spectra, a first for exoplanet spectroscopy with JWST. We also investigate sources of contamination that could bias the transit depth, due to overlapping spectral orders \citep{radica_applesoss_2022, darveau-bernier_atoca_2022} and field stars, and provide strategies to mitigate these effects. Finally, we compare the transmission spectrum of WASP-39 b to \cite{feinstein_early_2023}.

In what follows, we outline the NIRISS SOSS contamination problem in Section \ref{sec:Contamination_problem} and the JWST observations used in this work in section \ref{sec:obs}. We describe the data reduction in Section \ref{sec:data_reduction}, including our approaches for background subtraction, spatial profile estimation and spectrum extraction. In Section \ref{sec:contamination}, we describe our approach to modelling contamination from dispersed field stars. We outline the light curve analysis in Section \ref{sec:light_curve}, including the analysis of the white light curves, the spectroscopic light curves, and the covariance of the spectrum. Next, in section \ref{sec:transmission_spectroscopy}, we evaluate the effects of limb-darkening, extraction method, and contamination on the spectrum. Using the spectrum of WASP-39 b, we conduct a comparative assessment of NIRISS SOSS pipelines in section \ref{sec:pipeline_comparison}. The case studies of the transmission spectra of WASP-39 b and WASP-96 b are presented in Section \ref{sec:case_studies}. Finally, we summarise and discuss our findings in Section \ref{sec:summary}.

\section{The Contamination problem of NIRISS} \label{sec:Contamination_problem}

\begin{figure*}
	\includegraphics[width=2\columnwidth]{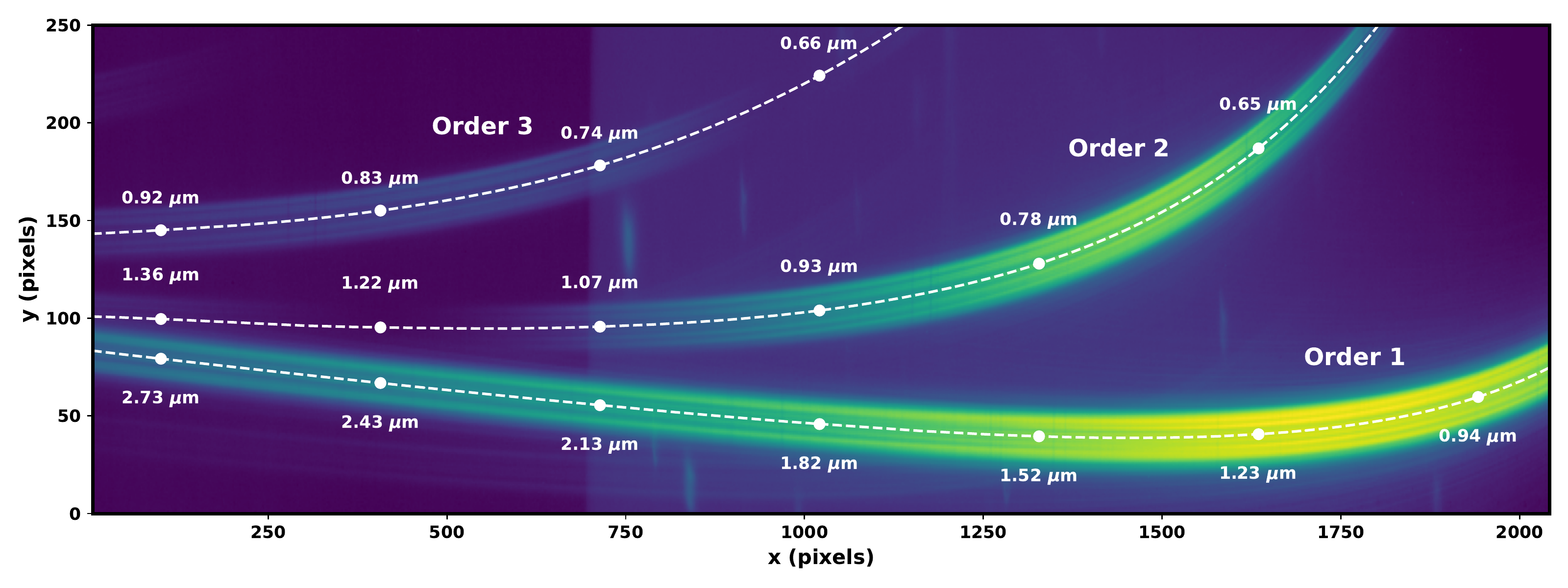}
    \caption{Detector view of the NIRISS SOSS observation of WASP-96 b, displaying the flux in log scale. The central trace positions of the three spectral orders are shown with white dashed lines (as identified by the algorithm described in section \ref{sec:trace}), with white dots marking a set of wavelengths for reference. The physical overlap of the 1st and 2nd orders is apparent on the left, corresponding to the red end of both orders. Outside the traces, we see the characteristic NIRISS SOSS background as well as some bright spots, which are due to 0th-order field star contamination. An additional faint spectral trace from a magnitude $G \sim 19$ field star can be seen at the bottom of the detector, causing a $\sim100$ ppm error in the transit depth at around 1.5 $\mu$m if not corrected.}
    \label{fig:detector}
\end{figure*}

To harness the full potential of NIRISS SOSS for atmospheric characterisation of exoplanets we need to understand and address sources of systematics that could otherwise bias or worsen the uncertainties of the measured transit depth. One known source of error comes from unwanted flux from nearby spectral orders and field stars that can contaminate the spectrum on the detector, presenting a challenge when extracting the spectrum \citep{radica_applesoss_2022}. 

\subsection{Order overlap}

The first source of contamination stems from the strong mechanical constraints of the G700XD grism, resulting in partial overlap of the first and second spectral orders at longer wavelengths \citep{darveau-bernier_atoca_2022}. This overlap is different from the overlap in wavelength between the two orders (around 0.9 to 1.4 $\mu$m) in that the orders physically overlap on the detector, which we illustrate in figure \ref{fig:detector}. A simple box extraction typically used for high signal-to-noise ratio (S/N) spectra would therefore extract a combination of order 1 and 2, particularly between 2.0 - 2.8 $\mu$m (order 1) and 1.0 - 1.4 $\mu$m (order 2). \cite{darveau-bernier_atoca_2022} showed that such cross contamination could bias the transit depth at the per cent level if not properly accounted for. We note that this is a second-order effect that only affects blended regions with different atmospheric signals, meaning that a flat transmission spectrum would remain unaffected. To mitigate this type of cross-contamination, we have developed an extension of the optimal extraction algorithm \citep{horne_optimal_1986} to allow for the simultaneous extraction of multiple overlapping spectral traces, including the background, which we describe in section \ref{sec:extraction}. Like the ATOCA algorithm \citep{darveau-bernier_atoca_2022}, our approach relies on accurate information about the spatial profiles of the PSF to disentangle the spectral orders.

\subsection{Field stars}

Another source of contamination comes from nearby stars in the field of view, so-called field stars, given the slitless nature of the instrument. These show up in the background as either 0th-order point sources or dispersed spectra that can overlap with the target spectrum. Assuming that a contaminating field star is not variable during the observation, the transit depth in the presence of contamination is reduced by a factor of $1/(1+F_{\text{cont}} / F_{\text{out}})$. From this, we see that the problem gets worse for faint targets and spectral regions with low throughput. Careful planning of NIRISS SOSS observations can ensures that contamination levels from bright nearby field stars are kept to a minimum, e.g. by utilising the ExoCTK Contamination Calculator tool \citep{bourque_exoplanet_2021}. However, some stars will unavoidably remain in the field of view, especially in crowded regions or if they are too faint to be detected by previous surveys while still observable with JWST due to its superior sensitivity. In the observations considered in this work, we found that the contamination level from field stars can reach several per cent, making it a large source of systematic error. We note that field stars can cause problems even away from the target spectrum given that contamination can interfere with the background subtraction and 1/f noise correction. In section \ref{sec:cont_mask} we outline a method for identifying the affected regions and in section \ref{sec:contamination} we present a strategy to model and correct for the contamination.

\section{JWST observations} \label{sec:obs}

Here we outline the JWST NIRISS SOSS observations (0.6 - 2.8 $\mu$m) of two transiting hot Jupiters: WASP-39 b and WASP-96 b. The transit of WASP-39 b was observed for 8.2 hours on July 26 2022, as part of the ERS program \#1366 \citep{batalha_transiting_2017}, consisting of 537 integrations (9 groups per integration) with an effective integration time of 49.4 seconds per integration. Next, the transit of WASP-96 b was observed for 6.4 hours on June 21 2022, as part of the JWST ERO \citep{pontoppidan_jwst_2022}, under program \#2734. The observation consists of 280 integrations (14 groups per integration) with an effective integration time of 76.9 seconds per integration. The total observing time for each target was chosen such that the pre-transit and post-transit duration was around half the full transit duration each, plus additional overhead. The transit duration of WASP-39 b and WASP-96 b is 2.8 and 2.4 hours, respectively.

The above TSOs all used the GR700XD + CLEAR configuration and recorded the data using the nominal 256$\times$2048 subarray (SUBSTRIP256) with the NISRAPID readout pattern (reading out all frames, that is, one frame per group). Each target was also observed with the F277W filter, blocking wavelengths $\lesssim$ 2.4 $\mu$m to help isolate the 1st order from the overlap region. As shown in section \ref{sec:cont_mask}, the inclusion of the F277W exposures proved useful in mapping the contamination from field stars.

Furthermore, to obtain an accurate background model and to derive a throughput function, we used additional observations from the JWST commissioning programs \#1541 and \#1091. The latter program includes the 5.4 hours TSO of the bright ($J = 9.6$) A1V standard star BD+60 1753, which we used to derive the relative throughput of the different spectral orders, as outlined in section \ref{sec:throughput}. All data used in this work was publicly available through the Mikulski Archive for Space Telescopes (MAST) archive at the Space Telescope Science Institute.

\section{Data reduction} \label{sec:data_reduction}

\begin{figure}
	\includegraphics[width=\columnwidth]{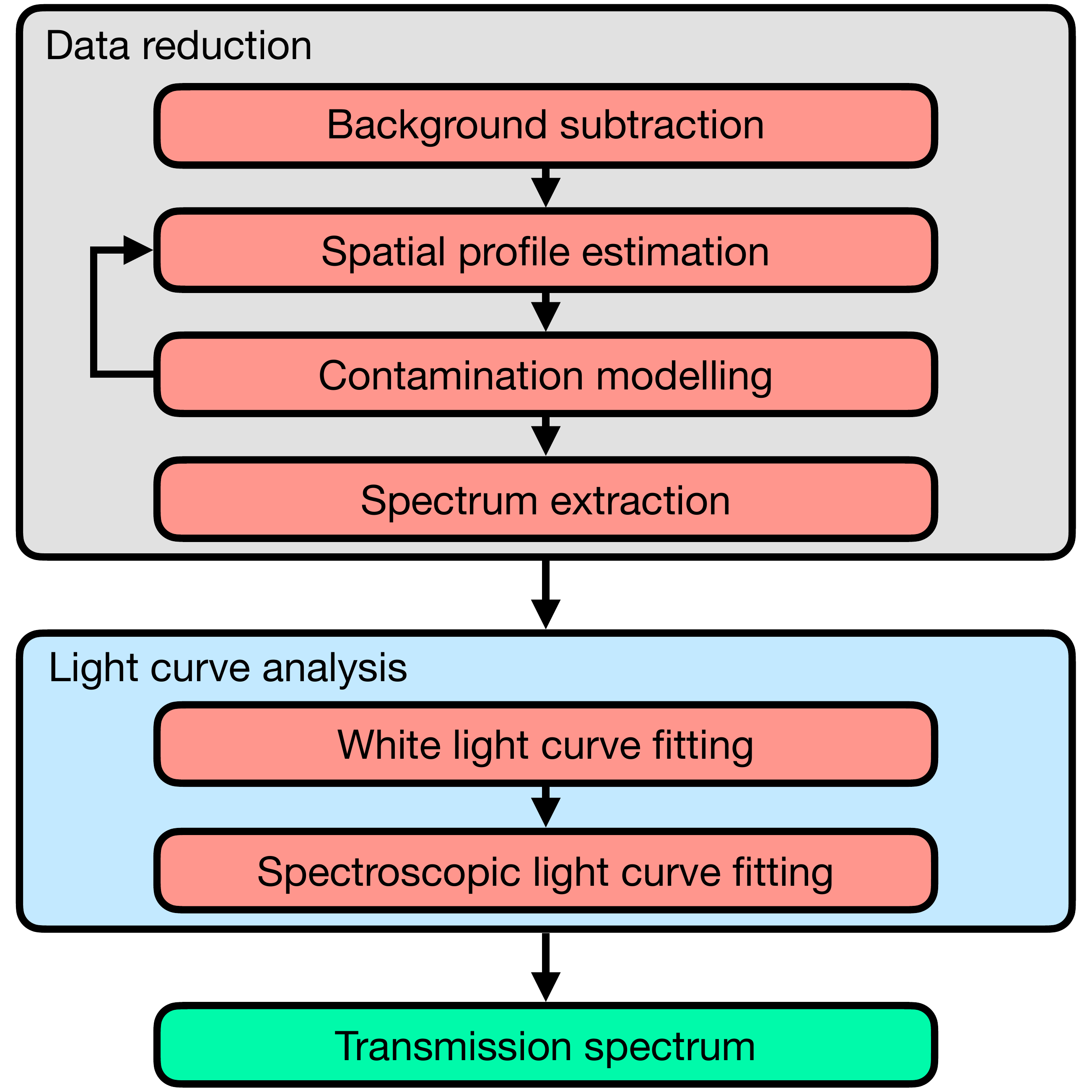}
    \caption{Schematic showing the main steps of the data analysis pipeline, JExoRES, for transmission spectroscopy with NIRISS SOSS used in this work.}
    \label{fig:pipeline}
\end{figure}

The first step of the JExoRES pipeline is the data reduction, that is, converting the raw detector data into a spectral time series. An outline of the pipeline is shown in figure \ref{fig:pipeline}. Before we extract the spectra we first mask regions affected by cosmic ray hits, identify contamination from field stars and subtract the background flux. Next, we refine the trace positions and go on to construct the spatial profile for each spectral order. Using these spatial profiles, we create a linear model of each detector column consisting of multiple orders (which can overlap). We then use weighted linear regression to extract the spectrum of each order simultaneously.  

Below we describe in detail the data reduction of the NIRISS SOSS TSOs of WASP-39~b and WASP-96~b. Our starting point is the Stage 2 calibrated files (CALINTS) that are publicly available from the MAST archive. At this stage, the data is superbias and dark subtracted as well as reference pixel corrected, linearity corrected and flat-field corrected. The ramp of each pixel has been fitted to determine the average count rate, in units of DN/s (digital numbers per second), where jumps due to cosmic rays were discarded from the fit. The data were calibrated with the JWST calibration software version 1.5.3 using reference data from the JWST Calibration Reference Data System (CRDS) with contexts 0916 (WASP-96~b) and 0937 (WASP-39~b). We note that the same reference files were used for both observations, even though different CRDS contexts were used. Throughout this work, we used the wavelength solution provided by the CRDS reference file jwst\_niriss\_spectrace\_0023.fits.

\subsection{Cosmic rays and snowballs}

Since the slope fitting in Stage 1 of the JWST Science Calibration Pipeline is performed on the jump-free segments of each ramp (which are then averaged together), we expect most pixels flagged by the jump detection to be corrected. Some pixels would however still show excess flux, for example if charge spills over from nearby saturated regions. We therefore start by searching for outliers that have not been properly corrected in order to mask these from further analysis. This include so called "snowball" events, caused by low-energy particles depositing much of their energy into the detector. Snowballs can affect hundreds of pixels and are currently not fully corrected for by the JWST Science Calibration Pipeline \citep{rigby_science_2023}. We start by considering pixels flagged as good, saturated or where a jump was detected by the JWST Science Calibration Pipeline as potentially useful pixels. Pixels flagged as low quality for other reasons were rejected, amounting to around 1.8 \% of the pixels (not counting reference pixels). To search for outliers we first subtracted a median-filtered time series from each pixel as well as the median from each detector column of each integration, to account for excess 1/f noise. With these residuals, we created a binary mask consisting of the union of all 2$\sigma$ outliers, saturated pixels and jump detections; and rejected all connected regions that included at least one 5$\sigma$ outlier. This way we mask both single-pixel cosmic ray hits and extended snowballs; which like other cosmic rays consists of large outliers and flagged pixels, but also a faint extended halo of increased flux. We capture this faint halo by considering the union 2$\sigma$ outliers. Using this method we rejected an additional 0.3 \% of the pixels during the observation of WASP-96~b, which is the target with the longest effective integration time, hence most affected by cosmic ray hits per integration. For the observation of WASP-39~b, we only rejected an additional 0.15 \% of the pixels.

\subsection{Background subtraction}

\begin{figure}
	\includegraphics[width=\columnwidth]{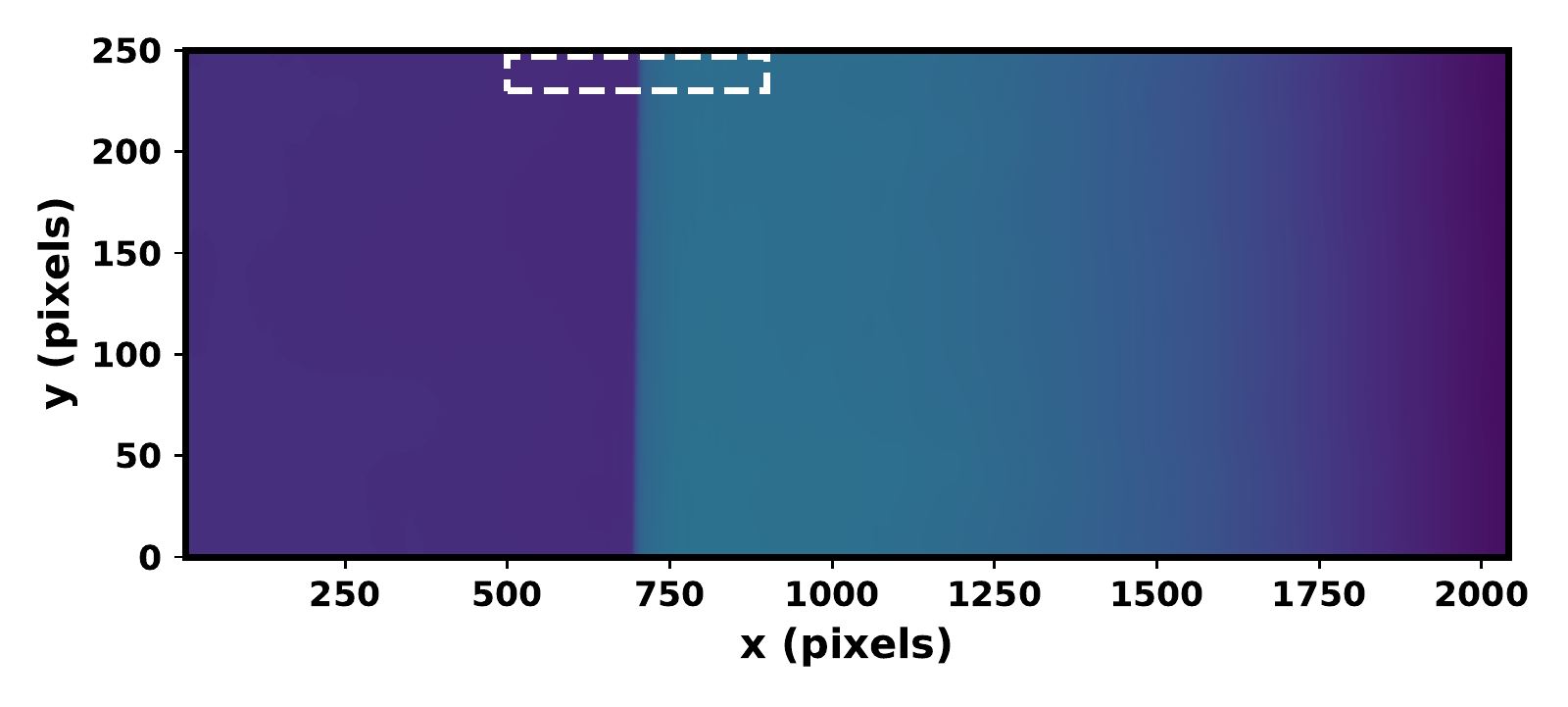}
    \caption{The NIRISS SOSS background, dominated by zodiacal light, as obtained from the GR700XD background observation in program \#1541. This model is re-scaled to the background of each observation and subtracted. We highlight the area used to fit the background model to the data in white.}
    \label{fig:bgr}
\end{figure}

Since the NIRISS SOSS background is non-trivial and difficult to estimate from the target observations alone, given the lack of suitable detector regions where the background can be sampled without interference from either the bright target spectrum or field star contamination, we use data from the GR700XD background observation in program \#1541 to construct a background model. This is similar to the background subtraction performed by \cite{feinstein_early_2023} and \cite{fu_water_2022}. We later refine the background level to reduce the effect of 1/f noise. However, this initial background estimate is important to correctly estimate the PSF in the spatial direction to be used for the spectrum extraction. In figure \ref{fig:detector}, we illustrate the peculiar shape of the NIRISS SOSS background. The dominant background source is zodiacal light \citep{2016jdox}, which is dispersed through the GR700XD grism, leading to a smoothly rising background towards longer wavelengths. The brighter component of the background likely comes from the 0th order of the background emission, which is coming in at an angle. Due to this angle, the pick-off mirror (POM) does not illuminate the whole detector, giving rise to the sharp cutoff in the background level at around the 700-pixel column. This would explain why we do not observe any 0th-order contamination from field stars below the $\sim$700 pixel column. 

The GR700XD background observation consists of an 18-panel mosaic that we used to extract a background model. Using the median of these exposures we removed the stellar sources. We then smoothed the resulting image in the spatial direction to reduce the noise level, given that most of the variation is observed in the dispersion direction. The resulting background model is shown in figure \ref{fig:bgr}. The idea is then to re-scale this model to each observation before subtraction. However, first we need to establish that the shape of the background remains stable. To test this, we construct a metric based on the ratio of the flux between two distinct regions of the background. For this purpose, we choose the regions $x \in [800, 850]$, $y \in [230, 250]$ and $x \in [600, 650]$, $y \in [230, 250]$, respectively, corresponding to two regions just above and below the sharp cutoff in background level (away from the spectral traces). We then measure the ratio of these two regions using the median image of each TSO and find that this ratio is consistent to within 1 \% among the observations of WASP-39 b, WASP-96 b and the GR700XD background observation. Therefore, due to the good agreement of the shape of the background among the target observations, we go on to subtract a scaled version of the background model from each observation. We re-scaled the model to the median integration, using the region $x \in [500, 900]$, $y \in [230, 250]$, and subtracted the result from each integration. Here we assume that the background is not changing on the time scale of hours \citep{fu_water_2022}. For WASP-39 b and WASP-96 b, we find that we had to re-scale the background model by a factor of 0.88 and 0.45, respectively, indicating that the background flux level differs among targets, whereas the background shape remains very similar. In the future, it may be more accurate to take new background observations for each target. Later, during the spectrum extraction, we refine the background level of each integration to combat 1/f noise. 

\begin{figure}
        \subfloat{
            \includegraphics[width=\columnwidth]{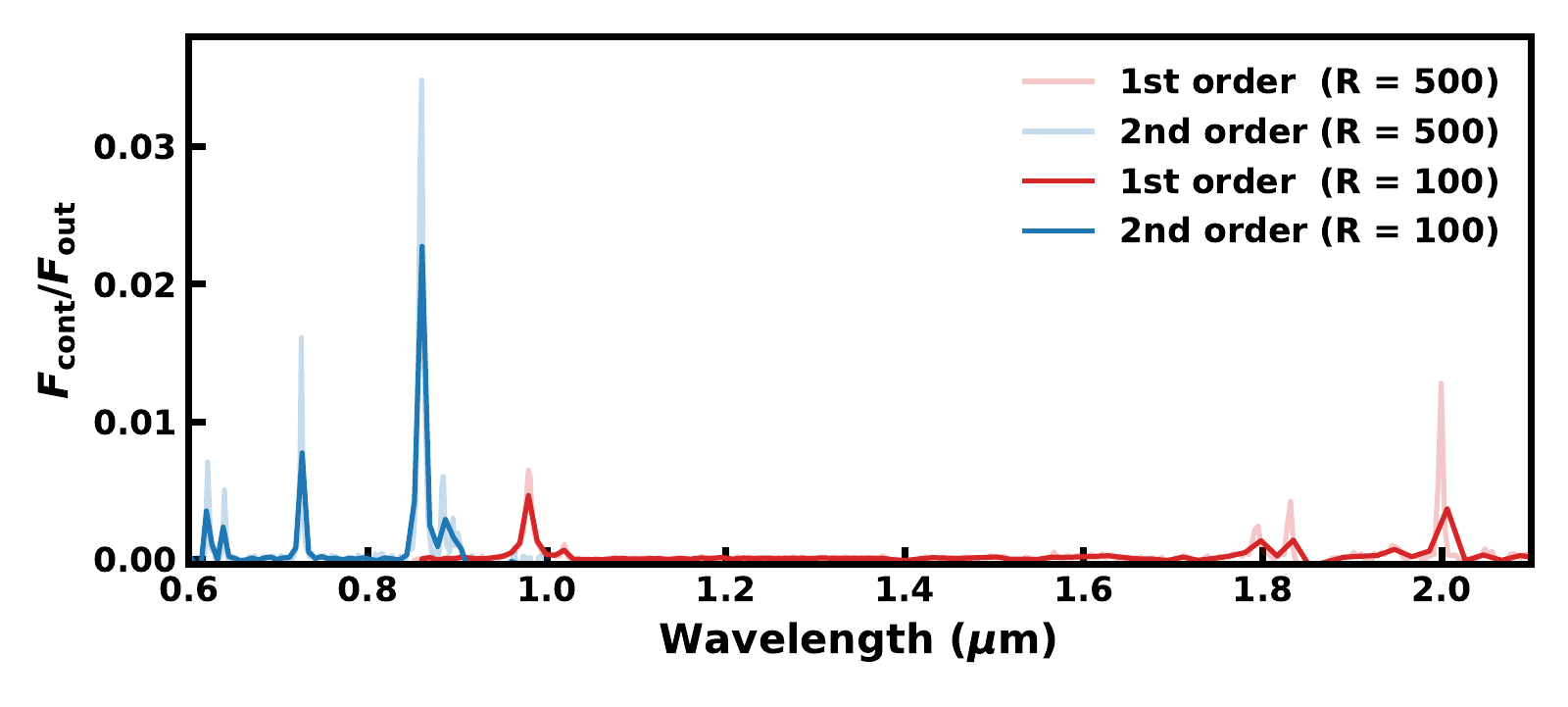}
        } \\
        \subfloat{
            \includegraphics[width=\columnwidth]{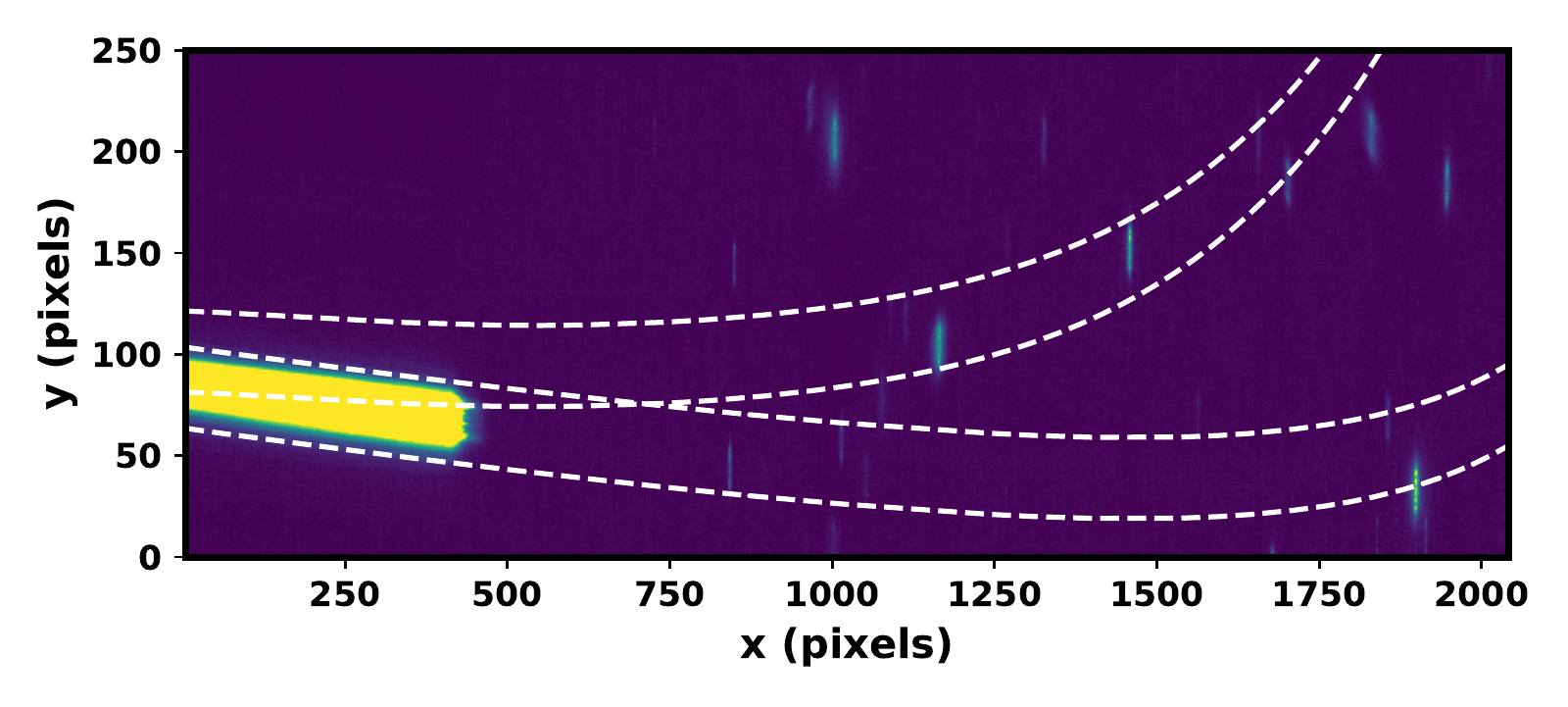}
        } \\
        \subfloat{
            \includegraphics[width=\columnwidth]{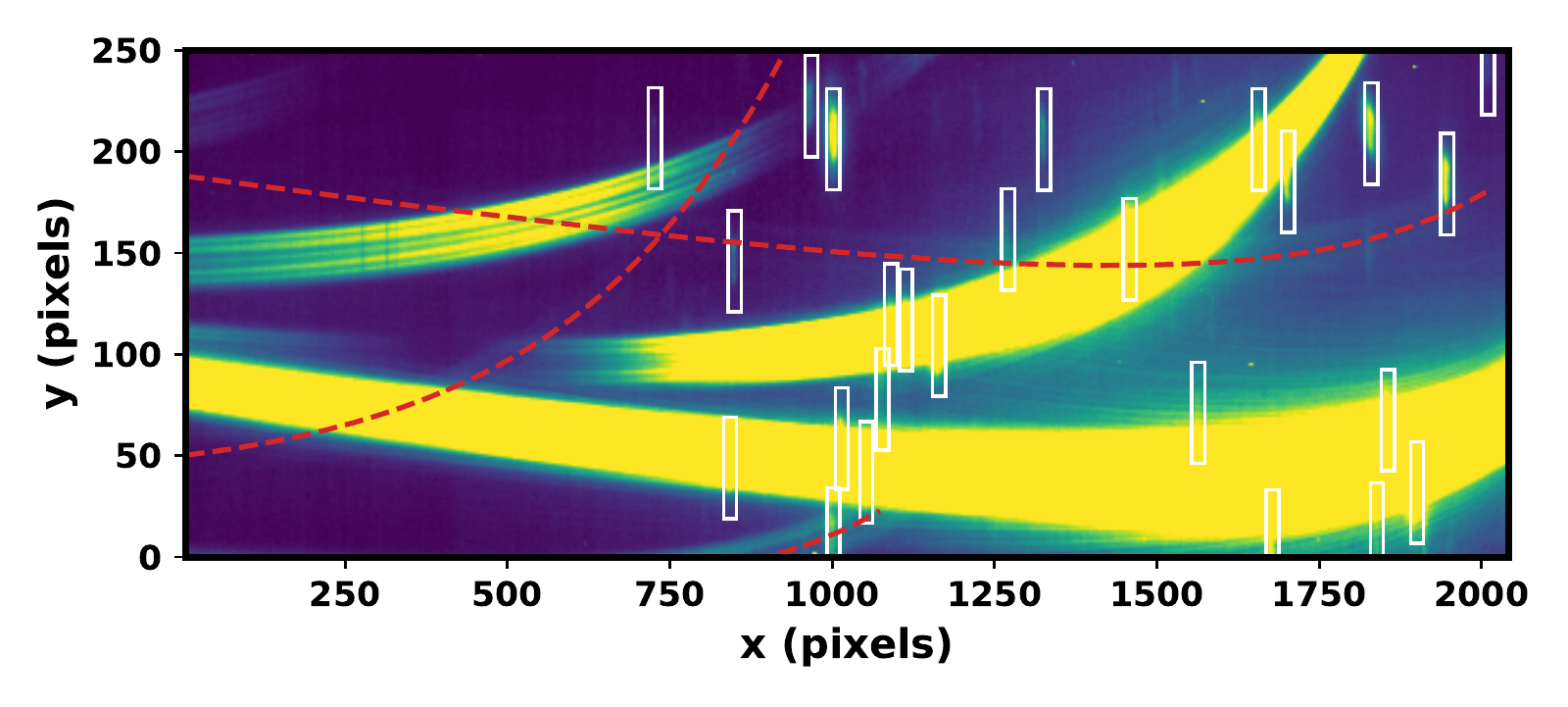}
        }
        \caption{Field star contamination for the observation of WASP-39 b. The top panel shows the ratio of the contamination flux $F_{\mathrm{cont}}$ and the out-of-transit flux $F_{\mathrm{out}}$ for the 0th-order contamination in the F277W bandpass (at two different resolutions to show the effect of binning), extracted using a 40-pixel box extraction centred on the traces. Given that the CLEAR filter lets in more light, this should be seen as a minimum estimate of the 0th-order contamination. The middle panel shows the F277W exposure overlaid with the 1st and 2nd-order traces (dotted white curves) to illustrate regions where the 0th-order contamination overlaps with the spectrum. The bottom panel shows the average (background-subtracted) integration in the CLEAR configuration, highlighting the 0th-order contamination in white and the relevant traces of the higher order contamination in red.}
        \label{fig:f277w}
\end{figure}

\subsection{Field star mapping} \label{sec:cont_mask}

Due to the slitless nature of NIRISS, light from field stars will be dispersed by the GR700XD grism and show up as additional spectral traces and 0th-order point sources that could contaminate the target spectrum. Again, we note we only observe 0th-order contamination in the brighter region of the background, i.e. above the $\sim$700 pixel column, corresponding to $> 2.1$ $\mu$m and $> 1.1$ $\mu$m for the 1st and 2nd order, respectively. This is in contrast to additional spectral traces from field stars that can appear anywhere on the detector. In section \ref{sec:contamination} we show how to address the contamination from field stars by modelling their contribution. To do this, we first need to create a mask of the affected regions. 

Similar to the nirHiss pipeline in \cite{feinstein_early_2023}, we utilise the F277W exposure to treat the 0th-order contamination. The F277W filter, allowing wavelengths between $2.4 < \lambda < 3.1$ $\mu$m, blocks light from the 2nd and 3rd order of the target spectrum, as well as most of the light from the 1st order, leaving only the spectral trace of the 1st order below around the 450-pixel column. The addition of this exposure presents a good opportunity to identify regions with 0th-order contamination from field stars, as we can easily identify regions showing excess flux. This is useful since the position of the target traces remains approximately the same between the F277W and CLEAR exposures. To assess the level of 0th-order contamination, we extract the spectrum of the average of the F277W integrations to obtain $F_{\text{cont}}$ and compare to the out-of-transit spectrum $F_{\mathrm{out}}$, extracted using the average of the out-of-transit integrations of the CLEAR exposure. In both cases we used a 40-pixel box extraction centred on the spectral traces, as constructed in \ref{sec:trace}. In figure \ref{fig:f277w}, we show the F277W exposures for the observations of WASP-39 b and the ratio between the contamination flux $F_{\mathrm{cont}}$ and the out-of-transit flux $F_{\mathrm{out}}$. We find that the contamination reaches per cent levels, given that the transit depth is reduced by a factor of $1/(1+F_{\text{cont}} / F_{\text{out}})$. Since these levels correspond to the contamination in the F277W bandpass, they are to be seen as a minimum level of contamination, meaning that these values cannot be directly used to correct the contamination in the CLEAR exposure. On average, we find that the flux of the 0th-order contamination is around three times brighter in the CLEAR exposure compared to the F277W exposure (similar to \cite{feinstein_early_2023}), however, this factor is not universal as it depends on the stellar parameters of the field stars. As can be seen in figure \ref{fig:f277w}, we find that binning can help mitigate the effect of 0th-order contamination due to their narrow profiles. The amount of 0th-order field star contamination is less for WASP-96 b due to a less crowded field of view.

We create a mask for each detected 0th-order field star by using a rectangle with a height of 50 pixels and a width of 20 pixels. The width corresponds to three times the average full width at half maximum (FWHM) of the contamination, obtained by fitting a Lorentzian to the extracted (F277W) flux of each 0th-order field star, which we found is a good fit to the broad wings of the contamination (in the wavelength direction). The extent of the masking needs to be re-evaluated if brighter contaminating sources are present in other observations, as in the case of HAT-P-18 b \citep{fu_water_2022}. We show the mask in the case of WASP-39 b in the bottom panel of figure \ref{fig:f277w}.

To identify additional spectral traces we turned to GAIA DR3 \citep{gaia_collaboration_gaia_2016, gaia_collaboration_gaia_2022}. Using the GAIA catalogue, we could estimate the position of the spectral traces from nearby stars with a magnitude brighter than $G \sim 21$. In the case of WASP-96 b, this allowed us to identify the $G = 19.0$ magnitude star (Gaia Source ID 4990044874536787328) responsible for the additional trace seen in figure \ref{fig:detector}, which overlaps with the main trace at around 1.5 $\mu$m and $>1$ $\mu$m for the first and second order, respectively. For the observation of WASP-39 b, we identify another $G = 19.3$ field star (Gaia Source ID 3643098772089343104) whose traces cross the 1st order at around 2.5 $\mu$m (and partially at 1.8 $\mu$m), and the 2nd order above $1$ $\mu$m. We also found another faint field star that overlapped with the 2nd order at 0.8 $\mu$m in the observation of WASP-39 b, which we were unable to identify using the GAIA catalogue. We note that the contamination level of the 2nd order above $0.9$ $\mu$m is large for both targets, resulting from the low throughput in this region. For this reason, we later mask the 2nd order above $0.9$ $\mu$m from further analysis. The remaining regions must be carefully examined to make sure that the contamination is low enough to be negligible, or some correction must be applied. In section \ref{sec:contamination}, we describe a novel approach to model the flux from these additional spectral traces. 

\subsection{Order tracing} \label{sec:trace}

\begin{figure}
	\includegraphics[width=\columnwidth]{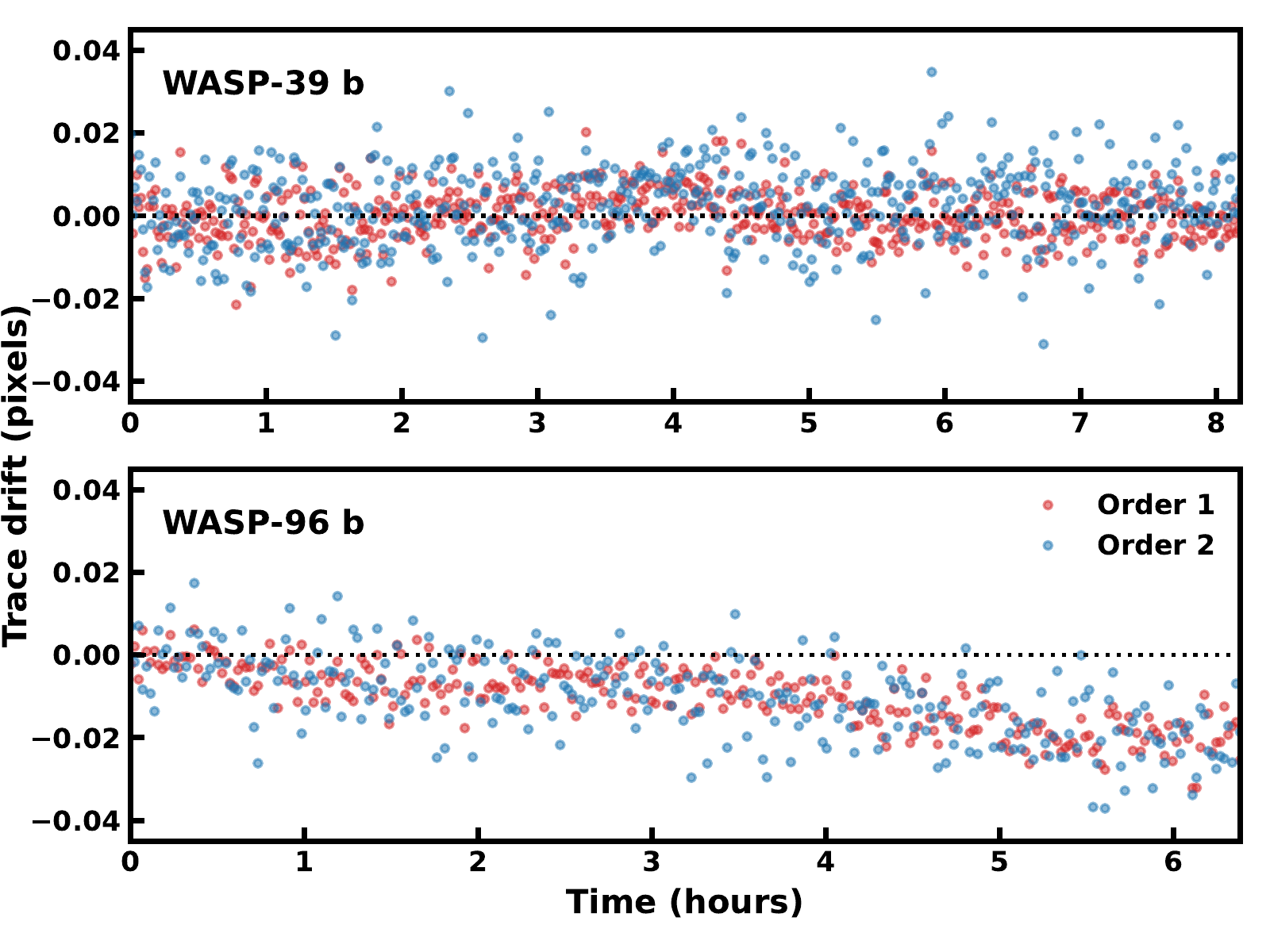} 
    \caption{Trace drift in the spatial direction for the observations of WASP-39~b and WASP-96~b. For WASP-96 b, we find a small linear trend in the trace positions of both orders. We note that the standard deviation of the drift (without removing the trend) is less than 1 mas for both targets, thanks to the excellent pointing stability provided by JWST.}
    \label{fig:trace_drift}
\end{figure}

Due to small changes in the angle of the pupil wheel (holding the GR700XD grism) and variations in the target acquisition, we expect the spectral trace to vary slightly between observations. Therefore, to accurately estimate the spatial profiles, we refine the position of the traces. Here we present a method to refine the traces similar to \cite{radica_applesoss_2022}. We start by finding the "edges" of the spectral orders by locating the maxima/minima of the gradient in the spatial direction. Before computing the gradient we applied a Gaussian filter with a standard deviation of 2 pixels in the spatial direction, which acts to suppress high-frequency variations in the core of the PSF while at the same time making the edge detection more sensitive in regions with low flux. Next, we use the reference trace from CRDS (jwst\_niriss\_spectrace\_0023.fits) to reject edges further away than 25 pixels from this reference trace. We then improve the position estimate of the edges by fitting a Gaussian to the maxima/minima of the gradient. Once we have measured the edge locations along the different spectral orders (although with some gaps), we reject outliers that deviate more than 5 pixels away from the median distance between the reference trace and the position of the bottom/top edges. We then use these edges to find the centre of the trace by averaging the position of the two edges. Given that the 1st order is much brighter than the 2nd order in the region where they overlap, we could not obtain an estimate of the bottom edge of order 2. To fill in this gap, as well as other regions where only one of the edges was found, we use the fact that the PSF does not depend on the spectral order. For this reason, we can use the width of the trace (i.e. the distance between the bottom and the top edges) from another order, at some particular wavelength, to estimate the location of the missing edges empirically. Finally, we fit a linear spline to a binned-down version of the trace of each order to even out local variations due to noise. This way we can robustly interpolate/extrapolate any remaining regions where no trace estimate was possible, for example, due to low flux.

Next, we investigate the trace stability over the course of the observation of WASP-39~b and WASP-96~b. In figure \ref{fig:trace_drift} we show the drift of the traces as measured in the spatial direction, computed as the median trace position. We find a small linear trend in the trace position of both order 1 and 2 for the observation of WASP-96 b. No such trend is observed for WASP-39 b. This is consistent with the finding of a linear trend in the white light curves of WASP-96 b, whereas no such trend is seen for WASP-39 b, as described in section \ref{sec:light_curve}. We note that even with this trend for WASP-96 b, the standard deviation of the drift is only 0.5 and 0.7 mas for order 1 and 2, respectively\footnote{Given that the pixel scale is 0.0658"/pixel in the spatial direction.}; while the overall drift is 1.4 mas over the observation period of 6.4 hours. These results are in line with the $\sim$ 1 mas ($1\sigma$ per axis) pointing stability reported by \cite{rigby_science_2023}, and showcase the superb pointing stability of JWST.

\subsection{Spatial profile estimation} \label{sec:prof}

\begin{figure}
	\includegraphics[width=\columnwidth]{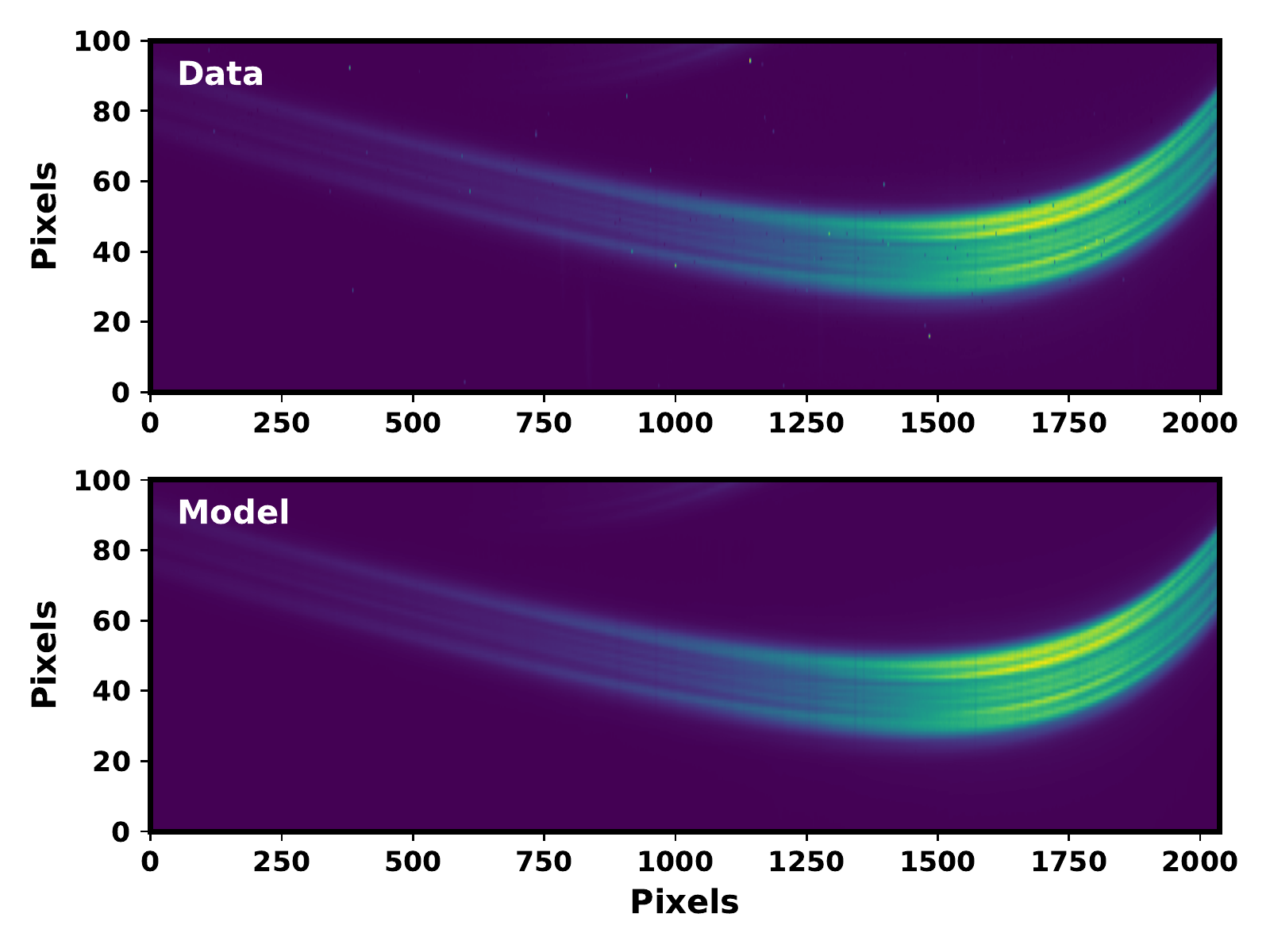}
 \caption{Detector region showing the trace of the 1st spectral order. The top panel shows a background subtracted integration of WASP-96, while the bottom panel shows our model of the same integration. A few bad pixels can be seen in the data which are not present in the model.}
    \label{fig:spectrum_model}
\end{figure}

To perform the spectrum extraction and disentangle the spectral orders, we require an accurate estimate of the spatial profile of each order, that is, the spatial part of the PSF. Recently, \cite{radica_applesoss_2022} proposed a method for generating the spatial profiles for each order, where they extracted the core of the profile from data and used simulated PSFs to reconstruct the wings. However, due to concerns about the observed NIRISS SOSS PSF not matching simulations, particularly in the wings, raised by the NIRISS team\footnote{JWST Technical report JWST-STScI-008270, SM-12, provided at \url{https://www.stsci.edu/files/live/sites/www/files/home/jwst/documentation/technical-documents/_documents/JWST-STScI-008270.pdf}}, we use a different approach that is fully empirical.

Given the stability of the traces, we construct the spatial profile using the average of all integrations. We start by building the spatial profiles in the regions where the cores of the spectral orders do not overlap. Before we "read off" the profiles, we first aim to subtract the wings from adjacent spectral orders. To do this, we extract the profile of the first order below 1 $\mu$m, given that this is the only region on the detector affected by only a single order (disregarding scattered light), allowing us to access the shape of the wings. We interpolate this part of the profile onto an oversampled grid centred on the midline of the trace determined in section \ref{sec:trace}. We then fit the average (in wavelength) of this profile to each spectral order simultaneously, one detector column at a time, while masking the central $\pm$10 pixels of the trace. To do this, we used the same weighted linear regression used for the spectrum extraction, described below in section \ref{sec:extraction}, and the trace information obtained in section \ref{sec:trace}. We note that the fitting of the wings is not very sensitive to the masking width of the core since the fitting routine rejects outliers, meaning that the core (which is the most wavelength-sensitive part of the PSF) will be masked regardless if it differs from the PSF at around 1 $\mu$m. Finally, with the wings of the spatial profiles at hand, we first subtract the wings from the 1st order and extract the core of the profile (within $\pm$25 pixels from the midline) of the 2nd order below 1.1 $\mu$m. Likewise, we subtract the wings from the 2nd order and extract the core of the profile of the 1st order below 2.1 $\mu$m. After merging these profiles with the wings obtained above we interpolate the profiles onto the same oversampled grid as before.

Next, we note that the shape of the PSF varies with wavelength, as determined by the JWST Optical Telescope Element, but does not change between spectral orders \citep{radica_applesoss_2022}. This means that we can reconstruct the spatial profile of the 2nd order where the two orders overlap by using the profile of the 1st order in this wavelength range (which we already constructed). By then masking the 1st order with an aperture of 40 pixels, we can extract the spectrum of the 2nd order in the region where the two orders overlap since we know the spatial profile of the 2nd order in this region. Using this spectrum, we can model and subtract off the 2nd order, thus allowing us to extract the profile of the 1st order above 2.1 $\mu$m. Finally, we linearly interpolate the profile in masked regions (in the dispersion direction) to not be affected by 0th-order contamination from field stars. We showcase the model used to extract the spectrum in figure \ref{fig:spectrum_model}, built with the empirically derived spatial profiles, as outlined above.

\subsection{1/f noise mitigation} \label{sec:1_f}

The JWST HgCdTe detectors used for NIRISS, NIRCam, and NIRSPec all suffer from 1/f noise -- correlated read noise with a power spectrum that increases with decreasing frequency. During the non-destructive readout of the ramp, pixels are digitised at a rate of 10 $\mu$s per pixel and, in the case of NIRISS, read column-by-column (fast scan direction), with a small pause between columns. It is to this time series that 1/f noise is added, which means pixels along a particular column have an almost constant offset added since the 1/f noise power is concentrated at low frequencies. As a result, the 1/f noise induces correlated noise among pixels read close in time, leading to the characteristic horizontal bands observed with HgCdTe detectors. Reference pixels at the boundary of the detector are used to sample and correct for offsets due to 1/f noise and other sources; however, this may not be sufficient, in which case further mitigation strategies are needed. If left uncorrected, we thus expect different spectral channels to share correlated noise. See \cite{schlawin_jwst_2020} for more information about 1/f noise and the JWST noise floor.

For NIRISS SOSS, correcting for residual 1/f noise is non-trivial given the lack of unilluminated regions that could be used as additional reference pixels. However, since we have extracted the spatial profiles, we can estimate residual background flux even in regions with non-zero target flux if we simultaneously fit for the background and the flux of the different orders. In section \ref{sec:extraction}, we build a linear model to do precisely this, thus addressing two problems at once. 

\subsection{Spectrum extraction} \label{sec:extraction}

\begin{figure}
	\includegraphics[width=\columnwidth]{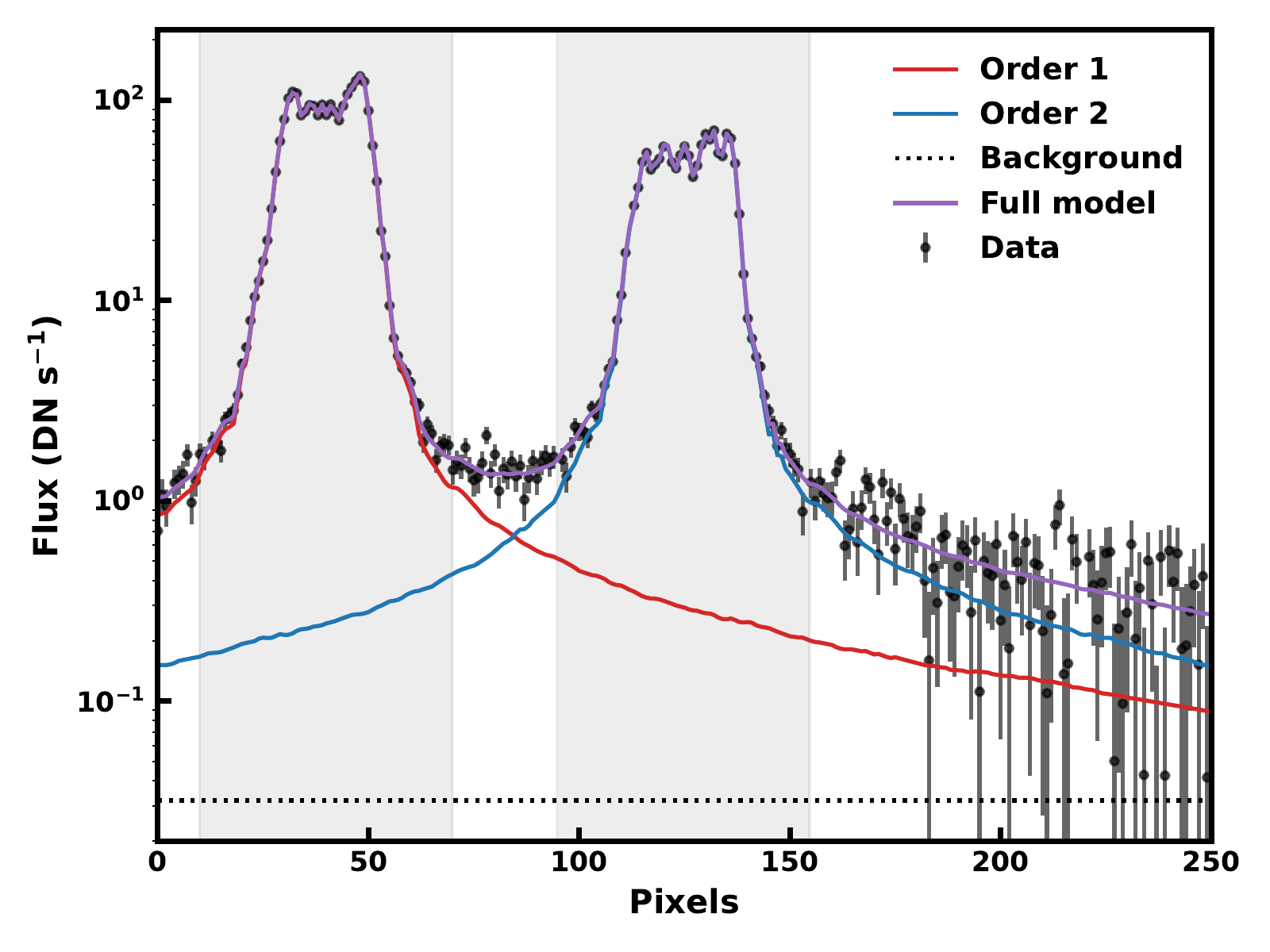}
    \caption{Model fit of a detector column of an integration of WASP-96 b, corresponding to 1.5 $\mu$m and 0.8 $\mu$m for orders 1 and 2, respectively. The full model is shown in purple, which is in good agreement with the data, even outside the extraction regions (shaded grey). In addition to the two spectral orders we also fit for a constant background level (dotted black line) to mitigate 1/f noise.}
    \label{fig:extraction_column}
\end{figure}

\begin{figure}
	\includegraphics[width=\columnwidth]{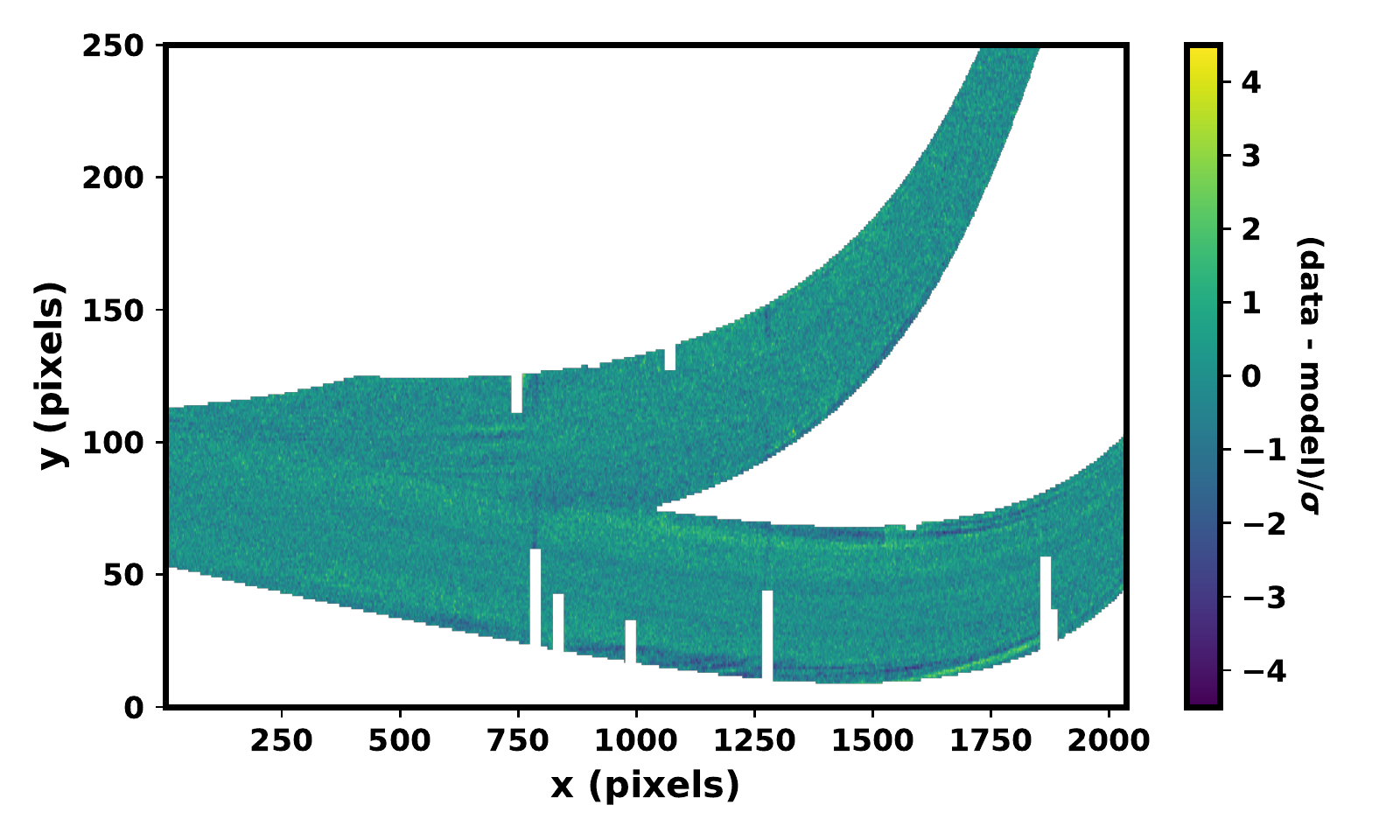}
    \caption{Normalised residuals after removing the extracted model spectra, the background, and the dispersed field star contamination, from one integration of WASP-96 b. Masked regions due to 0th-order contamination are shown in white. For clarity, we do not show rejected bad pixels and cosmic ray hits (amounting to $\sim 2 \%$ of pixels). We note that no large residuals can be seen and that the reduced $\chi^2$ is close to one.}
    \label{fig:extraction_residuals}
\end{figure}

\begin{figure}
	\includegraphics[width=\columnwidth]{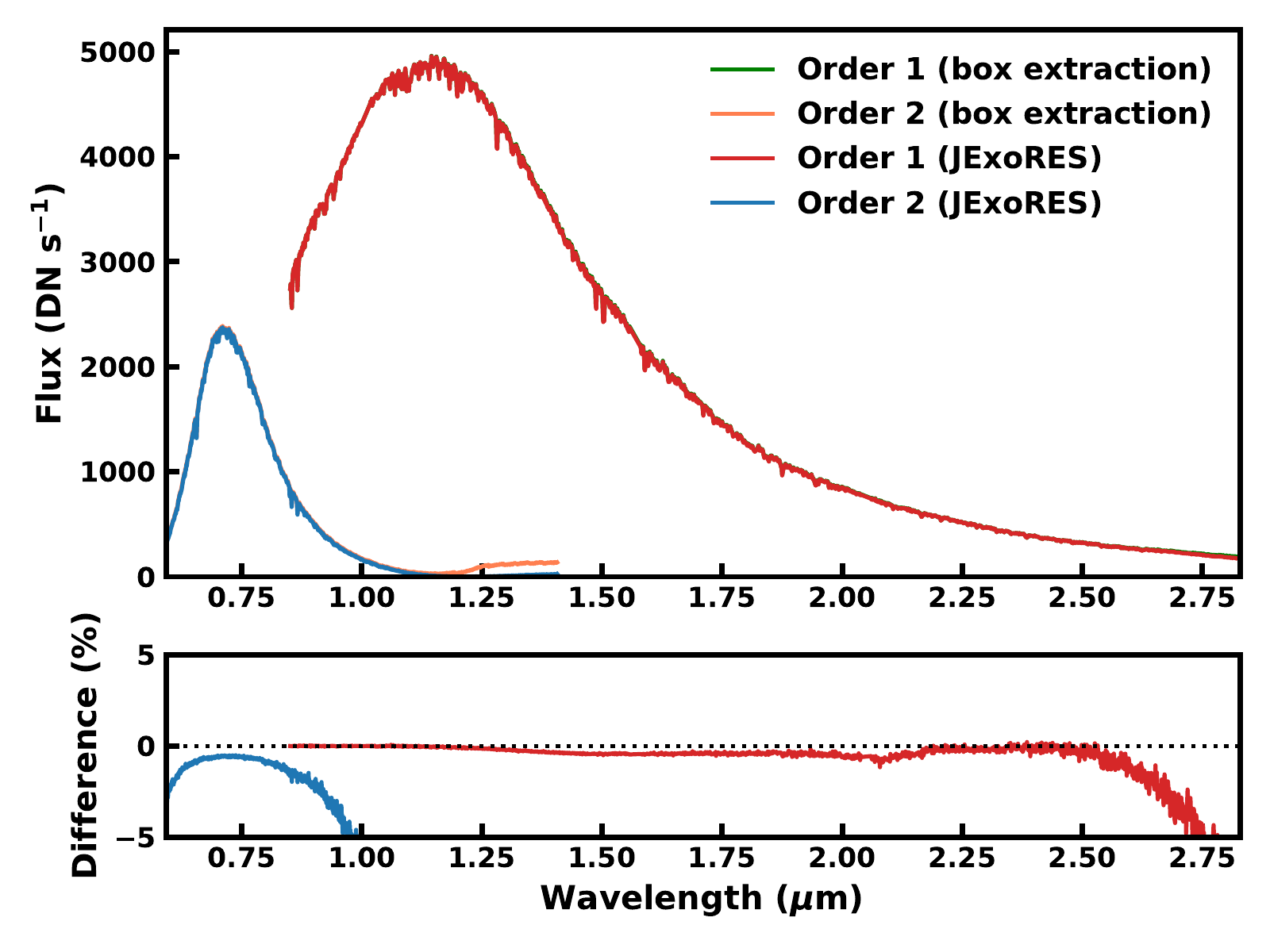}
 \caption{An extracted spectrum of WASP-96, comparing JExoRES to a simple box extraction (40-pixel aperture). The lower panel shows the relative difference between the two extraction methods ($(F_\text{optimal}-F_\text{box})/F_\text{optimal}$), illustrating that the contaminating level of the stellar flux from adjacent spectral orders can reach several per cent. The red end of both orders shows contamination due to the traces being close/overlapping, while the blue end of order 2 shows contamination due to a combination of low flux and the bright wings from order 1 (which is near the blaze peak in this region).}
    \label{fig:extracted_spectrum}
\end{figure}

In order to extract the spectrum we build a linear model of the flux map at the detector level which we fit to the data. Our method is a multi-order variant of the commonly used approach of optimal extraction of spectroscopic data \citep{horne_optimal_1986}. Like the classical optimal extraction method, JExoRES relies on the spatial profiles to extract the spectrum. However, unlike the ATOCA algorithm \cite{darveau-bernier_atoca_2022}, JExoRES makes no assumptions about the wavelength solution, the spectral resolution kernels, or the throughput function to perform the spectrum extraction. To extract the spectrum from the different spectral orders we assume that there is no tilt in the wavelength solution, which for NIRISS SOSS is small\footnote{\label{foot:report}JWST Technical report JWST-STScI-008270, SM-12}, allowing us to model each detector column independently.

For each detector column, with pixel number labelled by $i$, we build a model $m_i$ as a linear combination of the spatial profiles determined in section \ref{sec:prof}:
\begin{equation} \label{eq:model}
     m_i = \sum_j P_{ij} f_j \,,
\end{equation}
where $f_j$ is the integrated flux of spectral order $j$ with profile $P_{ij}$. To solve for the fluxes we fit the model $m_i$ to the data $d_i$, for a given detector column, by minimising
\begin{equation}
     \chi^2 = \sum_i \left( \frac{d_i - m_i}{\sigma_i} \right)^2\,,
\end{equation}
with respect to $f_i$ and where $\sigma_i$ are the pixel uncertainties (assuming the noise is uncorrelated for now). Masked regions and bad pixels are set to have zero weight. We can rewrite this as an unweighted least-squares problem as follows 
\begin{equation} \label{eq:chi2}
     \chi^2 = \sum_i \left( d'_i - \sum_j P'_{ij} f_j \right)^2\,,
\end{equation}
where $d'_i = d_i / \sigma_i$ and $P'_{ij} = P_{ij} / \sigma_i$. We then solve for the fluxes by minimising \eqref{eq:chi2} using \texttt{scipy}'s least-squares solver \texttt{lstsq}, allowing us to optimally extract the fluxes of multiple overlapping spectral orders. In order to reject outliers not previously masked, we perform the extraction iteratively by removing the largest $>5\sigma$ outlier (by setting their weights to zero) each iteration until convergence. Typically only one iteration is needed since we have already masked most cosmic rays. Finally, we obtain the uncertainty estimate of the extracted fluxes from the diagonal entries of the covariance matrix $(\bm{P}'^T\bm{P}')^{-1}$ (the non-diagonal entries are all close to zero, even in the region where the two spectral orders partially overlap). An example of the result of the model fitting is illustrated in figure \ref{fig:extraction_column}. We note that at the blue end of the 1st order, we only extract one profile due to the absence of order 2. For a single profile (and no background flux), i.e. an isolated spectral order, our extraction is equivalent to the optimal extraction method by \cite{horne_optimal_1986}, commonly used in HST transmission spectroscopy studies \citep[e.g.,][]{sing_hubble_2008} and beyond; with solution
\begin{equation}
    f = \frac{\sum_i p_i d_i / \sigma_i^2}{\sum_i p_i^2/ \sigma_i^2}\,.
\end{equation}

To reduce the effect of 1/f noise and to refine the background model, we choose to also fit a background level to each detector column by including an additional constant profile in \eqref{eq:model}. This way we are able to self-consistently mitigate the effect of 1/f noise.

When performing the fit, we only include a 60-pixel aperture around the traces of the 1st and 2nd order. We also mask a 60-pixel aperture mask around the 3rd order (that we are not extracting) and the 0th-order field star contamination, with the mask described in section \ref{sec:cont_mask}. Given that the spatial profiles are interpolated in the regions with 0th-order field star contamination, we can attempt to reconstruct the true flux in these regions, provided enough pixels are unaffected. Wavelength channels with a mask covering more than 40 \% of the target flux, either from 0th-order field star contamination, bad pixels, or cosmic ray hits, are rejected from further analysis. Since the extraction of the two orders is coupled, we cannot mask regions in one order without it also affecting the extraction of the other order. Thus, if the mask covers more than 40 \% of the flux of a wavelength channel, we temporarily remove the mask such that the wing of the profile reaching the other order remains accurate. 

As can be seen in figure \ref{fig:extraction_column}, our empirically constructed model is able to accurately describe the data. We further illustrate this in figure \ref{fig:extraction_residuals} for the whole detector. The reduced $\chi^2$ is on average $1.25$ and $1.28$ for the observation of WASP-39 b and WASP-96 b, respectively. This result justifies the use of same average spatial profiles for all integrations (but different between targets), owning to the excellent stability delivered by JWST. Some structure is seen within the core of orders 1 and 2 in figure \ref{fig:extraction_residuals}, which is mainly due to minor variability of the PSF, but this is typically a small effect. 

In figure \ref{fig:extracted_spectrum} we compare an out-of-transit spectrum of WASP-96 as extracted using the optimal method described here and a simple box extraction with a 40 pixel aperture\footnote{To isolate the effect of contamination from adjacent spectral orders we corrected bad pixels, subtracted the refined background before the box extraction.}. The optimally extracted spectrum is normalised such that the sum of the profiles within the same 40 pixel aperture is unity such that we can compare the two spectra directly. We find that the spectra are in very good agreement in regions of high flux. However, as can be seen in the lower panel of figure \ref{fig:extracted_spectrum}, we find that stellar flux can be contaminated by the adjacent order by several per cent, especially at the red end of each order. In section \ref{sec:transmission_spectroscopy}, we compare the resulting transmission spectra from both extraction methods and find that for both WASP-96 b and WASP-39 b, the difference is small.

\section{Modelling field star spectra} \label{sec:contamination}

Compared to 0th-order field star contamination that we can afford to mask due to their small footprint on the detector, higher-order field star contamination in the form of additional spectral traces is less forgiving in terms of their extent. Both the observations of WASP-39~b and WASP-96~b suffer from this type of contamination. By modelling the stellar spectrum of these additional spectral traces with the same spatial profiles derived for the main target, but shifted, we are able to assess the level of contamination and correct for it. To do this, we must first derive the relative throughput function to map the stellar model flux onto the detector.

\subsection{Throughput function} \label{sec:throughput}

We use the TSO of BD+60 1753 (program \#1091), a bright well-known A1V star ($J = 9.6$), to derive the throughput function. BD+60 1753 was observed for 5.4 hours on June 5 2022 with the CLEAR filter, consisting of 876 integrations (3 groups per integration), and with the F277W filter for an additional 40 integrations. The data reduction and spectrum extraction was performed in accordance with section \ref{sec:data_reduction}. 

To estimate a relative throughput function for our extraction method, we use the stellar model of BD+60 1753 provided by the CALSPEC catalog\footnote{\url{https://archive.stsci.edu/hlsps/reference-atlases/cdbs/current_calspec/}} \citep{bohlin_techniques_2014}, similar to the NIRISS team\footnote{JWST Technical report JWST-STScI-008270, SM-12}. We convolve this model with the average resolution of each spectral order and interpolate onto the NIRISS SOSS wavelength grid, before dividing the observed spectrum with the model spectrum (both in units proportional to number of photons per second per $\mu$m). Because we find some outliers in strong absorption lines, we mask these regions and interpolate the throughput. Finally, the throughput function was smoothed using a Gaussian filter.

\begin{figure}
	\includegraphics[width=\columnwidth]{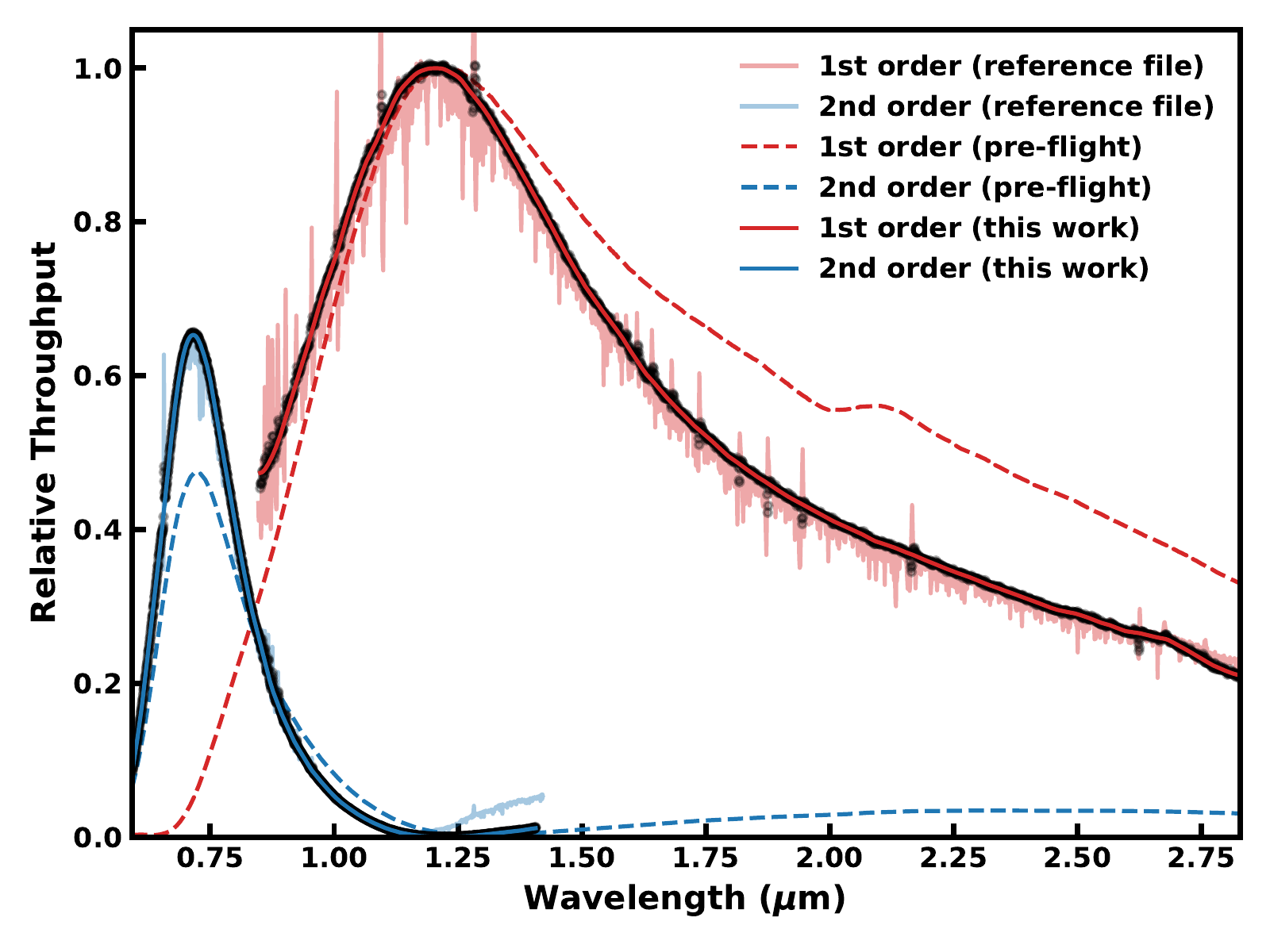}
    \caption{Relative throughput function of the 1st and 2nd order of NIRISS SOSS, comparing pre-flight expectations to the in-flight throughput derived using the standard star BD+60 1753. The throughput of the two orders is shown in red and blue for order 1 and 2, respectively. The pre-flight throughput is shown in the dashed curves (NIRISS team; see section~\ref{sec:throughput}). The throughput reference reported by the NIRISS team is shown in the light solid curves (jwst\_niriss\_photom\_0034.fits). The throughput derived in this work is shown in the dark solid curves. The grey points show our derived throughput function at the pixel-level without smoothing.}
    \label{fig:throughput} 
\end{figure}

The relative throughput function of NIRISS SOSS is shown in figure \ref{fig:throughput}, where we compared the throughput function obtained here to the reference file jwst\_niriss\_photom\_0034.fits (NIRISS team), and the  pre-flight expectation\footnote{\url{http://jwst.astro.umontreal.ca/wp-content/uploads/NIRISS_Throughput_SOSS_webpage.txt}} (NIRISS team). We note that the in-flight throughput function is somewhat different from pre-flight expectations, in particular for order 2, where the throughput is higher than expected. Furthermore, the reference file jwst\_niriss\_photom\_0034.fits differs from this work at the red end of the 2nd order, as they were unable deblend the two orders in this region\footnote{JWST Technical report JWST-STScI-008270, SM-12}.

\subsection{Stellar modelling}

\begin{figure}
	\includegraphics[width=\columnwidth]{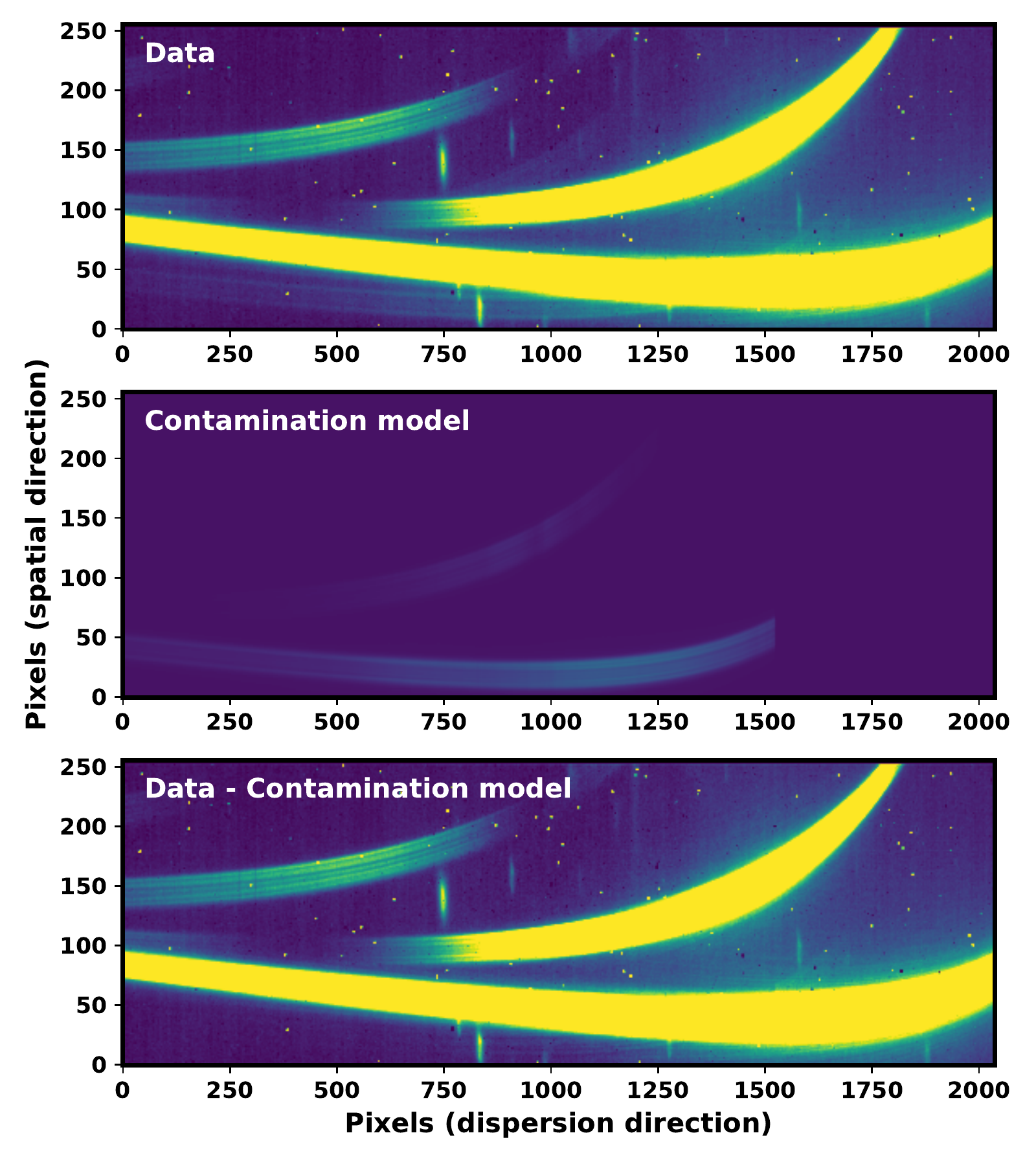}
    \caption{Field star spectrum modelling. The top panel shows an integration from the observation of WASP-96 b. Apart from WASP-96, two additional spectral traces are detected, which come from a $G = 19.0$ magnitude dispersed field star. The middle panel shows the best fit model of the field star spectrum, and the bottom panel shows the contamination subtracted data.}
    \label{fig:cont_model_WASP_96}
\end{figure}

As described in section \ref{sec:cont_mask}, we use GAIA to identify field stars down to $G \sim 21$. With this method, we identified two $G \sim 19$ magnitude stars affecting the considered observations, one per target. We also identified a fainter field star trace overlapping with the 2nd order of WASP-39, visible in the bottom panel of figure \ref{fig:f277w}. Once we have identified the field star spectra, we can move on to model their contribution. This is challenging given that we do not have access to the field star spectra in the regions where they overlap with the target spectra. We approach this by fitting a stellar atmospheric model to the data, away from target traces, allowing us to fill in the missing regions with the model. The fit is performed on the average of all integrations to boost the S/N of the faint field star spectra. To model the spectral traces we employ the same linear model used to extract the target spectrum, but now we allow the spatial profiles to be translated on the detector. However, instead of extracting the field star spectrum, we model the flux directly on the detector by adopting the stellar atmospheric models from the ATLAS9 catalogue \citep{castelli_new_2_2003} and the NIRISS SOSS throughput function. For each of the spectra in the ATLAS9 grid, limited to $[\text{M}/\text{H}] \in [ -1.0\,, +0.5]$, we fit a scaling parameter for the flux and two translation parameters to refine the position of the field star trace. We then selected the model that minimised the $\chi^2$. Before performing the fit, we reduced the effect of the bright wings from the target spectrum by subtracting the wing model we fitted to the data earlier, as described in section \ref{sec:prof}. During the fit, we masked regions further than $\pm$20 pixels away from the field star traces, together with the target traces, 0th-order contamination, and bad pixels. 

In figure \ref{fig:cont_model_WASP_96}, we show the result of the contamination model fitting for the observation of WASP-96 b. The best fit model indicates that the field star in question is an M dwarf with $T_\text{eff} \approx 3750$ K ($[\text{M}/\text{H}] \approx 0.0$, $\log (g) \approx 5.0$). Note that this is just the best fit and that we have not constrained the stellar parameters. To assess the plausibility of this result we compare the $G_{\text{BP}} - G_{\text{PP}}$ colour magnitude of the stellar model to that provided by GAIA. Using the best fit stellar model, we compute the $G_{\text{BP}} - G_{\text{PP}}$ colour magnitude to be 1.89. This is consistent with the measured (average) colour of $G_{\text{BP}} - G_{\text{PP}} = 1.88 \pm 0.03$ in GAIA DR3 \citep{gaia_collaboration_gaia_2022}, thus making us confident about the fit. The two field stars in the observation of WASP-39 b are also found to be best explained by M-type stars. 

We find that the contamination level resulting from the overlap with field star spectra reaches up to 250 ppm for WASP-39 b and about 100 ppm WASP-96 b, ignoring the higher level of contamination at the red end of order 2 which we later mask. We discuss the effect of contamination on the spectrum in section \ref{sec:transmission_spectroscopy}. A limitation of our modelling is that we do not know the spatial profiles nor the throughput function in wavelength ranges that falls outside the detector. This is responsible for the sharp cutoff in the model trace in the middle panel of figure \ref{fig:cont_model_WASP_96}. However, at this point (around 1.3 $\mu$m for order 1) the contamination is small and decreasing (towards shorter wavelengths). After subtracting the contamination model from each integration we revise the spatial profiles before extracting the final spectrum. 

\section{Light-curve analysis} \label{sec:light_curve}

With the reduced time-series spectra at hand -- we move on to the light-curve analysis. We perform the light-curve analysis in two steps. First, we integrate the flux of both orders to produce two high S/N white-light curves, which we use to obtain the wavelength-independent orbital parameters. Second, using these orbital parameters, we infer the transit depth as a function of wavelength from the spectroscopic light curves. This two-step process is common practice in transmission spectroscopy studies \citep[e.g.][]{nikolov_hubble_2014, kreidberg_clouds_2014}. To account for the wavelength-dependent limb darkening, we choose to fit the limb-darkening coefficients (LDCs) at low resolution, using binned-down light curves at $R = 15$, which we later interpolate onto the high-resolution grid. Below we describe the light-curve analysis in greater detail.

\subsection{White light curve fitting} \label{sec:wlc}

\begin{figure}
    \begin{subfigure}{\columnwidth}
    \includegraphics[width=\columnwidth]{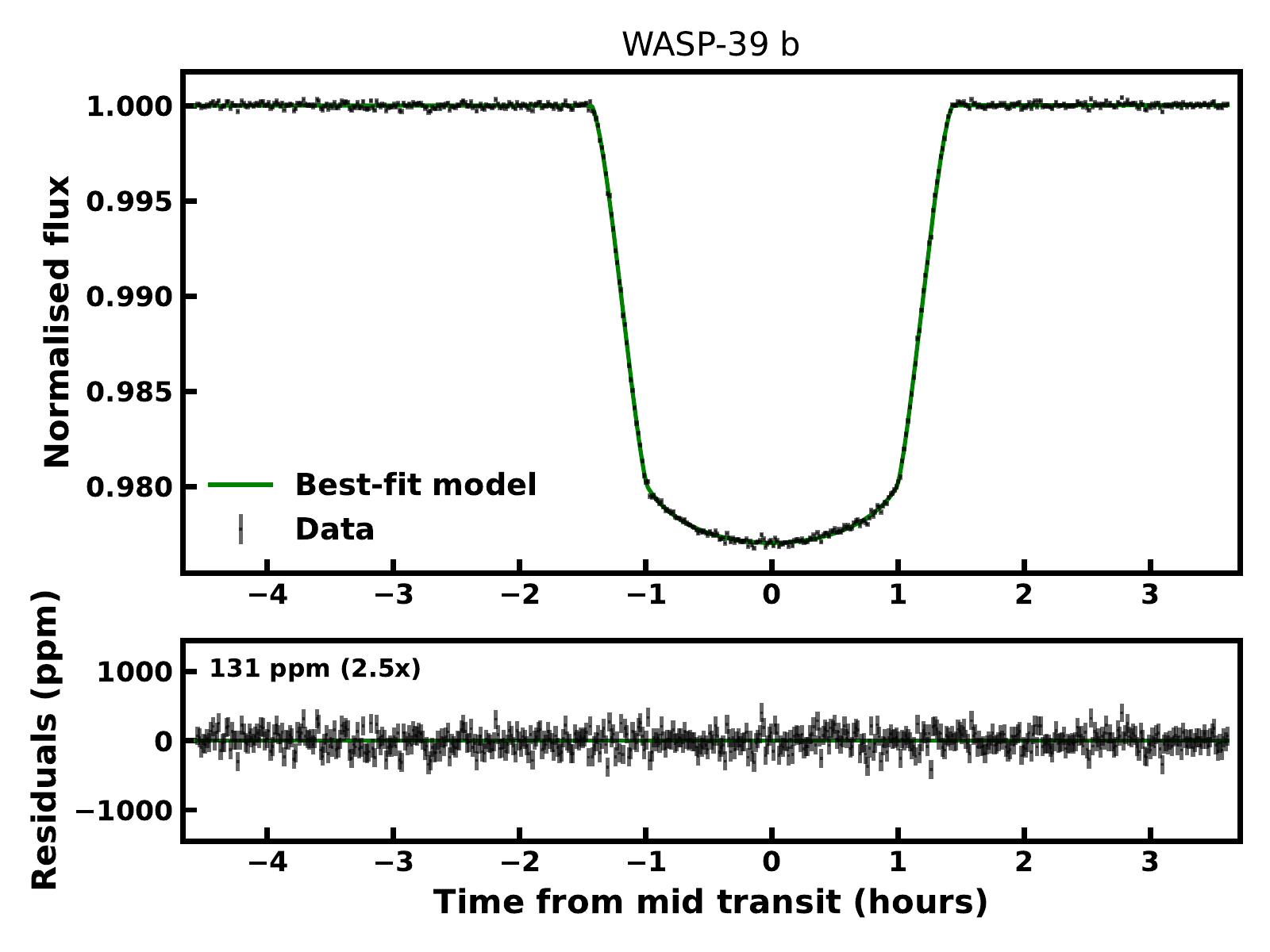}
    \end{subfigure}
    \begin{subfigure}{\columnwidth}
    \includegraphics[width=\columnwidth]{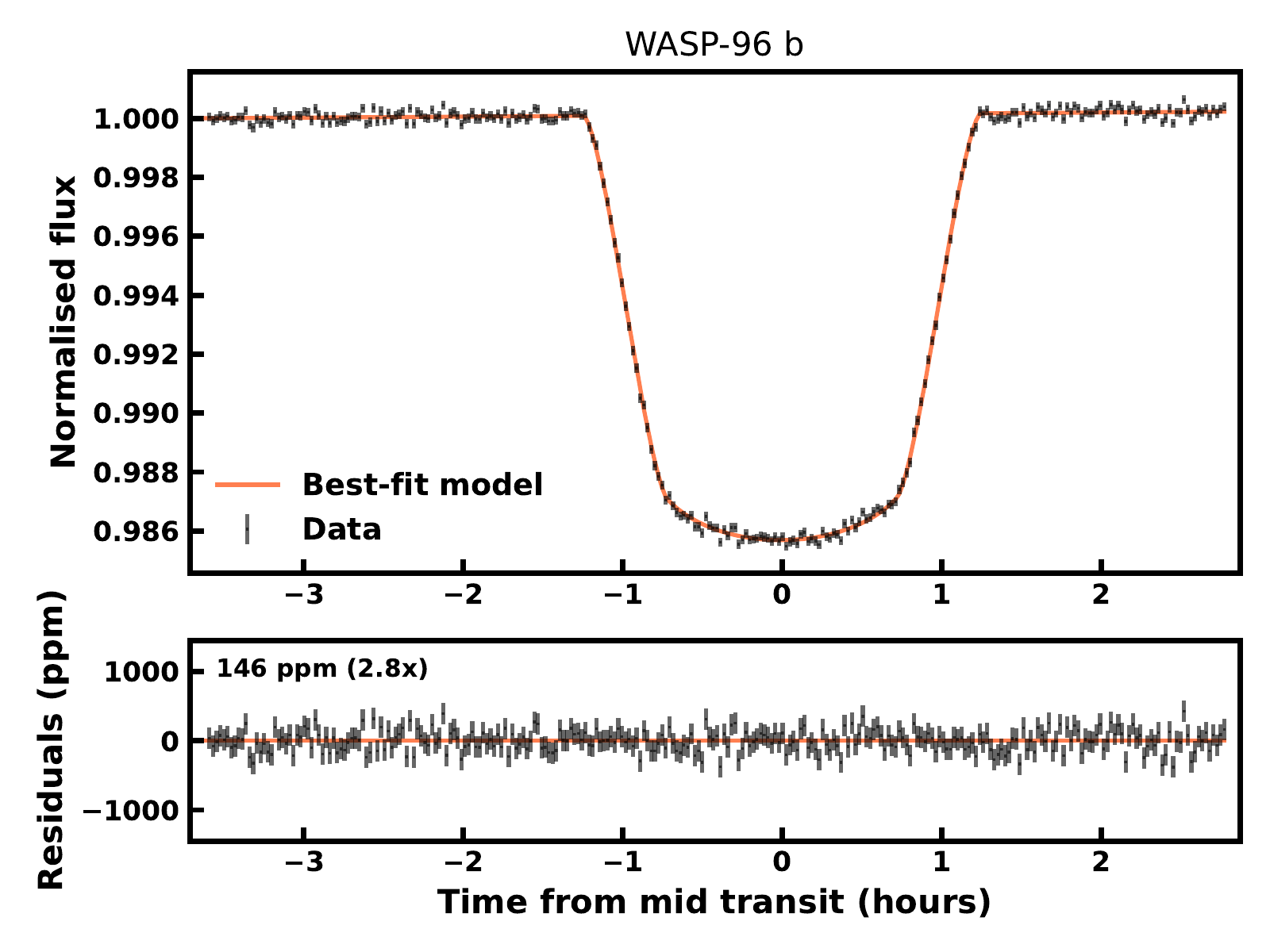}
\end{subfigure}

\caption{NIRISS SOSS white light curves of the 1st order for WASP-39~b and WASP-96~b, together with best-fit models. The value on each bottom panel is the RMS of the residuals in ppm (with native time resolution), where the value in the bracket is the ratio ($\beta$) of the RMS to the predicted noise level, assuming only white noise. We find that the light curves are free of large systematic trends.}
\label{fig:white_light_curves}
\end{figure}

We start by searching for additional outliers before making the white-light curves. To do this, we normalise each wavelength channel by dividing by the out-of-transit flux. We then identify pixels that deviate more than $4\sigma$ from the median light curve, of each order, and replace these by the linear interpolation of the neighbouring pixels in time. We only found a few outliers using this method, demonstrating that we rejected most bad pixels and cosmic ray hits during the data reduction. Finally, we created the two white-light curves by summing the flux of all unmasked wavelength channels for each spectral order (not including order 2 above 0.9 $\mu$m). We show the 1st order white light curves for the two targets in figure \ref{fig:white_light_curves}.

To model the transits light curves, we adopted the well-tested \texttt{batman} code \citep{kreidberg_batman_2015}. Given that we do not find any large systematic trends in the light curves of the observed targets, which is in line with previous JWST observations \citep[e.g.][]{jwst_transiting_exoplanet_community_early_release_science_team_identification_2022}, we only model a linear trend in the transits light curves as follows
\begin{equation} \label{eq:transit_model}
    F(t) = F_{\text{out}} (1 + \alpha (t - t_0)) T(t, \bm{\theta}) \,,
\end{equation}
where $F_{\text{out}}$ is the out-of-transit flux, $\alpha$ is the slope of the linear trend, $t_0$ is the time at the start of the observation, and $T(\bm{\theta})$ is the model describing the light curve; parameterised by $\bm{\theta} = (T_{\text{mid}}, P, i, a/R_*, R_\text{p}/R_*, \bm{u})$. Here, $T_{\text{mid}}$ is the mid-transit time, $P$ is the orbital period, $i$ is the orbital inclination, $a/R_*$ is the semi-major axis normalised by the stellar radius, $R_\text{p}/R_*$ is the planet-to-star radius ratio, and $\bm{u}$ is a vector containing the LDCs. The choice of detrending was driven by the observation of a small linear trend in the WASP-96 b white light curves, which may be connected to the linear drift in the trace positions on the detector, as shown in figure \ref{fig:trace_drift}. In both cases, we fixed the orbital period to the literature values given that we only observe one transit.

We adopted the quadratic limb-darkening model, described by two parameters $(u_1, u_2)$, as done in previous transmission spectroscopy studies of WASP-39~b and WASP-96~b \citep[e.g.][]{kirk_lrg-beasts_2019, nikolov_solar--supersolar_2022}. The motivation for using the quadratic limb-darkening law in contrast to more flexible models comes from the study of \cite{espinoza_limb_2016}, where they showed that the bias on the inferred transit parameters is small for systems similar to WASP-96 b and WASP-39 b when using the quadratic limb-darkening model.

To find the best-fit light curve models and to sample the posterior distributions of the fitted parameters we used the Markov Chain Monte Carlo (MCMC) routine implemented with the Python package \texttt{emcee} \citep{foreman-mackey_emcee_2013}. We choose to jointly fit the white light curves from both orders to account for correlated noise. To fit the quadratic LDCs we used the parameterization by \cite{kipping_efficient_2013} for efficient sampling, that is $q_1 = (u_1 + u_2)^2$ and $q_2 = u_1 / (2(u_1 + u_2))$. By then adopting uniform priors for $(q_1, q_2)$ over the interval $[0, 1]$, we uniformly sample only physically plausible values for $(u_1, u_2)$, allowing us to better constrain the LDCs. Uninformative priors were used for the other transit parameters. We initialise the MCMC using 100 walkers in a small volume near the maximum likelihood solution and run the chains until we reach 100 $\tau$ samples, where $\tau$ is the maximum integrated autocorrelation time. We chose a burn-in phase of 10 $\tau$ samples. Finally, we take the posterior medians to be the point estimates of the fitted parameters. 

To account for correlated noise among the light curves, from e.g. uncorrected 1/f noise, we fit the white light curves of the two spectral orders simultaneously using the following log-likelihood function
\begin{equation}
    \log P = \sum_k \log N(\bm{y}_k - \bm{F}(t_k), \bm{\Sigma}_k)\,,
\end{equation}
where $N(\bm{\mu}, \bm{\Sigma})$ denotes the two-dimensional normal distribution with mean $\bm{\mu}$ and covariance $\bm{\Sigma}$. For each integration, denoted by index $k$, the mean is given by the difference between the data $\bm{y}_k$ and the model $\bm{F}(t_k)$ (both of which are two-dimensional, corresponding to spectral order 1 and 2). Here we assume that the noise is not correlated in time. Next, we construct the covariance matrix of the kth integration as
\begin{equation}
\bm{\Sigma}_k = 
    \begin{pmatrix}
\beta_1^2 \sigma_{1,k}^2 & \rho \beta_1 \beta_2 \sigma_{1,k} \sigma_{2,k}\\
\rho \beta_1 \beta_2 \sigma_{1,k} \sigma_{2,k} & \beta_2^2 \sigma_{2,k}^2
\end{pmatrix}\,,
\end{equation}
where $\sigma_{1,k}$ and $\sigma_{2,k}$ are the corresponding uncertainties of the two light curves, and where we introduce the correlation coefficient $\rho$ and the parameters $\beta_1$ and $\beta_2$, to account for the uncertainties being different from expectations. By allowing these parameters to vary we can estimate the contribution of correlated noise to the overall noise budget, given that the pixel uncertainties obtained from the data reduction only include white noise sources. This way, we also make sure that the uncertainties of the derived parameters are not underestimated. 

\begin{table}
\caption{Parameters for WASP-39 b obtained from the joint white light curve analysis of order 1 and 2. The orbital period is held fixed and is taken from \protect\cite{mancini_gaps_2018}.}
\centering
\begin{tabular}{lc}
\hline \hline
Parameter & Value \\ \hline
Mid-transit time, T$_{\text{mid}}$ (BJD - 2400000.5) & $59787.056731_{-0.000016}^{+0.000016}$ \\ 
Inclination, $i$ ($^{\circ}$) & $87.735_{-0.032}^{+0.032}$ \\ 
Normalised semi-major axis, $a / R_*$ & $11.406_{-0.027}^{+0.026}$ \\ 
Orbital Period, $P$ (days) & 4.0552941 (fixed) \\
Eccentricity, $e$ & 0 (fixed) \\
Correlation coefficient, $\rho$ & $0.480_{-0.034}^{+0.033}$ \\
Order 1 \\
Planet-to-star radius ratio, $R_\text{p} / R_*$ & $0.14599_{-0.00020}^{+0.00020}$ \\ 
Linear LDC, $u_1$ & $0.222_{-0.023}^{+0.024}$ \\ 
Quadratic LDC, $u_2$ & $0.136_{-0.043}^{+0.042}$ \\ 
Slope, $\alpha$ (ppm/hour) & $3.1_{-2.6}^{+2.6}$ \\ 
Uncertainty re-scaling parameter, $\beta_1$ &  $2.508_{-0.076}^{+0.078}$ \\ 
Order 2 & \\
Planet-to-star radius ratio, $R_\text{p} / R_*$ & $0.14534_{-0.00024}^{+0.00025}$ \\ 
Linear LDC, $u_1$ & $0.373_{-0.028}^{+0.028}$ \\ 
Quadratic LDC, $u_2$ & $0.175_{-0.046}^{+0.046}$ \\ 
Slope, $\alpha$ (ppm/hour) & $-4.0_{-4.9}^{+4.9}$ \\ 
Uncertainty re-scaling parameter, $\beta_2$ & $2.211_{-0.067}^{+0.068}$ \\ \hline
\end{tabular}
\label{tab:parameters_WASP_39}
\end{table}

\begin{table}
\caption{Parameters for WASP-96 b obtained from the joint white light curve analysis of order 1 and 2. The orbital period is held fixed and is taken from \protect\cite{hellier_transiting_2014}.}
\centering
\begin{tabular}{lc}
\hline \hline
Parameter & Value \\ \hline
Mid-transit time, T$_{\text{mid}}$ (BJD - 2400000.5) & $59751.324674_{-0.000037}^{+0.000037}$ \\ 
Inclination, $i$ ($^{\circ}$) & $85.356_{-0.048}^{+0.049}$ \\ 
Normalised semi-major axis, $a / R_*$ & $8.988_{-0.045}^{+0.045}$ \\ 
Orbital Period, $P$ (days) & 3.4252602 (fixed) \\
Eccentricity, $e$ & 0 (fixed) \\
Correlation coefficient, $\rho$ & $0.497_{-0.047}^{+0.044}$ \\
Order 1 \\
Planet-to-star radius ratio, $R_\text{p} / R_*$ & $0.11930_{-0.00068}^{+0.00079}$ \\ 
First LDC, $c_1$ & $0.24_{-0.15}^{+0.16}$ \\ 
Second LDC, $c_2$ & $0.17_{-0.20}^{+0.19}$ \\ 
Slope, $\alpha$ (ppm/hour) & $34.7_{-5.2}^{+5.2}$ \\ 
Uncertainty re-scaling parameter, $\beta_1$ & $2.78_{-0.12}^{+0.12}$ \\ 
Order 2 & \\
Planet-to-star radius ratio, $R_\text{p} / R_*$ & $0.12108_{-0.00085}^{+0.00086}$ \\ 
First LDC, $c_1$ & $0.72_{-0.14}^{+0.13}$ \\ 
Second LDC, $c_2$ & $-0.17_{-0.17}^{+0.17}$ \\ 
Slope, $\alpha$ (ppm/hour) & $69.1_{-9.5}^{+9.4}$ \\ 
Uncertainty re-scaling parameter, $\beta_2$ & $2.62_{-0.11}^{+0.11}$ \\ \hline
\end{tabular}
\label{tab:parameters_WASP_96}
\end{table}

We summarise the result from the joint white light curve fitting of both orders for WASP-39 b in table \ref{tab:parameters_WASP_39}. We used a circular orbit with a fixed period of 4.0552941 days \citep{mancini_gaps_2018}. We find that the orbital parameters are mostly consistent with previous studies \citep{faedi_WASP-39b_2011, ricci_multifilter_2015, maciejewski_new_2016, fischer_hst_2016, nikolov_vlt_2016, mancini_gaps_2018, kirk_lrg-beasts_2019}, and agree to within 1$\sigma$ of the parameters from all pipelines in \cite{feinstein_early_2023}. We find little evidence of a linear trend in the white light curves given that the slope $\alpha$ is consistent with zero for both orders within $\sim 1 \sigma$. The scatter after subtracting the best-fit model is 131 ppm for order 1 and 257 ppm for order 2, which is a factor of $\sim 2.5$ times higher than the predicted noise level from the data reduction. This result is similar to the findings by \cite{espinoza_spectroscopic_2023}, who found that the white light curve scatter was about three times higher than predicted by the JWST pipeline for NIRSpec G395H data. On the other hand, we find that the scatter of the out-of-transit pixel-level light curves are about 1.1 and 1.2 times higher than expected for the 1st and 2nd order, respectively; which is not enough to explain the residual scatter of the white light curves after subtracting the models. We also find that the noise of the two white light curves is moderately correlated, with $\rho \sim 0.5$. A likely explanation for both phenomena is residual 1/f noise -- introducing correlations among pixels that are read close in time. Hence, 1/f noise would introduce correlations among the two spectral orders, given that they share the same detector columns, as well as among nearby wavelength channels, resulting in the errors being larger than expected when binning. We observe both of these effects.

We find that the best fit LDCs from the white light curve of WASP-39 b are close to the modelled LDCs computed with the Synthetic-Photometry/Atmosphere-Model (SPAM) algorithm \citep{howarth_stellar_2011}, which come out to be $u_1 = 0.238$ and $u_2 = 0.166$ for order 1, and $u_1 = 0.409$ and $u_2 = 0.133$ for order 2. We used ExoCTK \citep{bourque_exoplanet_2021} with a PHOENIX ACES stellar atmospheric model \citep{husser_new_2013} and the four-parameter non-linear limb-darkening law \citep{claret_new_2000} to compute the SPAM LDCs. For robustness we also performed the white light curve fitting of WASP-39 b using the logarithmic limb-darkening law (employing the Kipping-like priors by \cite{espinoza_limb_2016}) and found the parameters to be consistent within less than 1$\sigma$ of the quadratic limb-darkening results in table \ref{tab:parameters_WASP_39}.

Similarly, we provide the result from the joint white light curve fitting for WASP-96 b in table \ref{tab:parameters_WASP_96}. Again, we find that the system parameters are in agreement with \cite{hellier_transiting_2014, patel_empirical_2022}, and within $2\sigma$ of \cite{nikolov_absolute_2018}. As in the case of WASP-39 b, we used a circular orbit and fixed the period (to 3.4252602 days) \citep{hellier_transiting_2014}. We find that the derived LDCs for WASP-96 are consistent with that of WASP-39 to within 1$\sigma$ for order 1, which is what we expect given the similarity of the two stars, as outlined in table \ref{tab:stellar_parameters}. However, we find that the LDCs differs at the 2$\sigma$ level for order 2. We note that the uncertainties of the LDCs for WASP-96 are much larger than that of WASP-39, likely due to the difficulty of measuring accurate LDCs in systems with high impact parameters \citep{muller_high-precision_2013}, which for WASP-96 b is $b = 0.7277 \pm 0.0085$ (compared to $b = 0.4508 \pm 0.0065$ for WASP-39 b). After subtracting the best-fit model the residual scatter is 146 ppm and 293 ppm for order 1 and 2, respectively. In contrast to WASP-39 b, we find a linear trend in the white light curves of both orders for WASP-96 b, which may be related to the linear trend seen in the position of the traces, as shown in figure \ref{fig:trace_drift}. We again find $\beta \sim 2 - 3$ and $\rho \sim 0.5$, suggesting a common source responsible for the correlated noise.

\subsection{Spectroscopic light curves}

\begin{table}
\caption{Stellar parameters for WASP-39 and WASP-96}
\begin{tabular}{lcc}
\hline \hline
Property & WASP-39 $^{\mathrm{(a)}}$ & WASP-96 $^{\mathrm{(b)}}$ \\ \hline
H band magnitude & 10.3 & 10.9\\
$T_{\text{eff}}$ (K) & $5485 \pm 50$ & $5540 \pm 140$ \\
$[\text{M}/\text{H}]$ (dex) & $0.01 \pm 0.09$ & $0.14 \pm 0.19$ \\ 
$\log g$ ($\log_{10}$(cm s$^{-1}$)) & $4.41 \pm 0.15$ & $4.42 \pm 0.02$ \\ \hline
\end{tabular}
\newline
\footnotesize{\textbf{Notes}.
$^{(\mathrm{a})}$\protect\cite{mancini_gaps_2018}. }
$^{(\mathrm{b})}$\protect\cite{hellier_transiting_2014}. 
\label{tab:stellar_parameters}
\end{table}

\begin{figure}
    \begin{subfigure}{\columnwidth}
    \includegraphics[width=\columnwidth]{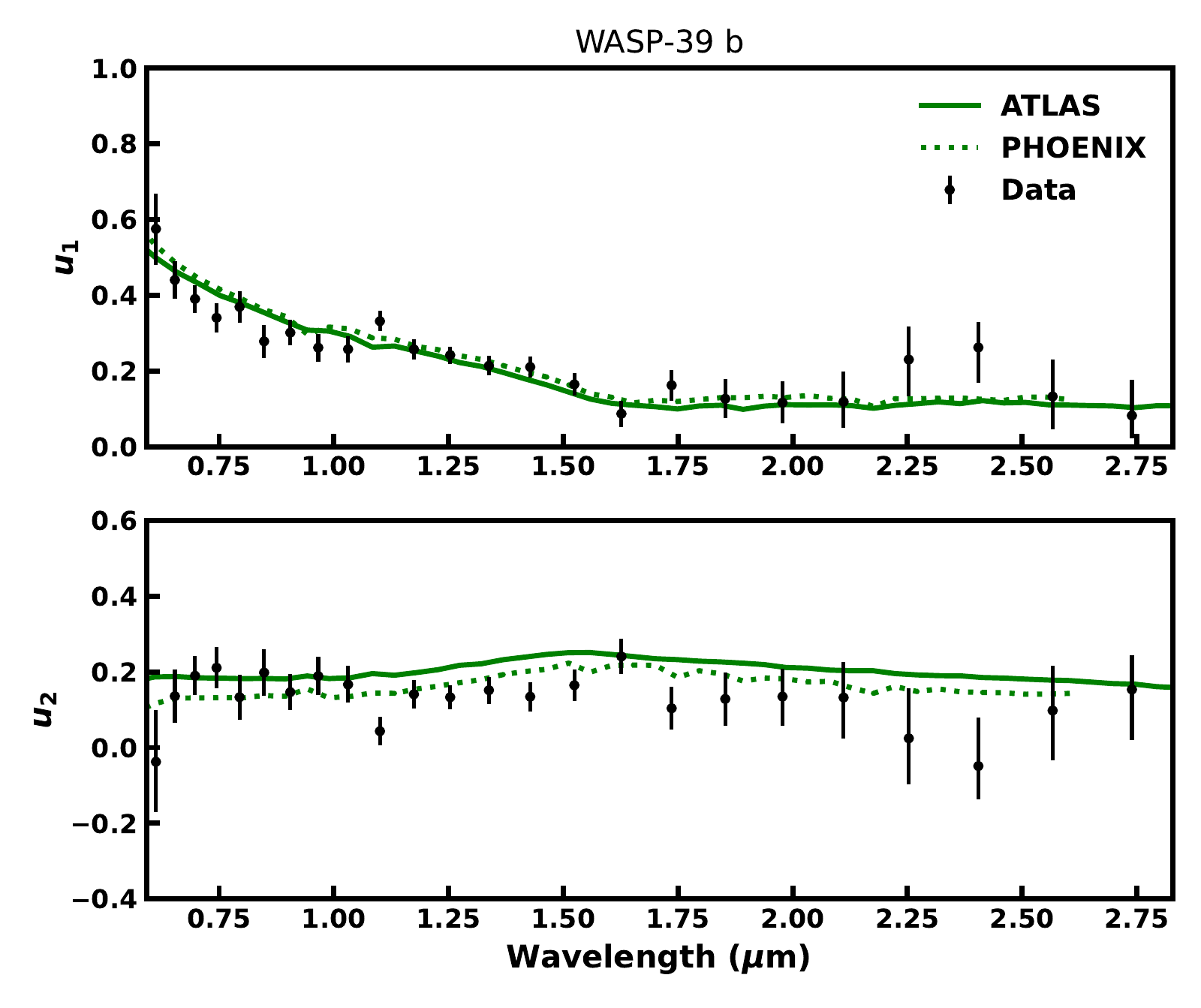}
    \end{subfigure}

    \begin{subfigure}{\columnwidth}
    \includegraphics[width=\columnwidth]{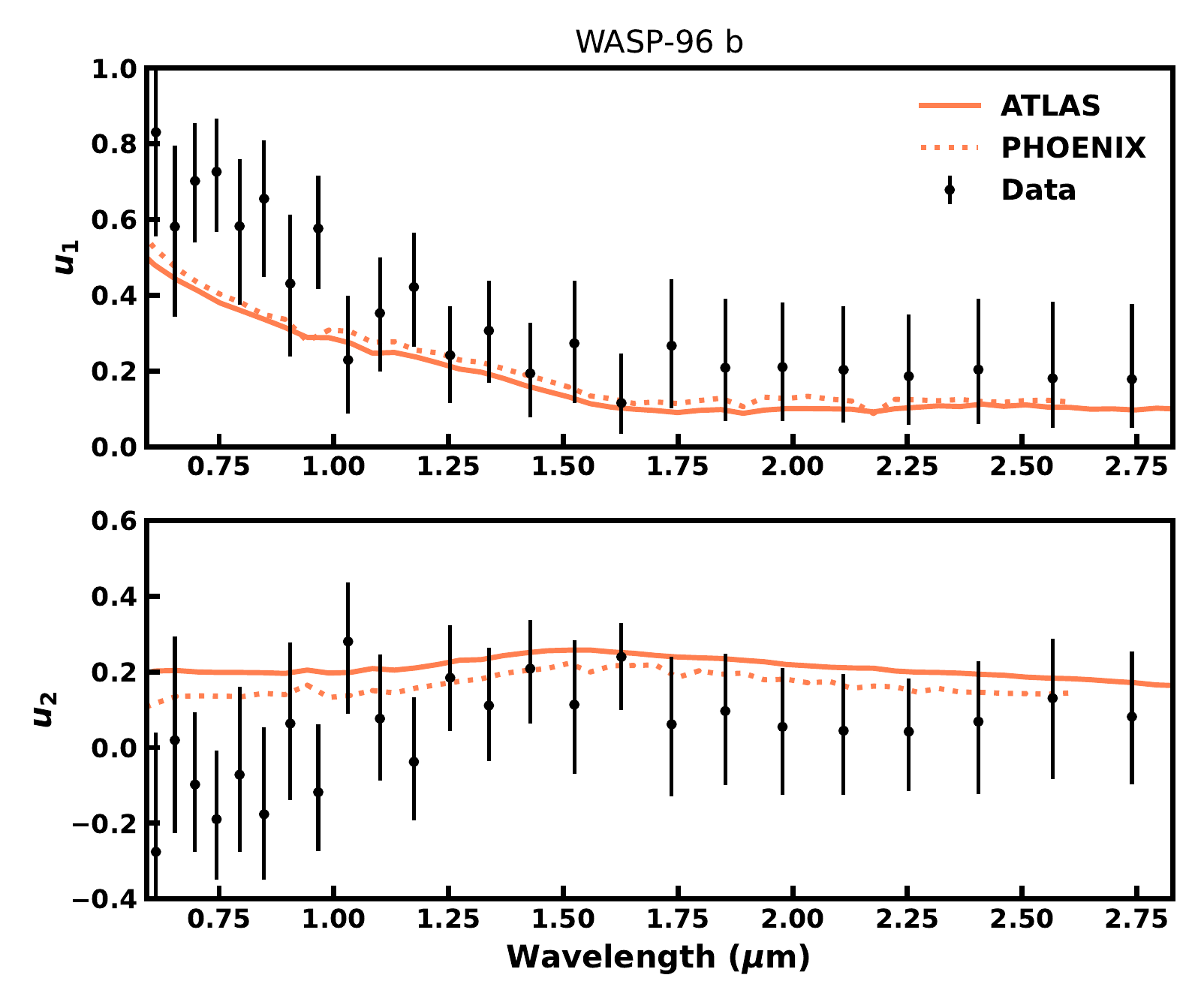}
\end{subfigure}

\caption{Quadratic LDCs, $u_1$ and $u_2$, for WASP-39 (top) and WASP-96 (bottom), as a function of wavelength. The black points show the fitted LDCs inferred from the data, binned at $R = 15$. For comparison, we show two sets of modelled LDCs, obtained with ExoCTK using ATLAS9 and Phoenix ACES stellar atmospheric models. The modelled LDCs were reprocessed according to the SPAM method for a like-for-like comparison.}
\label{fig:fitted_limb_darkening}
\end{figure}

Before we create the spectroscopic light curves, we first need to establish the level of binning, if any. Up front, we wish to keep the resolution as close as possible to native since we do not want to lose any information that could aid atmospheric retrievals. On the other hand, we risk biasing the transit depth due to contamination if not properly corrected or masked, especially in regions of low flux. With this in mind, and because the wavelength solution is only accurate to about a pixel, we select two nominal resolutions for this work, $R = 500$ and $R = 100$. We choose $R = 500$ in particular as this gives a minimum bin width of 2 pixels and since \cite{constantinou_characterizing_2022} found this resolution to provide atmospheric constraints approaching those obtained with native resolution (albeit for mini-Neptunes). We use $R = 100$ when comparing spectra with different data reduction strategies due to visual clarity. After binning the two orders onto a common wavelength grid, we combined the two orders through a weighted average at the light curve level. Note that we normalise the spectroscopic light curves by dividing the light curves with the out-of-transit flux before binning. Moreover, we masked the 2nd order above 0.9 $\mu$m to protect against contamination.

We fit the spectroscopic light curves individually, which amount to around 750 bins at $R = 500$, using the same method as for the white light curves, although now with $\Sigma_k= (\beta \sigma_k)^2$ for each wavelength channel. During the fit of the spectroscopic light curves, we let $R_p/R_*$, $F_{\text{out}}$, $\alpha$, and $\beta$ vary. We fixed the mid-transit time, inclination, and normalised semi-major axis to the values obtained from the white light curves. As for limb-darkening, we derived the wavelength-dependent LDCs from the data. To do this, we binned the light curves at $R = 15$ and combined the two orders at the light curve level before fitting the light curves with the priors by \cite{kipping_efficient_2013}. We show the derived quadratic LDCs in figure \ref{fig:fitted_limb_darkening}. We achieved an average uncertainty of the LDCs of 0.06 and 0.17 for WASP-39 and WASP-96 (at $R = 15$), respectively. The larger errors in the case of WASP-96 stem from the high impact parameter of WASP-96 b and the lower S/N of the dataset, making the LDCs harder to infer. The low resolution was motivated by the fact that modelled quadratic LDCs of WASP-39 and WASP-96 appear to vary smoothly with wavelength. Later, we also investigate the effect of fitting the LDCs directly at $R = 100$. 

As seen in figure \ref{fig:fitted_limb_darkening}, we find that in the case of WASP-39 b, the modelled SPAM LDCs are close, but appear not to be fully consistent, to the fitted LDCs. For WASP-96 b, the difference between the model and fitted LDCs is even larger, especially towards the optical. We also note that the two stellar atmospheric models differ in the predicted LDCs. We appreciate that to definitively say that the model LDCs are inconsistent with the data, we need to propagate the uncertainty from the system parameters and estimate the uncertainty of the model LDCs, for example by propagating the uncertainty of the stellar parameters, which is beyond the scope of this study. For the above reasons, and since the fitted LDCs for the two targets differ when we expect them to be close due to very similar stellar parameters, we emphasise the importance of fitting the LDCs to the data when possible, given that the transit depth can be highly dependent on the LDCs.

\subsection{Covariance estimation} \label{sec:covariance}

\begin{figure}
	\includegraphics[width=\columnwidth]{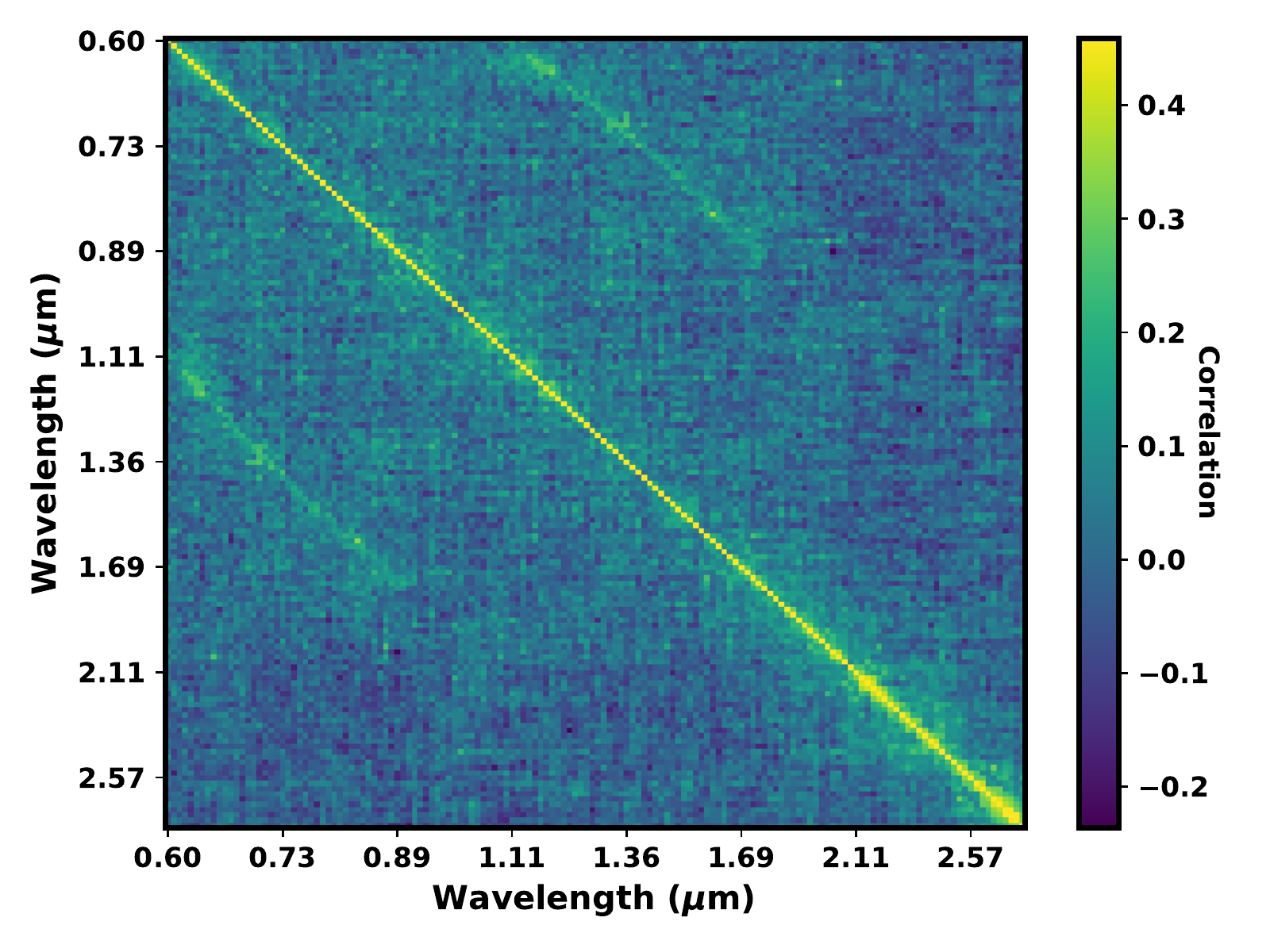}
    \caption{The empirical correlation matrix for the spectrum of WASP-96 b, calculated at $R = 100$. We find two distinct patterns: first, that nearby spectral channels are correlated and second, that a region between $0.6 - 0.9$ $\mu$m is correlated with another at $1.1 - 1.8$ $\mu$m. The latter finding indicates that light curves from the 1st and 2nd orders are correlated when they are extracted from nearby (or the same) detector columns.}
    \label{fig:corr}
\end{figure}

\begin{figure}
	\includegraphics[width=\columnwidth]{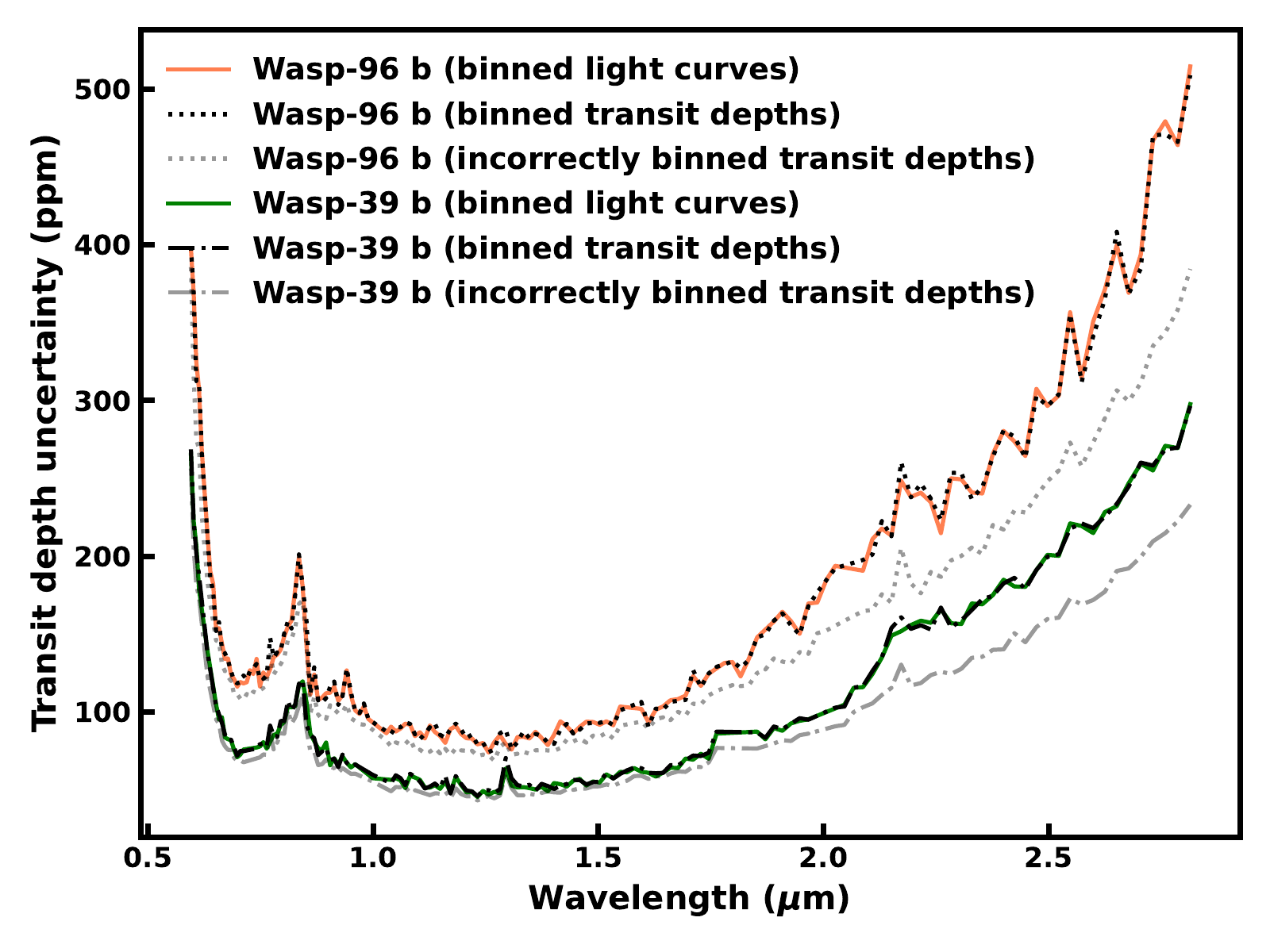}
    \caption{Transit depth uncertainties for the spectra of WASP-39~b and WASP-96~b, at $R = 100$, when fixing the system parameters and the LDCs. We compare the uncertainties binned at the light curve level (orange and green) to the uncertainties initially binned at $R = 500$ and then further binned to $R = 100$ after the light curve fitting (black). The latter strategy took into account the covariance of the data points -- otherwise, the binned uncertainties would have been underestimated (grey). Note that these uncertainties only make sense in isolation given that the spectral channels, both before and after binning, are correlated.}
    \label{fig:spectrum_uncertainty}
\end{figure}

We know from the white light curve analysis that different spectral channels share correlated noise -- something we are unable to probe directly by fitting the spectroscopic light curves individually. Instead of performing a computationally expensive joint fit of all spectroscopic light curves, we go on to show that correlated noise among the light curves carries over to the parameters derived from these light curves, allowing us to estimate the covariance matrix of our spectra. 

Assume we have $m$ observed light curves, corresponding to different spectral channels, each with $n$ integrations. The noise in these light curves is assumed to be Gaussian with a standard deviation that is constant in time. These light curves share correlated noise such that at each integration, the covariance matrix is $\bm{\Sigma}$. Given this setup, we can show that the covariance of the transit depths derived from these light curves is proportional to $\bm{\Sigma}$, assuming uninformative (Gaussian) priors and fixed system parameters. Let the observed light curves be described by the $n\times m$ matrix $\bm y = \bm F\,+\, \mathrm{noise}$, where $F_{ij} = F(t_i, \bm x_{\lambda_j})$ is the model from \eqref{eq:model} with the wavelength-dependent parameter vector $\bm x_{\lambda_j}$ of length $l$. To describe the noise with a multivariate Gaussian distribution we reduce the dimension by one by considering the vectorizing of the transpose of $\bm y$, that is $\text{vec}(\bm{y}^T) = (y_{t_1, \lambda_1}, y_{t_1, \lambda_2}, \ldots, y_{t_1, \lambda_m}, y_{t_2, \lambda_1}, \ldots)^T$. It follows that
\begin{equation} \label{eq:vec_y}
    \text{vec}(\bm{y}^T) = \text{vec}(\bm{F}^T) + \varepsilon\,,
\end{equation}
where $\bm{\varepsilon} \sim N(0, \mathbb{I}_n \otimes \bm \Sigma)$, and $\otimes$ is the Kronecker product. Next, we linearize $\bm{F}$ around the average best-fit parameters $\bm{\bar x} = \sum_i \bm x_{\lambda_i} / m$, such that $\bm{F} \approx \bm{J} \bm \phi + \ldots$ (ignoring constant terms), where $J_{ij} = \partial F(t_i, \bm{x}) / \partial x_j$ is the $n\times l$ Jacobian of the model evaluated at $\bm{\bar x}$ and $\phi_{ij} = x_{i,\lambda_j}$ are the entries of the $l\times m$ parameter matrix. We can thus approximate \eqref{eq:vec_y} as
\begin{equation} \label{eq:linear}
    \text{vec}(\bm{y}^T) \approx (\bm{J} \otimes \mathbb{I}_m) \text{vec}(\bm{\phi}^T) + \varepsilon + \ldots\,,
\end{equation}
where $\text{vec}(\bm{\phi}^T) = (x_{1, \lambda_1}, x_{1, \lambda_2}, \ldots, x_{1, \lambda_m}, x_{2, \lambda_1}, \ldots)^T$. Recognising \eqref{eq:linear} as a linear inverse problem, we write the inverse covariance of $\text{vec}(\bm{\phi}^T)$ as
\begin{equation}
\begin{aligned}
   \bm{\Sigma}_{\text{vec}(\bm{\phi}^T)}^{-1} &= (\bm{J} \otimes \mathbb{I}_m)^T \bm{\Sigma}_{\bm{\varepsilon}}^{-1} (\bm{J} \otimes \mathbb{I}_m) + \bm{\Sigma}_{\mathrm{priors}}^{-1} \\
   &= (\bm{J}^T \bm{J}) \otimes \bm{\Sigma}^{-1} + \bm{\Sigma}_{\mathrm{priors}}^{-1}\,,
\end{aligned}
\end{equation}
where $\bm{\Sigma}_{\mathrm{priors}}$ describes the covariance of the (Gaussian) priors \citep{rodgers_inverse_2000}. We thus assume that the posterior probability distribution is also Gaussian. With uninformative priors and assuming that $\bm{J}^T \bm{J}$ is invertible, this simplifies to
\begin{equation} \label{eq:cov_relation}
   \bm{\Sigma}_{\text{vec}(\bm{\phi}^T)} = (\bm{J}^T \bm{J})^{-1} \otimes \bm{\Sigma}\,.
\end{equation}
Therefore, in the limit of uninformative priors and fixed system parameters, correlations among any spectroscopically resolved parameter $x_k$ carries over from the correlated noise among the spectroscopic light curves, since $\bm{\Sigma}_{x_k} = (\bm{J}^T \bm{J})^{-1}_{kk}\, \bm{\Sigma}$. And even with the use of informative priors, there is no reason for $\bm{\Sigma}_{x_k}$ to remain diagonal when $\bm{\Sigma}$ is not. It is beyond the scope of this work to estimate the covariance matrix in the more general case, e.g. with more realistic priors and when propagating the covariance coming from fitting the system parameters, as done in section \ref{sec:wlc} for the two white light curves. However, in principle, by propagating the covariance due to parameters shared among multiple spectral channels, such as the inclination or semi-major axis, the correlations in the final spectrum may be larger than derived here, depending on the priors.

Next, we test the assumption that \eqref{eq:transit_model} is approximately linear near $\bm{\bar x}$ (i.e. \eqref{eq:linear}) by comparing the uncertainty of the transit depths from the marginalised posterior distributions, produced by the MCMC sampling, to the uncertainty predicted by the diagonal entries of \eqref{eq:cov_relation}. Using the transmission spectra of WASP-39 b and WASP-96 b (obtained with fixed system parameters and LDCs), we find that the two methods produce uncertainties that agree to within 2 \% (at $1\sigma$), making us confident that \eqref{eq:cov_relation} is indeed a good approximation.

To estimate $\bm{\Sigma}$, we calculate the sample covariance using the residuals of the data and the best-fit model,
\begin{equation} \label{eq:sample_cov}
    \Sigma_{ij} = \frac{1}{n} \sum_k^n [y_{ki} - F(t_k, \bm{x}_{\lambda_i})] [y_{kj} - F(t_k, \bm{x}_{\lambda_j})]\,.
\end{equation}
Note that the sample covariance matrix is only invertible when the number of integrations $n$ is larger than the number of spectral channels $m$, which in our case is satisfied for the spectra binned at $R = 100$. We illustrate the correlation matrix in the case of WASP-96 b in figure \ref{fig:corr} (which is qualitatively the same as the correlation matrix for WASP-39 b). As suspected, we find that light curves of different spectral channels are correlated, which due to \eqref{eq:cov_relation} means that the derived spectrum shares the same correlations, assuming fixed system parameters and with uninformative priors. We can now directly see why we found $\beta > 1$ and $\rho > 0$ in the white light curve analysis, given that we find that nearby spectral channels are indeed correlated and that wavelength regions in order 1 and 2 are correlated. Since both of these effects are expected from 1/f noise, and the fact that the level of correlated noise is much worse (both for neighbouring columns and between orders) if we skip the background refinement during spectrum extraction, we conclude that 1/f noise is the most likely explanation.

Another consequence of the fact that correlated noise carries over from the light curves to the derived transit depths is that the covariance must be taken into account when choosing to bin the transit depths instead of the light curves; otherwise, the uncertainties may be underestimated. This is in contrast to the proposed strategy by \cite{espinoza_spectroscopic_2023}, suggesting the transit depth to be binned in a post-processing stage to avoid increasing the uncertainties when binning at the light curve level without mentioning the covariance. This is a problem if the noise is correlated. Instead, when accounting for the covariance, we find that the transit depth uncertainties do not strongly depend on the order of the binning, that is, whether we perform the binning at the light curve level or afterwards on the derived transit depths. We demonstrate this finding in figure \ref{fig:spectrum_uncertainty}, where we show the uncertainties of the transmission spectra of both targets in figure \ref{fig:spectrum_uncertainty}, obtained by fixing the system parameters and LDCs. 

We find that the median transit-depth uncertainty is 78 ppm and 125 ppm for WASP-39 b and WASP-96 b at $R = 100$, respectively, assuming perfect knowledge of the system parameters and LDCs. 

The realisation that individual data points in the spectrum are not independent has implications for atmospheric retrievals, which may end up biasing the estimate of atmospheric parameters or overestimating the precision if the full covariance is not accounted for. As a worst case scenario, we illustrate the effect of correlated noise on the spectrum of WASP-39 b in figure \ref{fig:wasp_39_spectrum_corr}, resulting from not performing the background refinement step to correct for 1/f noise (as described in section \ref{sec:extraction}). Future studies can assess the robustness of atmospheric retrievals in the presence of correlated noise. Likewise, future reduction strategies may be able to reduce the amount of covariance further. For now, atmospheric retrievals with JWST/NIRISS SOSS observations should be conducted with caution.

\section{Factors affecting the transmission spectrum} \label{sec:transmission_spectroscopy}

In this section we use the JWST/NIRISS SOSS transmission spectra of WASP-39~b and WASP-96~b to evaluate the effects of limb-darkening, extraction method, and contamination on the resulting spectrum and its uncertainty. 

\subsection{Effects of limb darkening} \label{sec:LDCs_effects}

\begin{figure}
	\includegraphics[width=\columnwidth]{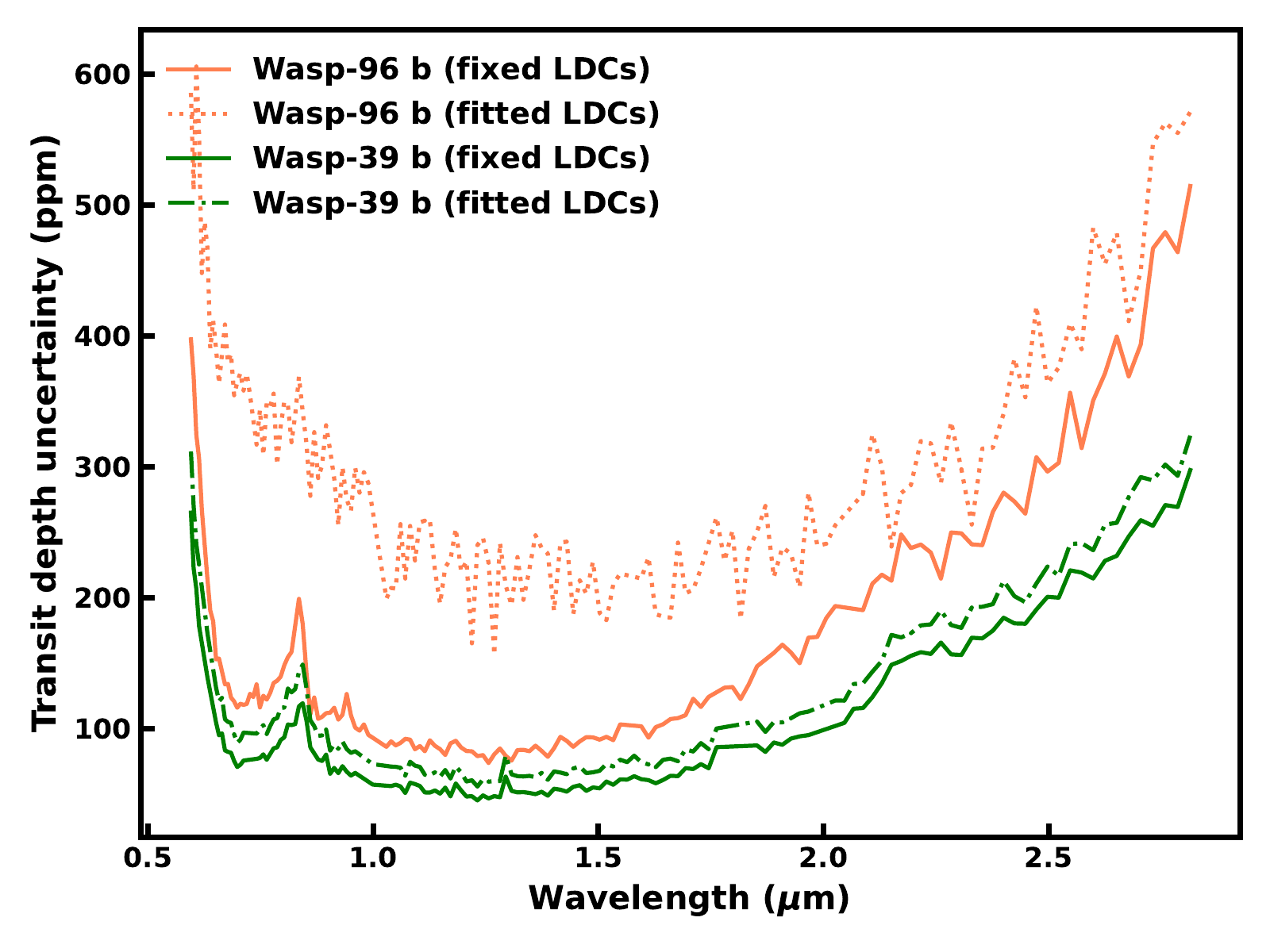}
    \caption{Transit depth uncertainties for the spectra of WASP-39~b and WASP-96~b at $R = 100$, comparing fixed LDCs to fitted LDCs. The large uncertainty for WASP-96 b when fitting the LDCs likely comes from the high impact parameter of the planet, making the LDCs hard to constrain.}
    \label{fig:error_LDCs}
\end{figure}

\begin{figure}
	\includegraphics[width=\columnwidth]{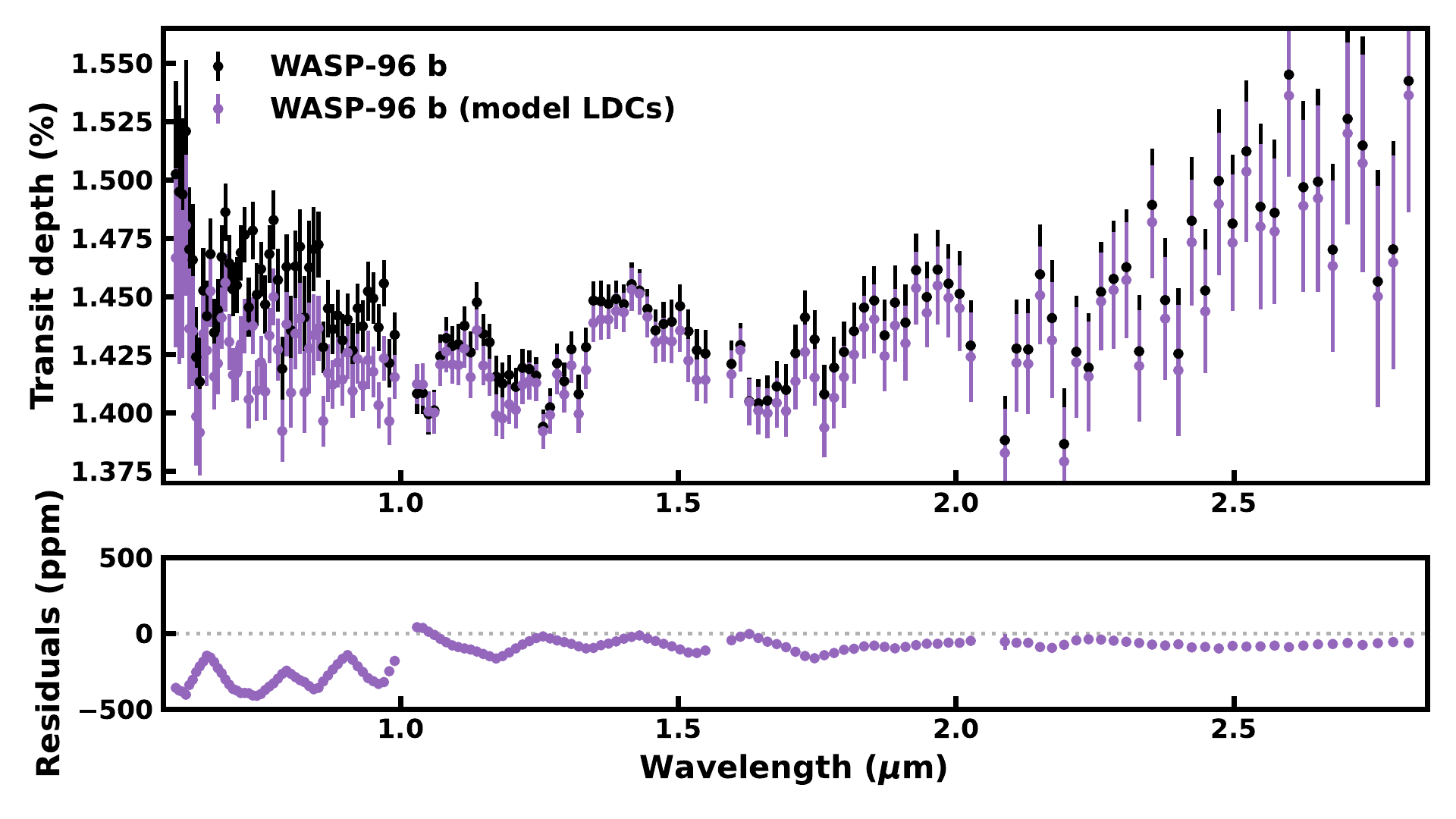}
    \caption{The effect of limb-darkening on the transmission spectrum of WASP-96 b, comparing the model LDCs from PHOENIX (purple) to the fitted LDCs at low resolution (black).} 
    \label{fig:spectrum_model_LDCs}
\end{figure}

We begin by analysing the effects of limb-darkening on the uncertainty of the transmission spectra of WASP-39~b and WASP-96~b. We investigate the increase in the transit depth uncertainty coming from fitting the LDCs, at $R = 100$ with the same priors as before, as compared to fixing them (to the derived values at $R = 15$). In figure \ref{fig:error_LDCs}, we illustrate the spectrum uncertainty coming from the two approaches for both WASP-39~b and WASP-96~b. We find that fitting the LDCs can significantly increase the spectrum uncertainty. In the case of WASP-96 b, being a planet with a high impact parameter ($b = 0.7277 \pm 0.0085$), the uncertainty is more than twice as large in some regions when fitting the LDCs compared to using fixed LDCs. Therefore, we conclude that the LDCs may be a limiting factor for the precision of the spectrum of WASP-96 b, especially since neither the modelled LDCs nor the LDCs from WASP-39 appear to match the fitted LDCs of WASP-96. In contrast, the uncertainty of the spectrum of WASP-39 b differs by 10 - 25 \% between the two approaches, meaning that the uncertainty coming from limb-darkening is much less of a problem for WASP-39 b.

To highlight the effect of different limb-darkening assumptions we show the spectrum of WASP-96 b using modelled LDCs in figure \ref{fig:spectrum_model_LDCs}. Here we used the same (SPAM reprocessed) PHOENIX LDCs as shown in figure \ref{fig:fitted_limb_darkening}, which we note differ from the fitted values. Because of this difference, the resulting spectrum is significantly altered\footnote{The jagged shape of the residuals is a result of the interpolation of the low-resolution fitted LDCs.}, especially towards the optical. Therefore, given the uncertainty of the LDCs, the true slope of the WASP-96 b transmission spectrum remains poorly constrained in the optical.

\subsection{Comparison of extraction methods}

\begin{figure}
    \begin{subfigure}{\columnwidth}
    \includegraphics[width=\columnwidth]{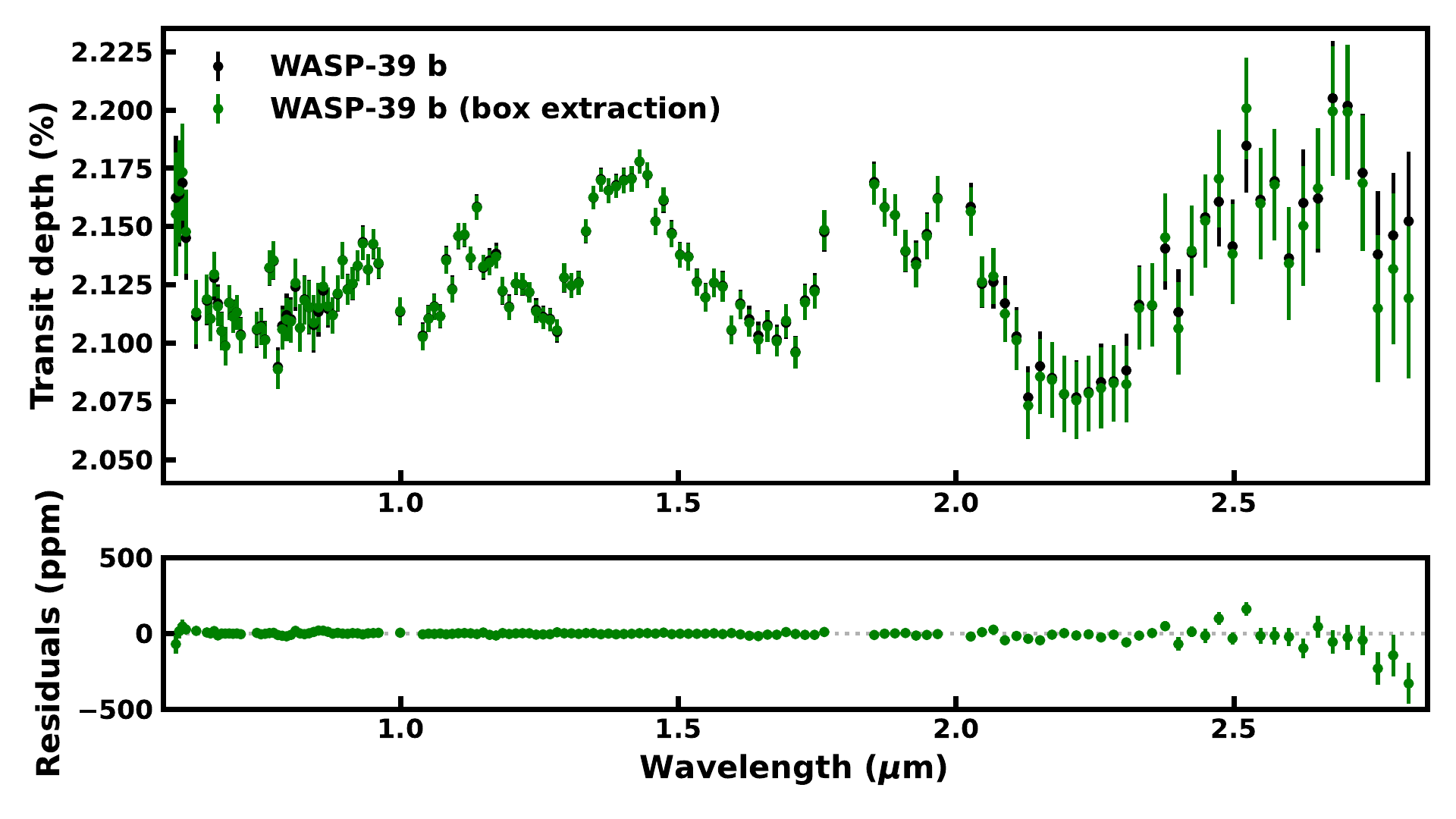}
    \end{subfigure}
    \begin{subfigure}{\columnwidth}
    \includegraphics[width=\columnwidth]{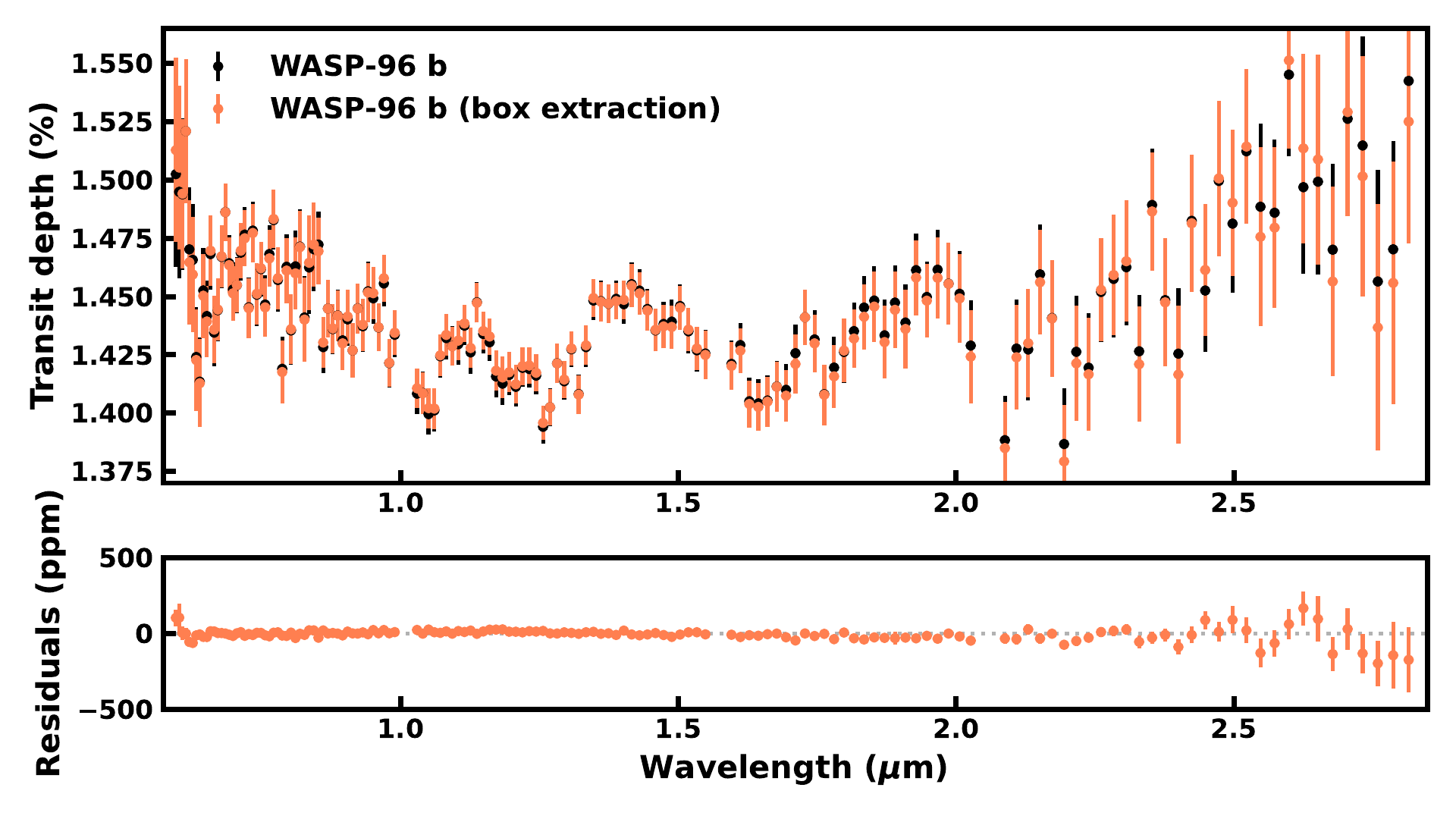}
\end{subfigure}
\caption{Transmission spectra of WASP-39~b and WASP-96~b at $R = 100$  -- comparing our extraction method to a box extraction (40-pixel aperture) while fixing the system parameters and the LDCs. The top panel show the spectrum of WASP-39~b, and the bottom panel shows the spectrum of WASP-96~b. We find no significant difference between the two extraction methods, indicating that the contamination due to overlapping spectral orders is small for these targets.}
\label{fig:spectra_box}
\end{figure}

\begin{figure}
    \begin{subfigure}{\columnwidth}
    \includegraphics[width=\columnwidth]{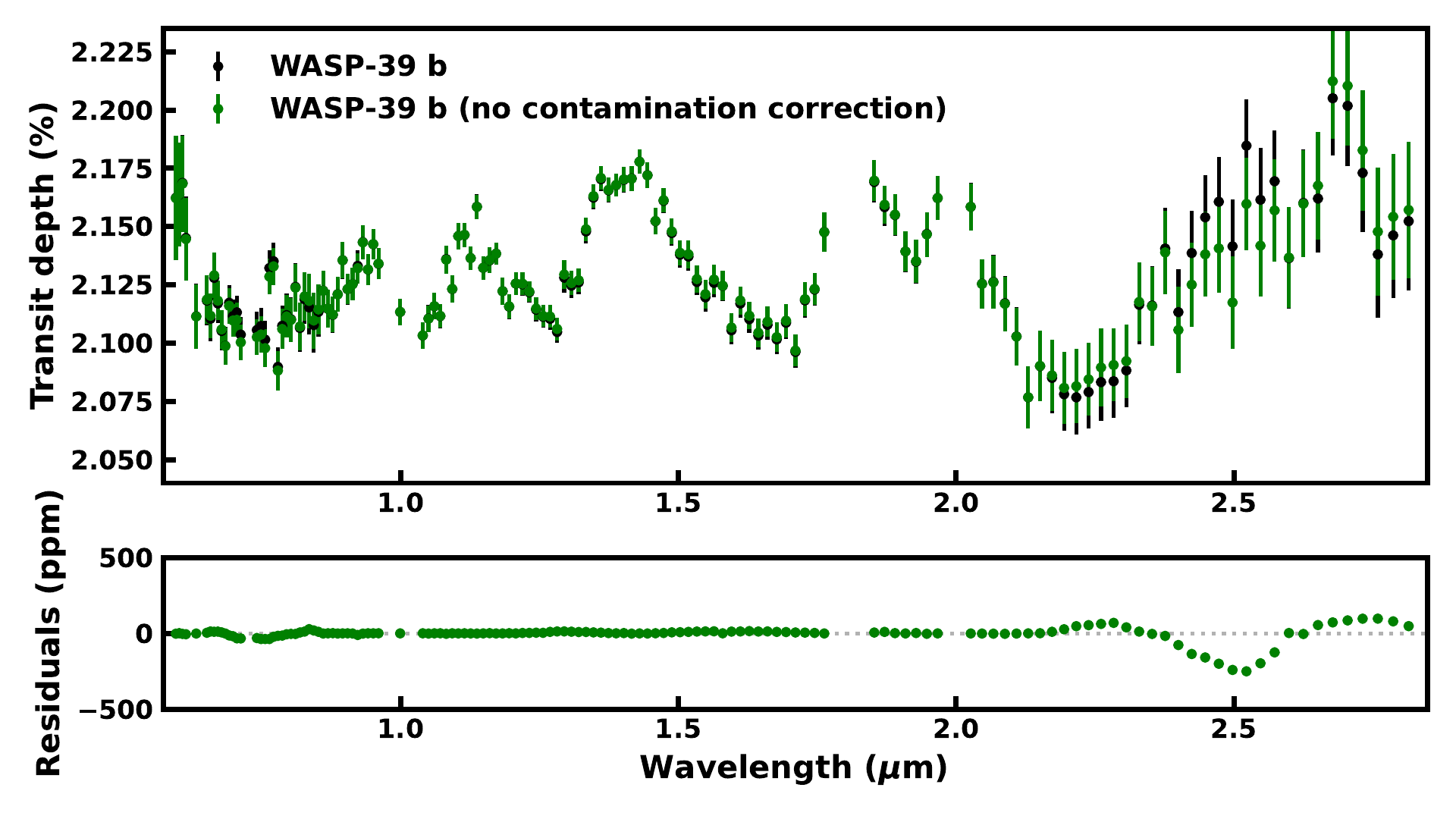}
    \end{subfigure}
    \begin{subfigure}{\columnwidth}
    \includegraphics[width=\columnwidth]{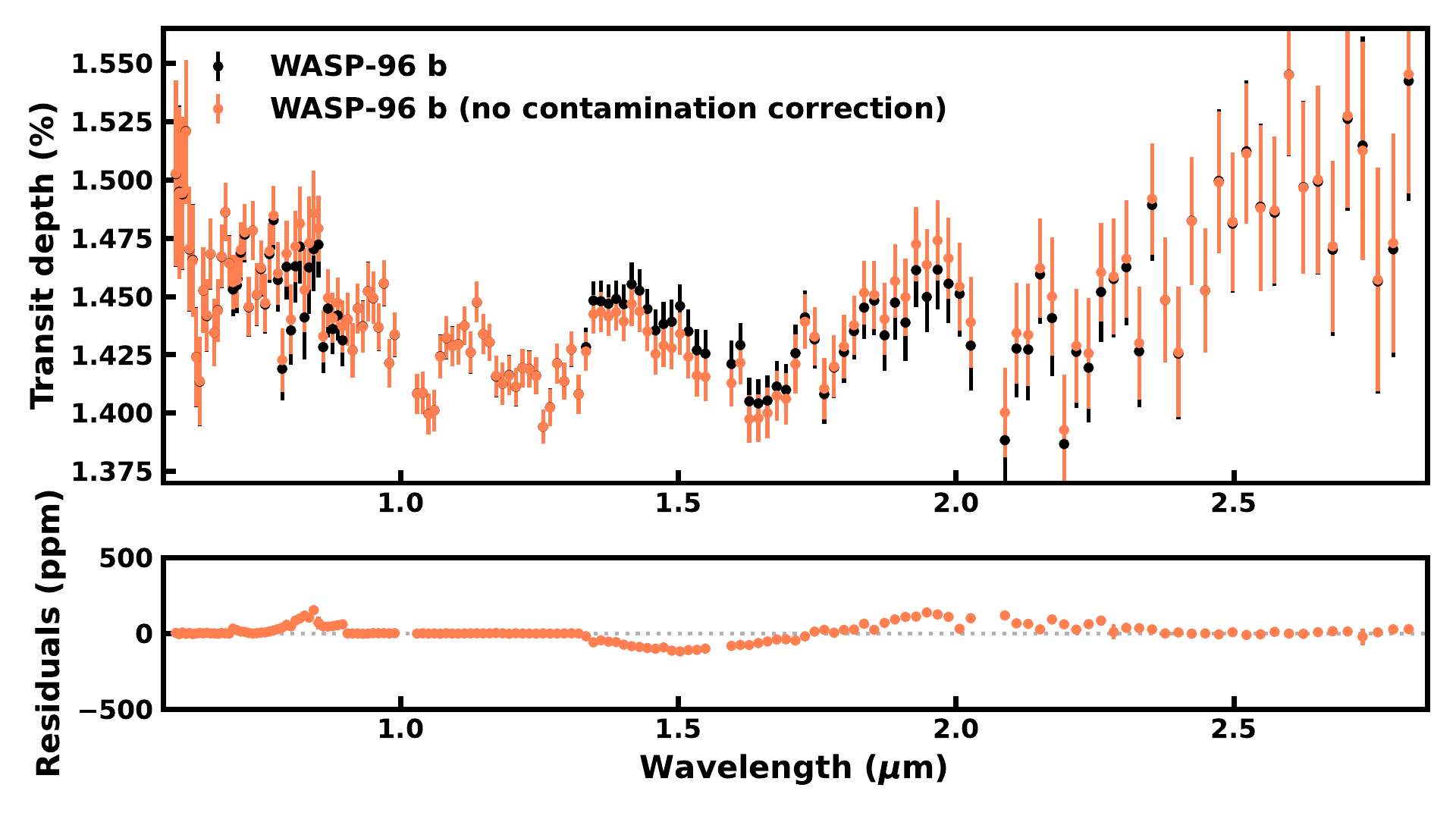}
\end{subfigure}
\caption{Transmission spectra of WASP-39~b and WASP-96~b at $R = 100$ -- with and without correcting for contamination due to additional spectral traces (i.e., dispersed field stars). The top panel shows the spectrum of WASP-39~b, and the bottom panel shows the spectrum of WASP-96~b. We find that the contamination level is about 100 ppm for WASP-96 b, coinciding with the 1.4 $\mu$m H$_2$O feature, and up to 250 ppm for WASP-39 b at 2.5 $\mu$m.}
\label{fig:spectra_cont}
\end{figure}

We previously showed, in section \ref{sec:extraction}, that the stellar spectrum can come out $>5\%$ higher using a box extraction as compared to extracting both orders simultaneously, as illustrated in figure \ref{fig:extracted_spectrum}. However, the transit depth is not affected to the same degree given that the contamination, in contrast to field star contamination, also contains the transit signal. To investigate what effect the overlapping spectral orders have on the resulting transmission spectrum we compare our extraction method to a simple 40-pixel box extraction. To isolate this effect we used the same bad pixel correction and refined background subtraction (to reduce 1/f noise) as used in JExoRES. We also fixed the system parameters and LDCs to their nominal values derived from the white light curve analysis and the $R = 15$ light curve fitting (all using our extraction method). We illustrate the spectra resulting from the two methods in figure \ref{fig:spectra_box}, where the largest difference is seen at the red end of the spectrum, where the two orders partially overlap on the detector. However, within the uncertainty, the spectra appear to be consistent. Here we estimated the variance of the difference between the spectra as $\sigma^2 = \sigma_1^2 + \sigma_2^2 - 2 \sigma_1 \sigma_2 \rho$, where $\sigma_1$ and $\sigma_2$ are the variances of two spectra and where $\rho$ is the correlation coefficient between them. To obtain $\rho$ we note that the transit depths from the two extractions use the same light curve model and that the resulting parameters are very similar, which means the arguments in section \ref{sec:covariance} apply. We thus estimate $\rho$ by computing the sample covariance using the light curve residuals (i.e., the difference between the light curve model and the data) from the two extraction methods. As expected, $\rho$ is close to one since the same data is used.

These findings suggest that contamination from overlapping spectral orders is by itself negligible for the transmission spectrum of WASP-39~b and WASP-96~b at the current noise level, assuming fixed system parameters and LDCs. This does not mean that a stand-alone box extraction performs equally well to the proposed optimal extraction, considering that there are other metrics that can influence the S/N of the derived spectrum. In particular, we used the bad pixel and 1/f noise correction from JExoRES, which we find to be important factors. We note that our optimal extraction method reduces the noise level above 2.5 $\mu$m by more than 10 \%.

\subsection{Effects of contamination} \label{sec:eff_contamination}

Next, we demonstrate the effect of the contamination correction outlined in section \ref{sec:contamination}. We compute the transmission spectra of both targets with and without the correction of dispersed field stars while fixing the system parameters and LDCs to their nominal values. For the uncorrected spectra we used the initial spatial profiles and not the refined profiles. The results are shown in figure \ref{fig:spectra_cont}, where we obtained the uncertainty of the residuals as before. We find clear differences in the spectra of both WASP-39~b and WASP-96~b, especially in the 1.4 - 2.0 $\mu$m range for WASP-96 b and around 2.5 $\mu$m for WASP-39 b. Note that the contamination causes a decrease in the transit depth, as expected, where it overlaps with the spectral traces of the target, while an increase in the transit depth is found elsewhere. The increase in transit depth is caused by an increase in flux of the background which when subtracted acts to inflate the transit depth. This principle also applies to 0th-order field star contamination. An interesting example of this phenomena is seen in the spectra of WASP-96 b at around 0.8 $\mu$m, where no contamination is crossing the 2nd order (see figure \ref{fig:cont_model_WASP_96}). Instead, the contamination is acting on the 1st order, indirectly affecting the background level of the 2nd order given that the two orders are coupled. 

For WASP-96 b, the contamination is caused by a magnitude $G = 19.0$ star, changing the transmission spectrum around the 1.4 $\mu$m H$_2$O feature by 100 ppm. Likewise, the contaminating star for WASP-39 b is of similar magnitude, $G = 19.3$, which changes the transmission spectrum by up to 250 ppm at around 2.5 $\mu$m. On the other hand, the contamination crossing the 2nd order for the observation of WASP-39 b is caused by a fainter star, which we failed to identify with GAIA, resulting in a 30 ppm decrease in the transmission spectrum at around 0.7 $\mu$m. Given these findings, we conclude that higher-order contamination due to field stars, if left uncorrected, can be a significant source of systematics for transmission spectroscopy with NIRISS SOSS.

\section{Comparative assessment of NIRISS SOSS pipelines} \label{sec:pipeline_comparison}

Because of the novelty of the NIRISS instrument and the data reduction, we first compare the transmission spectrum of WASP-39~b derived in this work to spectra from reductions performed in other recent work \citep{feinstein_early_2023}. First, we note that the out-of-transit scatter of the white light curves in \cite{feinstein_early_2023} are similar to what we obtain. We, therefore, conclude that the presence of correlated noise in NIRISS SOSS observations is not specific to our analysis, as we would expect the scatter to be about twice as low if the spectral channels were uncorrelated. For the transmission spectrum, we compare the JExoRES spectrum to the spectra obtained using the following pipelines in \cite{feinstein_early_2023}: nirHiss, supreme-SPOON, transitspectroscopy, NAMELESS, and FIREFly, as the transmission spectra from these pipelines are made available at high-resolution. We can thus bin these high-resolution spectra to a common wavelength grid, selecting $R = 100$, for a like-to-like comparison. 

\begin{figure}
	\includegraphics[width=\columnwidth]{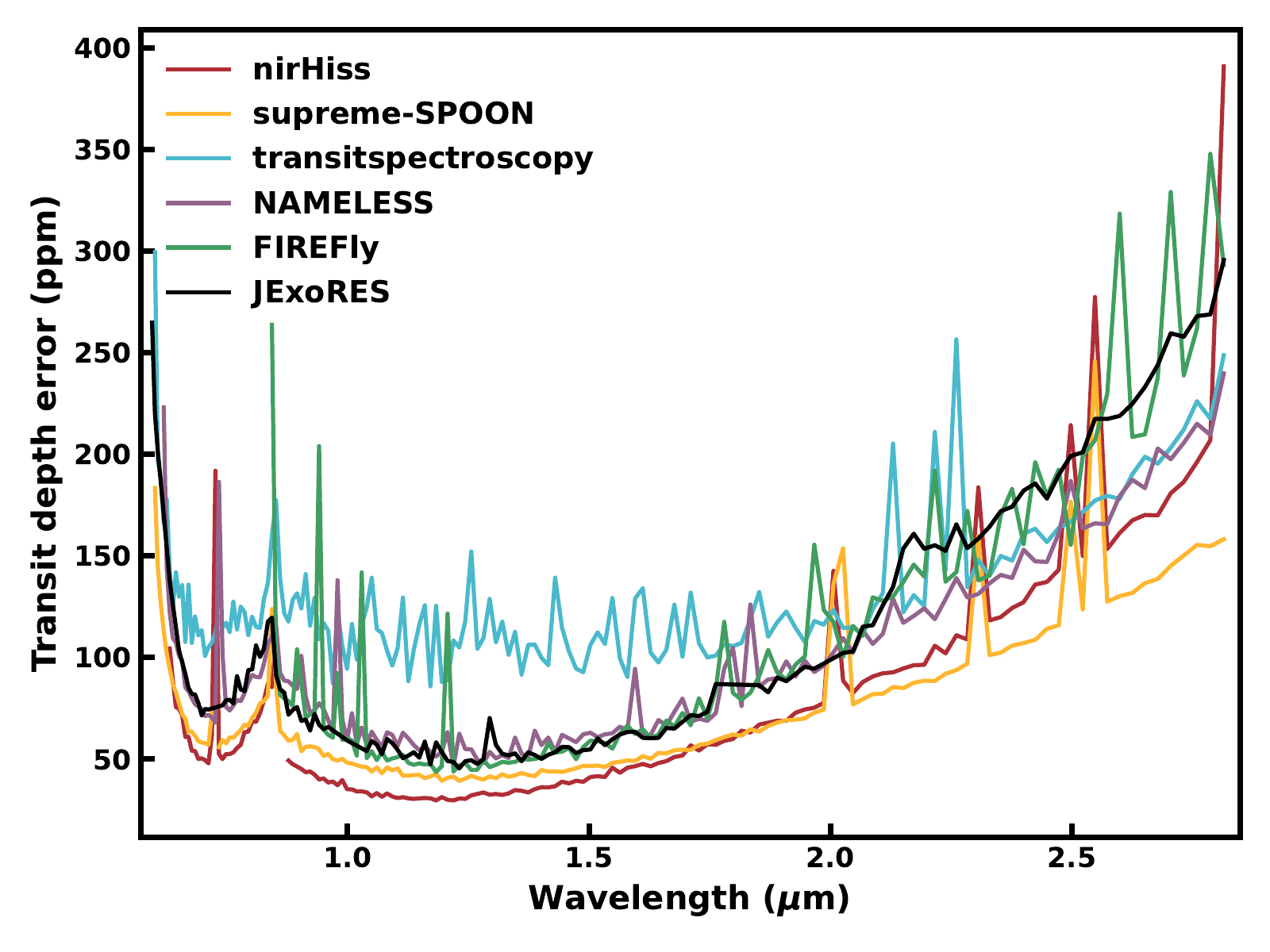}
    \caption{Uncertainty of the transmission spectrum of WASP-39 b, binned to $R = 100$, from multiple NIRISS SOSS reduction pipelines. Here we compare the uncertainties from JExoRES to that of the pipelines in \protect\cite{feinstein_early_2023}. The uncertainty estimates differ by more than a factor of two.}
    \label{fig:error_comparison}
\end{figure}

We show the uncertainties of the spectra from the considered pipelines, binned to $R= 100$, in figure \ref{fig:error_comparison}. Note that the nirHiss spectrum is available at both native instrument resolution and pixel-level resolution, and that we choose to use the latter in the above comparison. Apart from some outliers, we find that the FIREFly pipeline provides the closest match to our pipeline in terms of the uncertainty estimate. Moreover, we have not accounted for correlated noise when performing the binning since we do not have access to this information from these pipelines, which may play a role in the different uncertainty estimates.

\begin{figure}
	\includegraphics[width=\columnwidth]{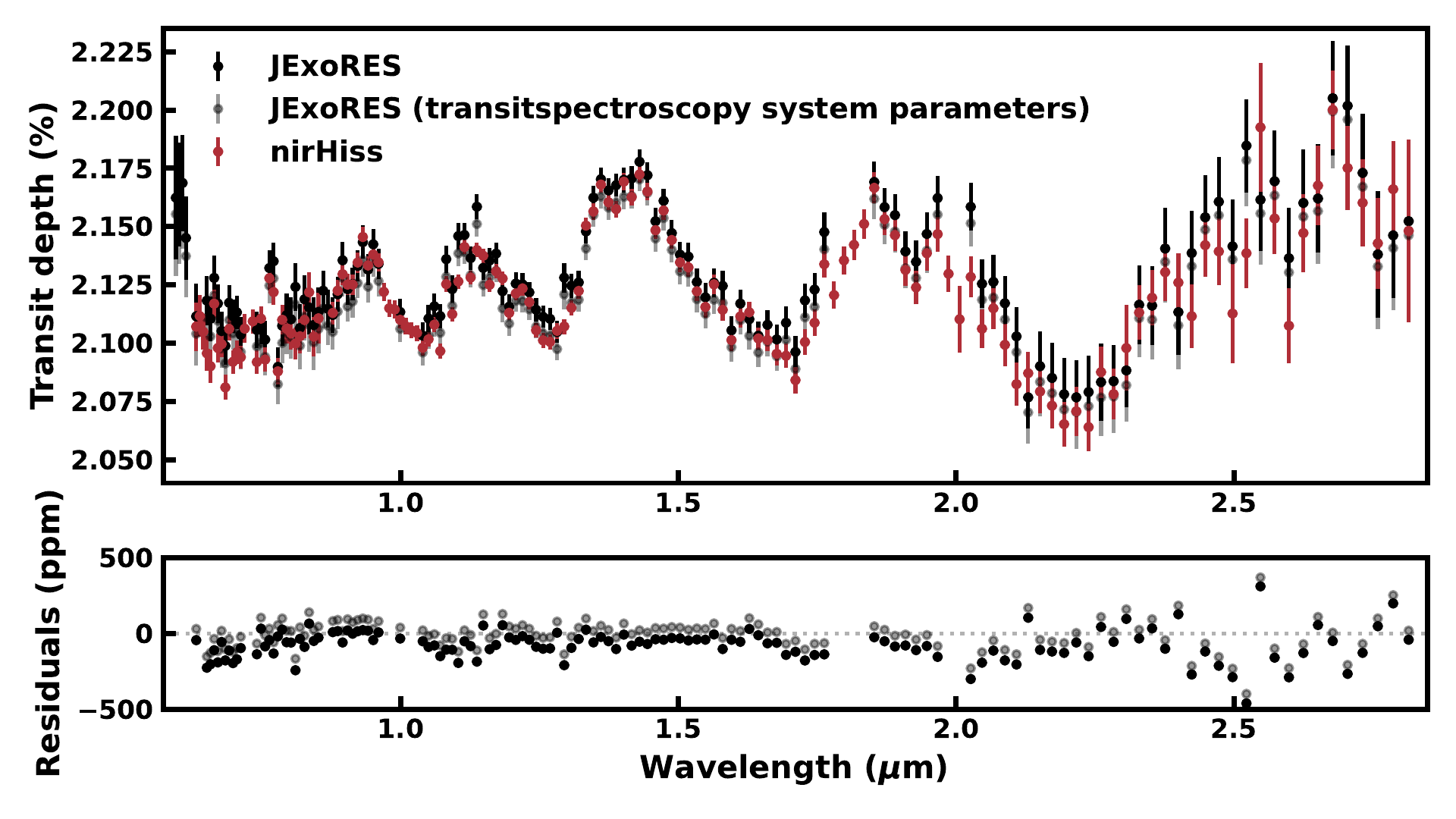}
    \caption{Comparison of the transmission spectrum of WASP-39 b in this work (black) to that of the nirNiss pipeline (red) from \protect\cite{feinstein_early_2023}. To first order, the difference between the spectra is a constant shift, which is reduced when using the same system parameters (grey). Other broadband variations and single data point outliers remain, even when taking into account the difference in system parameters.}
    \label{fig:nirHiss}
\end{figure}

\begin{figure*}
        \subfloat{
            \includegraphics[width=\columnwidth]{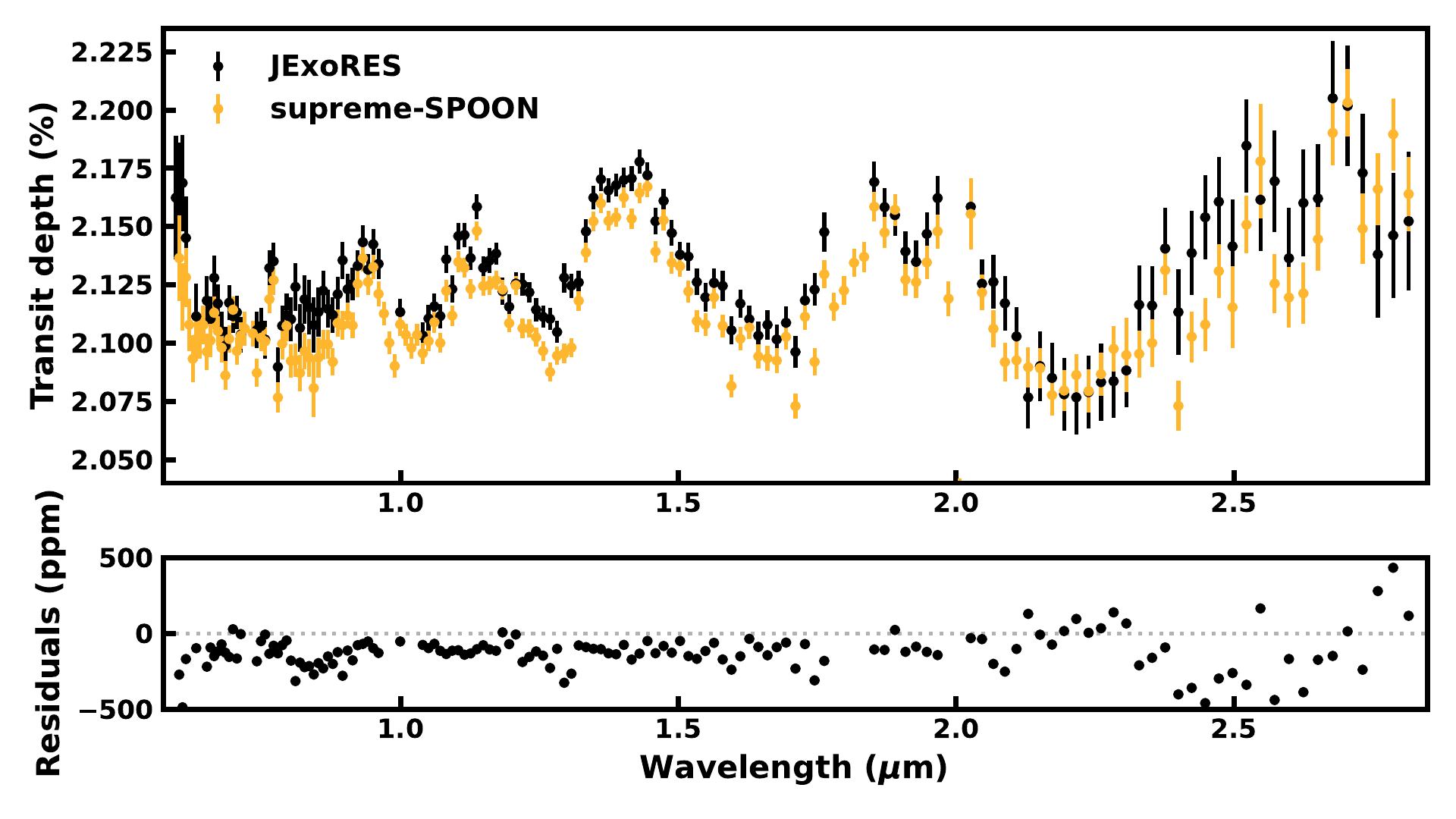}
        } \hfill
        \subfloat{
            \includegraphics[width=\columnwidth]{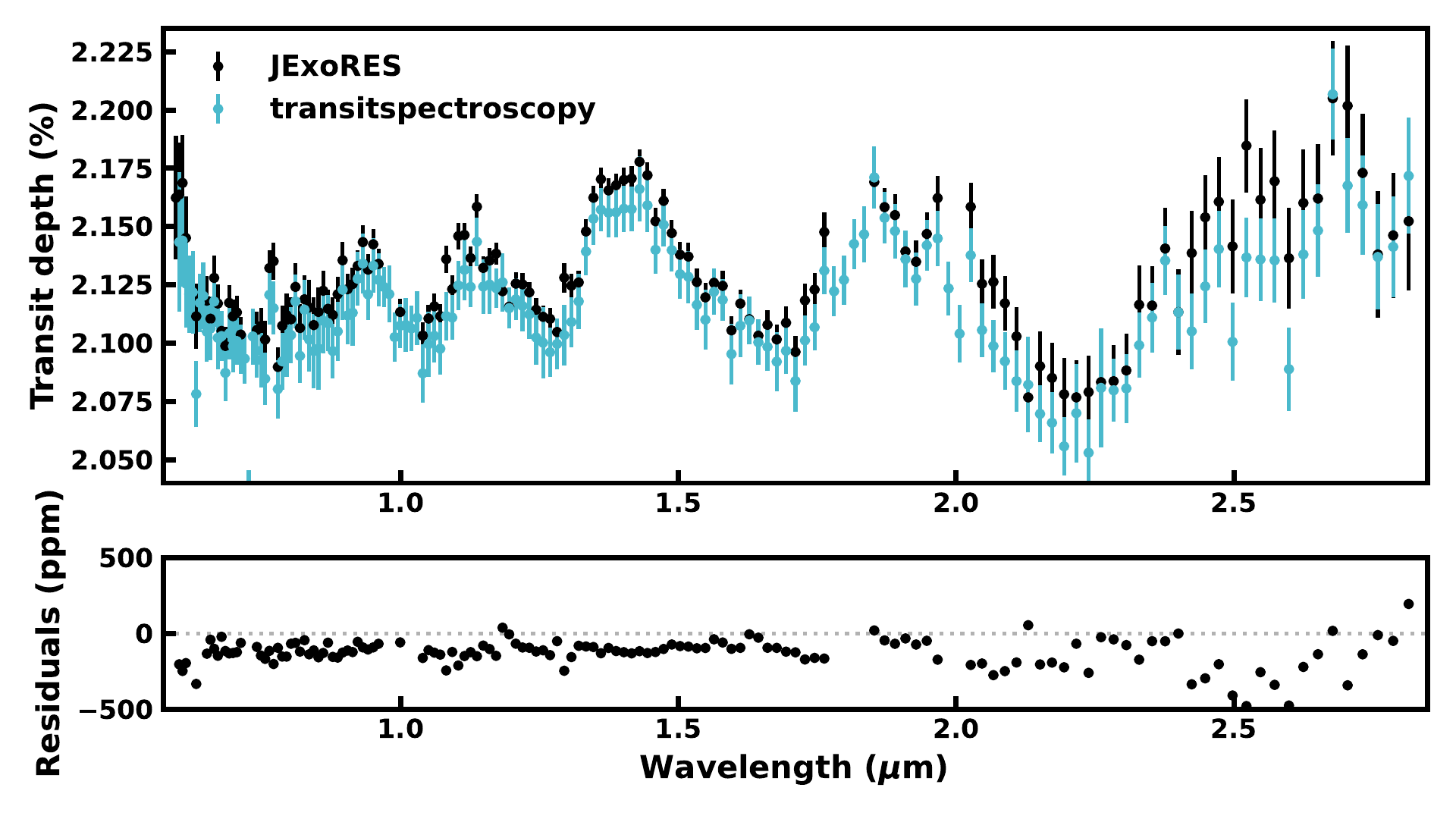}
        } \\
        \subfloat{
            \includegraphics[width=\columnwidth]{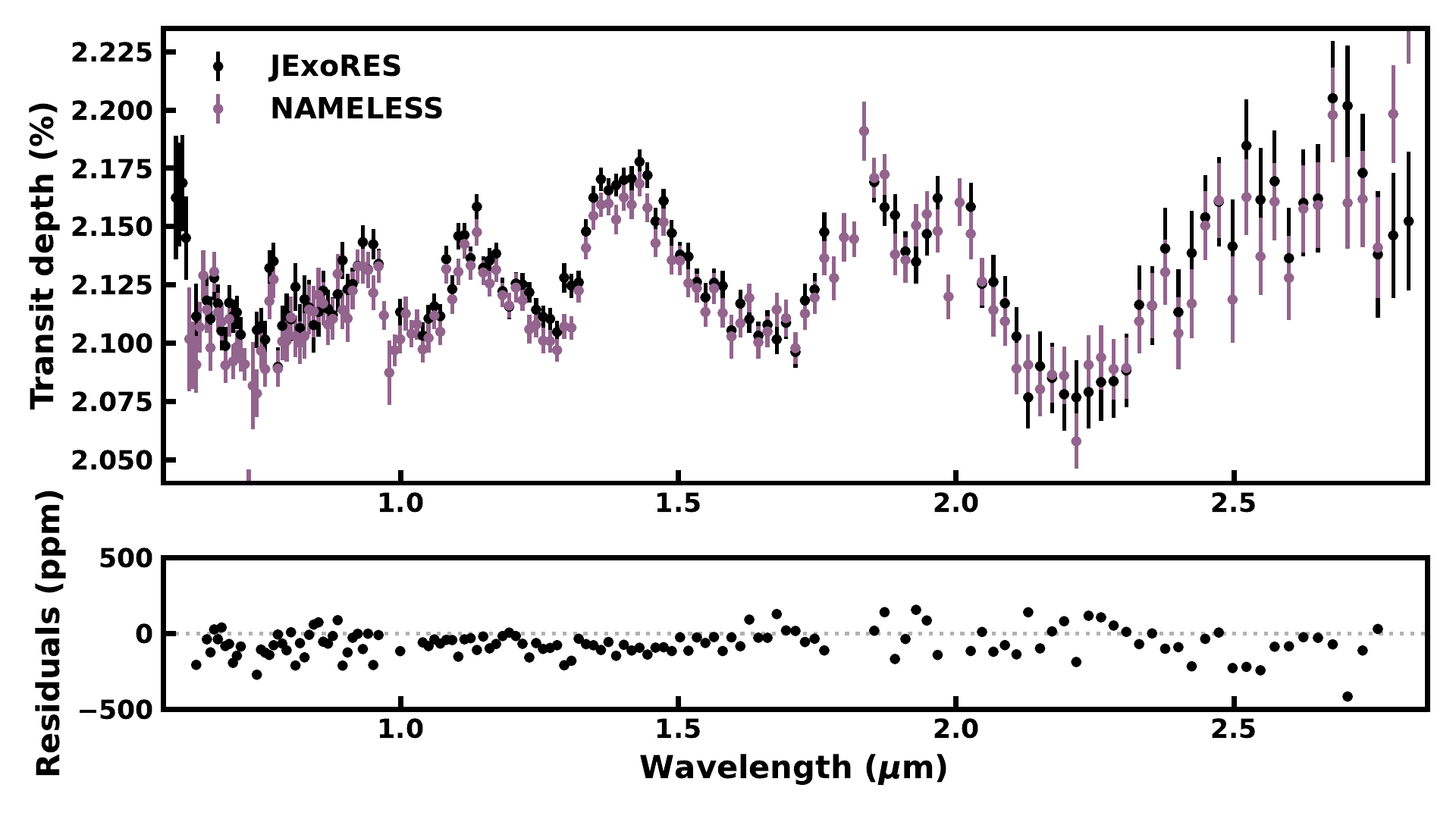}%
        } \hfill
        \subfloat{
            \includegraphics[width=\columnwidth]{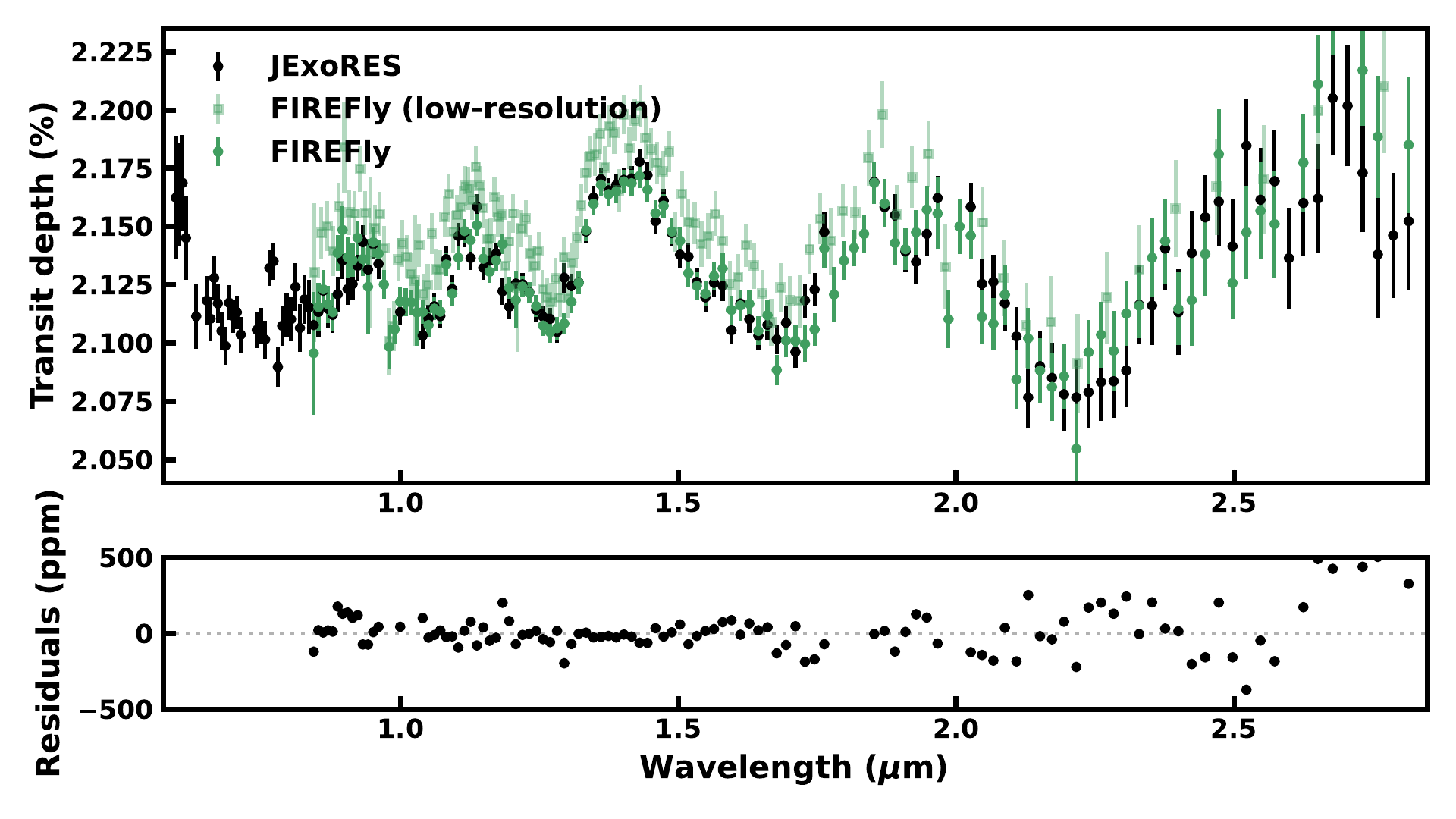}
        } 
        \caption{Comparison of the transmission spectrum of WASP-39 b in this work (black) to that of the supreme-SPOON (yellow), transitspectroscopy (cyan), NAMELESS (violet), and FIREFly (green) pipelines from \protect\cite{feinstein_early_2023}. In all cases, for consistency, we use the reported high-resolution spectrum from each pipeline and bin it down to R = 100 to compare with our spectrum at the same resolution. We again find a shift relative to our spectrum in the case of the spectra from supreme-SPOON, transitspectroscopy, and NAMELESS -- all using the same system parameters from the transitspectroscopy pipeline. The FIREFly pipeline uses different system parameters giving a baseline more similar to our spectrum. However, the reported low-resolution FIREFly spectrum is inconsistent with its high-resolution spectrum binned to $R = 100$. The residuals show that the region around 2.5 $\mu$m differs the most from the spectrum in this work, which we know is a region where field star contamination affects the spectrum.} 
        \label{fig:pipeline_comparison}
\end{figure*}

In figure \ref{fig:nirHiss}, we compare the binned pixel-level resolution nirHiss spectrum of WASP-39 b to our reduction. We find that to a first approximation, there is a shift in the baseline of the nirHiss spectrum compared to our spectrum, which we partly attribute to the slightly different system parameters used. nirHiss used the system parameters from the white light curve fitting of the transitspectroscopy pipeline, that is $a/R_* = 11.388_{-0.027}^{+0.028}$ and $b = 0.4496_{-0.0060}^{+0.0057}$ (giving $i = 87.737 \pm 0.031$$^{\circ}$), which are different enough from our fitted parameters in table \ref{tab:parameters_WASP_39} to induce a difference in the fitted LDCs which translates to a different spectrum, even though the parameters are consistent within 1$\sigma$. Therefore, the discrepancy is likely a result of the system parameter uncertainties not being propagated to the transmission spectrum. Overall, even when accounting for the difference in system parameters, we find broadband variations between the nirHiss spectrum and ours, something an atmospheric retrieval could attempt to fit and obtain different estimates for the atmospheric properties. Coupled with the smaller estimated uncertainties from nirHiss, this could lead to inconsistent retrieval results between the two pipelines. In particular, JExoRES appears to produce a larger transit depth at around $\sim$2.5 $\mu$m which, as seen in section \ref{sec:eff_contamination}, can occur if the field star contamination is not adequately corrected in the nirHiss reduction.

We now compare the transmission spectra from the other pipelines to that in this work as shown in figure \ref{fig:pipeline_comparison}. We again find a small shift in the spectra when compared to JExoRES, in particular from supreme-SPOON, transitspectroscopy, and NAMELESS. This effect is likely due to the shared system parameters used among these pipelines, coming from the white light curve analysis in transitspectroscopy. However, we note that differences in LDCs could also play a role. On the other hand, the FIREFly pipeline uses its own set of system parameters which do not appear to result in the same shift. We also note that the lower-resolution spectrum from FIREFly is significantly different from all other spectra, including the high-resolution spectrum from the same pipeline binned to $R = 100$, as illustrated in figure \ref{fig:pipeline_comparison}. This difference manifests itself as a larger transit depth at shorter wavelengths. For the other pipelines, we find various broadband variations when comparing to the transmission spectrum derived in this work, most notably at around 2.5 $\mu$m, where several of the pipelines give a lower transit depth than JExoRES, which could be a result of inadequate treatment of the contamination present in that region. This can be seen in figure \ref{fig:pipeline_comparison}. 

In summary, three important factors can influence exoplanet transmission spectra observed with JWST/NIRISS SOSS: (a) treatment of contamination from field stars, (b) methods used for estimating the uncertainties in the transit depths, and (c) consideration of the system parameters which to first order can affect the continuum level of the transmission spectrum. A combination of these factors could lead to significant differences in transmission spectra obtained from data reductions with different pipelines. Due to the sensitivity of atmospheric retrievals, such variations could potentially lead to inconsistent parameter estimates when performing retrievals on the spectra from different pipelines.

\section{Case studies} \label{sec:case_studies}

\begin{figure*}
    \begin{subfigure}{2\columnwidth}
    \includegraphics[width=\columnwidth]{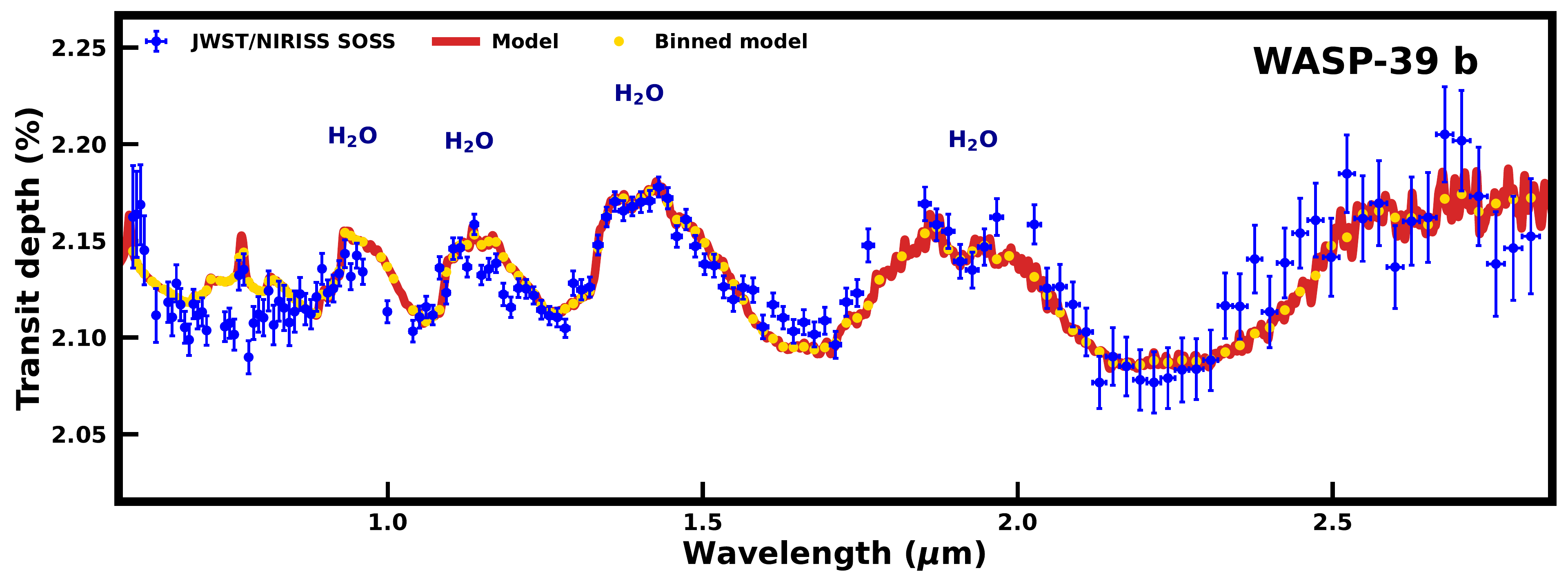}
    \end{subfigure}
    \begin{subfigure}{2\columnwidth}
    \includegraphics[width=\columnwidth]{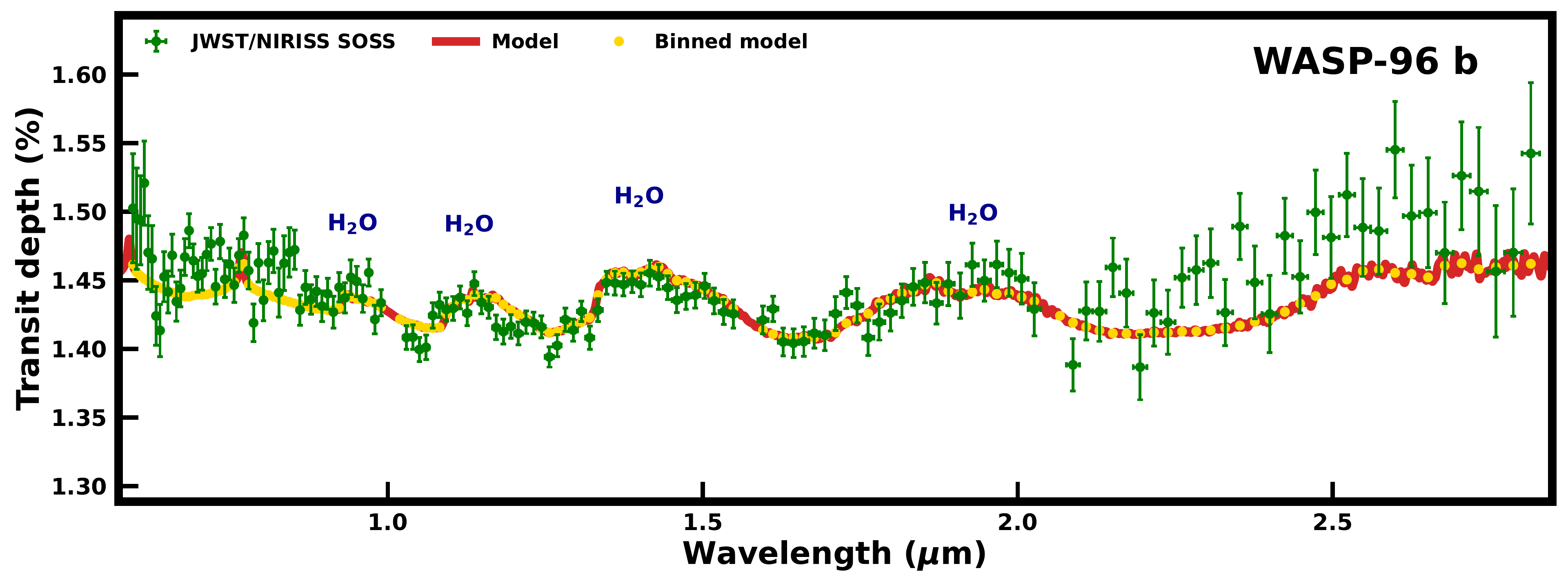}
\end{subfigure}
\caption{JWST/NIRISS SOSS transmission spectra of WASP-39~b (top) and WASP-96~b (bottom), binned at $R = 100$, shown together with nominal models. The most prominent spectral features in both cases are due to H$_2$O.}
\label{fig:spectra_case_studies}
\end{figure*}

In this section we present the JWST/NIRISS SOSS transmission spectra of WASP-39~b and WASP-96~b (0.6 - 2.8 $\mu$m) and assess their broad implications for the atmospheric properties. Both planets have been observed previously with HST and large ground based facilities in the optical and near-infrared, as discussed below. Here, we compare the JWST NIRISS spectra obtained in this work against nominal model transmission spectra to assess the prominent spectral characteristics and compare with inferences from previous studies. The derived transmission spectra for both planets are shown at $R=100$ in figure \ref{fig:spectra_case_studies}, and at $R=500$ in figure \ref{fig:spectra_case_studies_500} in the Appendix.

We model the transmission spectra using a variant of the AURA atmospheric modelling and retrieval code for transmission spectroscopy \citep{pinhas2018,constantinou_early_2023}. The model computes radiative transfer through a plane parallel atmosphere in transmission geometry under the assumption of hydrostatic equilibrium and local thermodynamic equilibrium. The chemical composition, temperature structure, properties of possible clouds/hazes are free parameters in the model; see \citet{constantinou_early_2023} for a recent implementation. Our goal here is not to conduct detailed atmospheric retrievals but instead to compute nominal forward models of transmission spectra that are in general agreement with the observed spectra and previous studies. The models reported here for both planets include an isothermal temperature structure and opacity contributions primarily from H$_2$O, Na and K, as discussed below. This choice of chemical species is motivated by their recent inferences from NIRISS observations of WASP-39~b \citep{feinstein_early_2023}. The molecular cross sections are obtained using the approach of  \cite{gandhi_genesis_2017} from the following sources of line lists: H$_2$O  \citep{barber2006,rothman2010}, Na \citep{allard_new_2019}, and K \citep{allard_kh_2016}. We also include continuum opacity due to H$_2$-H$_2$ and H$_2$-He collision-induced absorption \citep{borysow1988,orton2007,abel2011,richard2012}, and Mie scattering from ZnS \citep{Querry1987}, following \citet{pinhas2017} and \cite{constantinou_early_2023}.

\subsection{WASP-39 b} \label{sec:WASP_39}

WASP-39 b is a highly inflated hot Jupiter-sized planet with a radius of $1.27 R_\text{J}$ and a mass of only $0.28 M_\text{J}$ orbiting a bright G8 star with a period of 4.06 days \citep{faedi_WASP-39b_2011}. With an equilibrium temperature of 1116 K, this makes WASP-39 b an outstanding target for transmission spectroscopy given its low pressure and large atmospheric pressure scale height of $\sim 600$ km at the terminator, which translates to an atmospheric signal of about 250 ppm per scale height. Prior to JWST, the transmission spectrum of WASP-39 b has been observed in the optical and NIR with HST/STIS and Spitzer/IRAC \citep{fischer_hst_2016, sing_continuum_2016, barstow_consistent_2016}, VLT/FORS \citep{nikolov_vlt_2016}, HST/WFC3 \citep{wakeford_complete_2017, tsiaras_population_2018}, and the William Herschel Telescope (WHT)/ACAM \citep{kirk_lrg-beasts_2019}. In the optical, \cite{fischer_hst_2016, sing_continuum_2016} and \cite{nikolov_vlt_2016} detected a Rayleigh scattering slope as well as Na and evidence of K, indicating a largely cloud-free atmosphere. With the addition of the HST/WFC3 data, a clear H$_2$O feature was observed \citep{wakeford_complete_2017, tsiaras_population_2018}. 

With the advent of JWST, WASP-39 b was selected to be one of the main targets of the JWST Transiting Exoplanet Community ERS program \citep{batalha_transiting_2017, stevenson_transiting_2016, bean_transiting_2018}, due to its evident alkali metal absorption and multiple large H$_2$O features revealed with previous space- and ground-based observations. As part of this program, WASP-39 b was recently observed with all NIR instruments on JWST across the 0.5 - 5.5 $\mu$m wavelength range \citep{jwst_transiting_exoplanet_community_early_release_science_team_identification_2022, ahrer_early_2023, alderson_early_2023, rustamkulov_early_2023, feinstein_early_2023}. So far, these observations have led to the detection of multiple chemical species at high significance, including Na, K, H$_2$O, CO$_2$ and CO, as well as evidence of a SO$_2$ feature at 4.05 $\mu$m, which has been suggested to be a tracer for photochemistry \citep{tsai_direct_2022, rustamkulov_early_2023}.

We show the observed spectrum of WASP-39 b in the bottom panel of figure \ref{fig:spectra_case_studies} together with a nominal model. Similar to the model of WASP-96 b, the model of WASP-39 b comprises of opacity sources predominantly from H$_2$O, Na and K, with an isothermal temperature profile at 900 K at the day-night terminator. For the sources of extinction in the atmosphere we consider a model composition consistent with recent findings using the same dataset \citep{feinstein_early_2023}, suggesting a metallicity of 10-30$\times$ solar and inhomogeneous clouds with over $\sim$50\% coverage fraction. Our model assumes volume mixing ratios of 3$\times$10$^{-2}$ for H$_2$O and 5$\times$10$^{-5}$ for Na and 3$\times$10$^{-6}$ for K. These abundances correspond to expectations for $\sim$30$\times$ solar metallicity at the slant photosphere in thermochemical equilibrium, assuming most of the O is in H$_2$O and Na and K are in gaseous form \citep{Asplund2009,Welbanks2019}. The cloud extinction is contributed by Mie scattering due to ZnS particles with a mixing ratio of 10$^{-7}$, modal particle size of 0.01$\micron$ and a scale height of 0.7 relative to the bulk atmospheric scale height, see e.g. \cite{pinhas2017} and \cite{constantinou_early_2023}, and a 50\% cloud coverage. For the planet radius we use $R_p/R_*$ from the order 2 white light curve and $R_* = 0.939\,R_\odot$ \citep{mancini_gaps_2018}, and a reference pressure of 0.35 mbar. Finally, we note that different abundances, cloud coverage, and temperatures can give qualitatively similar fits to the data, as illustrated in figure \ref{fig:wasp_39_model_comp}, something a comprehensive atmospheric retrieval could explore in greater detail.

As shown in figure \ref{fig:spectra_pre_JWST}, the observed JWST spectrum overlaps with previous HST, VLT and WHT observations, compiled by \citep{kirk_lrg-beasts_2019}, with strong H$_2$O features dominating the spectrum as well as a resolved K feature at 0.77 $\mu$m. These studies have indicated that WASP-39 b has a supersolar metalicity atmosphere, albeit with varied estimates. The robustness of these analyses may be investigated by future retrieval studies, by accounting for the covariance of the NIRISS SOSS spectrum and by comparing this observation to the rich set of additional data available of WASP-39 b through the JWST Transiting Exoplanet Community ERS program.

\subsection{WASP-96 b} \label{sec:WASP_96}

WASP-96 b is a hot Jupiter with a mass of $0.48 M_\text{J}$, a radius $1.2 R_\text{J}$, and an equilibrium temperate of 1285 K \citep{hellier_transiting_2014}. The planet orbits a G8 star with a period of 3.43 days. WASP-96 b is of particular interest given that previous observations, both in the optical \citep{nikolov_absolute_2018} and the near-infrared \cite{yip_compatibility_2021}, showed evidence for a relatively cloud-free atmosphere. The first transmission spectrum of WASP-96 b was obtained in the optical with the Very Large Telescope (VLT)/FORS2, revealing a clear signature of the pressure-broadened profile of Na absorption. The planet was later observed with HST/WFC3 to reveal a strong water signature in the near-infrared \citep{yip_compatibility_2021}, however the HST observation was found to be inconsistent with the previous VLT observation given a large offset in transit depth. \cite{nikolov_solar--supersolar_2022} later explained this offset and combined both datasets, confirming a cloud-free atmosphere. The optical part of the WASP-96 b spectrum was then revisited with Magellan/IMACS, confirming the broad Na absorption feature \citep{mcgruder_access_2022}. 

A nominal model consistent with the observed spectrum is shown in the top panel of figure \ref{fig:spectra_case_studies}. As for WASP-39 b, the model spectrum includes opacity contributions primarily from H$_2$O, Na and K. The model assumes an isothermal temperature structure at 1000 K for the average atmospheric profile at the day-night terminator and 30\% cloud coverage. The model assumes volume mixing ratios of 10$^{-3}$ for H$_2$O, 5$\times$10$^{-5}$ for Na and 3$\times$10$^{-6}$ for K. The Na and K abundances are similar to those assumed for WASP-39 b, corresponding to $\sim$30$\times$ solar elemental abundances of Na and K, and the H$_2$O abundance corresponds to a solar 
oxygen abundance at the slant photosphere in thermochemical equilibrium \citep[e.g.,][]{Madhusudhan2012}. We assume the same parameters for cloud extinction by Mie scattering as for WASP-39 b, i.e. ZnS particles with a mixing ratio of 10$^{-7}$, modal particle size of 0.01 $\micron$ and a scale height of 0.7 relative to the bulk atmospheric scale height. For the planet radius we use $R_p/R_*$ from the order 2 white light curve and $R_* = 1.05\,R_\odot$ \citep{hellier_transiting_2014}, and a reference pressure of 0.25 mbar. We note that the data shows some excess absorption and spectral sub-structure in the 2-2.8 $\mu$m range that is not fully explained by our model. This may indicate the presence of hitherto unconsidered chemical species or unmitigated correlated noise in the data (as this is the range most affected, as illustrated by figure \ref{eq:sample_cov}). Detailed retrievals in the future may be able to robustly assess and constrain such contributions. 

The 30\% partial cloud cover in the model still allows for the presence of significant spectral features due to Na, K and H$_2$O. These species have also been detected with ground-based and HST observations in previous studies which argued for a relatively cloud-free atmosphere \citep{nikolov_absolute_2018,nikolov_solar--supersolar_2022}. Considering that we did not conduct formal atmospheric retrievals, as discussed above, it is possible that cloud-free models may also explain the present JWST spectrum. It is also possible that atmospheric variability may cause different levels of cloud cover with time. Finally, as discussed in section \ref{sec:LDCs_effects}, we find that the slope of the spectrum in the optical is strongly influenced by the treatment of limb darkening in the transit light curves. If we use model LDCs instead, then the slope is less steep and could require less contribution from clouds/hazes. This is also consistent with the fact that \cite{nikolov_absolute_2018, nikolov_solar--supersolar_2022}, using VLT and HST observations, adopt model values for either one or both of the quadratic LDCs (which are strongly correlated) and infer a cloud-free atmosphere.

\section{Summary and Discussion} \label{sec:summary}

We are entering a new era of atmospheric characterisation of exoplanets with the advent of the JWST. Transmission spectroscopy of exoplanets with JWST has the potential to provide unprecedented constraints on a wide range of atmospheric properties across the exoplanet landscape. However, accurate and precise retrievals of such atmospheric properties rely critically on the uncertainties in the observed spectra which, in turn, rely on accurate characterisation of the instrument systematics and noise sources. Among the JWST suite of instruments, NIRISS SOSS offers a unique opportunity for exoplanet transmission spectroscopy \citep{doyon_jwst_2012}, thanks to its wide wavelength coverage in the optical to near-infrared (0.6-2.8 $\mu$m) and medium resolution (R $\sim$ 700). We present diagnostics and case studies of exoplanet transmission spectroscopy with JWST/NIRISS SOSS, using observations from the ERS and ERO programs of giant exoplanets WASP-39~b and WASP-96~b. We develop and use a new end-to-end JWST data reduction pipeline, JExoRes, to investigate the instrument systematics, prominent sources of contamination and noise, and mitigation measures, along with different aspects of the data reduction and analysis procedures. Here we summarise our key findings. 

\subsection{Data reduction}

Multiple approaches have been suggested in recent literature for data reduction of exoplanet transit spectroscopy with JWST/NIRISS SOSS \citep{darveau-bernier_atoca_2022, radica_applesoss_2022, feinstein_early_2023, fu_water_2022}. We present a different spectrum extraction algorithm to address the issue stemming from overlapping spectral orders. The main idea is to build a linear model of the flux of each pixel on the detector, using empirically derived spatial profiles, which allows us to extract the flux from multiple overlapping orders and refine the background level to combat 1/f noise, all at once. Our method reduces to the well-tested optimal extraction method \citep{horne_optimal_1986} in the limit of a single spectral order. We note that our method does not require knowledge about the wavelength solution, the spectral resolution kernels, or the throughput function to perform the spectrum extraction.

\begin{itemize}
  \item Using empirical spatial profiles, we show that the 1st and 2nd spectral orders of JWST/NIRISS SOSS can be self-consistently extracted for SUBSTRIP256 observations, recovering the deblended spectrum, while at the same time allowing for accurate cosmic ray and bad pixel correction.
  \item Although the stellar flux is significantly contaminated by overlapping spectral orders, reaching a level of several per cent, we find little evidence that this type of contamination is affecting the transmission spectrum of either WASP-39~b or WASP-96~b at the noise levels of these observations.
\end{itemize}

\subsection{Field star contamination}

Using GAIA DR3, we identify one $G \sim 19$ field star each in the observations of WASP-96 b and WASP-39 b, giving rise to an additional spectral trace which overlap with the target spectra. In this work we develop a new approach to quantify and correct extended contamination from dispersed field stars. We also utilise the F277W exposure to mask the footprint of the 0th-order field star contamination, affecting only a small portion of the spectra.

\begin{itemize}
    \item To model the contamination, we derive the throughput function of NIRISS SOSS from the observation of the standard star BD+60 1753. We find the throughput function to be somewhat different from pre-flight expectations (NIRISS Team). In particular, the throughput of the 2nd order is higher than previous estimates.
    \item For the observations of WASP-39~b and WASP-96~b, dispersed field star contamination can alter the spectra by up to 100 - 250 ppm over a broad wavelength range. This is particularly apparent for WASP-96 b, where a dispersed field star overlap with the spectral trace at around 1.4 $\mu$m, artificially increasing the amplitude of the H$_2$O feature at that wavelength. Such field star contamination, if left uncorrected, can be a significant source of systematics for transmission spectroscopy with JWST/NIRISS SOSS.
\end{itemize}

\subsection{Light curve analysis}

Below we describe results from the light curve analysis of the NIRISS SOSS observations of WASP-39~b and WASP-96~b.

\begin{itemize}
  \item We constrain the covariance of the white light curves from the two spectral orders -- enabling us to find evidence for moderate levels of correlated noise and accurately derive the system parameters without underestimating their uncertainties.\\
  \item We also derive the covariance matrix of the transmission spectrum, a first for exoplanet spectroscopy with JWST, and find that correlated noise among the spectroscopic light curves carries over to the final spectrum. We show that in the limit of uninformative Gaussian priors and fixed system parameters, the correlation matrix of the spectrum is the same as the correlation matrix of the light curves. As a result, the error bars alone are not enough to characterise the covariance, which has implications for atmospheric retrievals. \\
  \item We conjecture that the covariance comes from residual 1/f noise, given that the correlation is stronger among spectral channels from nearby detector columns, which is further supported by the fact that the level of correlated noise is considerably more severe when not performing the background refinement step.\\
  \item Due to correlated noise, the covariance among data points must be accounted for if the binning is performed at the transit depth level. Otherwise, we risk underestimating the errors. When accounting for the covariance, we find that the spectrum uncertainties remain similar regardless of the order of the binning.  \\
  \item The LDCs of WASP-96 are poorly constrained, likely due to the high impact parameter of WASP-96~b, resulting in large transit depth errors when allowing the LDCs to vary. We conclude that limb-darkening may be a limiting factor for the precision of the spectrum of WASP-96~b, especially since neither the modelled LDCs nor the LDCs from WASP-39 (being very similar in stellar properties) appears to match the LDCs of WASP-96. On the other hand, the uncertainty coming from limb darkening is less of a problem for WASP-39~b. \\
  \item Future studies can assess the robustness of atmospheric retrievals in the presence of correlated noise and investigate reduction strategies to minimise these correlations further. For now, atmospheric retrievals with JWST/NIRISS SOSS observations should be done with caution, given that the data points may be correlated.
\end{itemize}

\subsection{Pipeline comparison}

We perform a comparative assessment of NIRISS SOSS pipelines using the observation of WASP-39 b by contrasting the transmission spectrum from JExoRES to those obtained using different pipelines in \cite{feinstein_early_2023}. We bin the reported high-resolution spectra to a common $R = 100$ wavelength grid for a like-to-like comparison.

\begin{itemize}
    \item Although the general shape of the spectra is in agreement, we find both broadband variations and isolated outliers. In particular, we notice the largest difference between the spectra in \cite{feinstein_early_2023} and this work at around 2.5 $\mu$m, an area we show is affected by field star contamination. Given the sensitivity of atmospheric retrievals, such variations and different uncertainty estimates could potentially lead to inconsistent atmospheric parameter estimates when using spectra from different pipelines. \\
    \item To a first approximation, we find an offset in the spectra from the nirHiss, supreme-SPOON, transitspectroscoy, and NAMELESS pipelines, compared to JExoRES. We show that this shift can be partly attributed to a difference in system parameters, thus affecting the fitted LDCs, despite the system parameters used being consistent within 1$\sigma$.
\end{itemize}

\subsection{Case studies}

We perform case studies of WASP-39~b and WASP-96~b, where we compare the JWST/NIRISS SOSS spectra obtained in this work against nominal model spectra to assess prominent spectral characteristics and compare with inferences from previous observations using HST and ground-based facilities. We reiterate that we do not conduct formal retrievals, see e.g. \cite{constantinou_early_2023}, on the spectra in this work. Thus, a broader set of chemical compositions, cloud properties, and temperature structures may be explored to better explain these observations. Future atmospheric retrievals on these datasets would be able to provide important constraints on these properties.

\begin{itemize}
    \item The spectrum of WASP-39 b predominantly shows strong H$_2$O features and a resolved K feature at 0.77 $\mu$m. We find that a model with 30$\times$solar metallicity and a 50\% cloud coverage mostly explains the observed spectrum, in line with the suggestions by \cite{feinstein_early_2023}, and previous HST and ground-based observations. However, a wider range of compositions and other atmospheric properties may be compatible with the data, as shown in figure \ref{fig:wasp_39_model_comp}.

    \item Similar to WASP-39 b, the spectrum of WASP-96 b is dominated by H$_2$O features, with potential contributions from Na and K in the optical. The spectrum can mostly be explained using a model with 30$\times$solar elemental abundances of Na and K, an H$_2$O mixing ratio corresponding to solar O abundance, and a 30\% cloud coverage. A relatively cloud-free atmosphere, as argued by \cite{nikolov_absolute_2018,nikolov_solar--supersolar_2022}, may also explain the present JWST spectrum, considering that we do not conduct a thorough atmospheric retrieval. We further note that the observed optical slope is uncertain, given that it depends on the limb darkening, which is difficult to constrain in the case of WASP-96 b.
\end{itemize}

\subsection{Concluding Remarks}
As already shown by several recent studies \citep{jwst_transiting_exoplanet_community_early_release_science_team_identification_2022, ahrer_early_2023, alderson_early_2023, rustamkulov_early_2023, feinstein_early_2023}, the phenomenal quality of JWST spectra allows an unprecedented discovery space for atmospheric characterisation of exoplanets. This includes high-confidence detections of prominent chemical species and key constraints on various atmospheric properties, all of which in turn provide important constraints on the physical and chemical processes in exoplanetary atmospheres. In several cases, especially for giant exoplanets, the high S/N of JWST spectra can even allow the detections of prominent molecules such as H$_2$O or CO$_2$ empirically, without much aid from model spectra. However, estimating the atmospheric properties, e.g. chemical abundances, temperature profiles and/or properties of cloud parameters, from the spectra would require robust atmospheric retrieval approaches. 

Retrieval algorithms are optimised to extract maximal information from a given data set \citep{madhu2018} and, hence, are highly sensitive to the uncertainties in the observed spectra. This is all the more important in the limit of high-S/N JWST spectra which may encode a treasure trove of atmospheric information and enable high-precision abundance estimates with implications for various planetary processes and formation mechanisms. At the same time, it also increases the risk of retrievals deriving inaccurate, but highly precise, atmospheric properties in cases where the uncertainties in the spectra are not robustly estimated, including underestimated and/or correlated noise sources. Therefore, accurate characterisation of instrumental effects and noise sources across the data reduction and analysis procedures, and the ability to account for correlated noise in the retrievals, is of high importance for atmospheric retrievals in the JWST era. Our present work is a step in this direction. We hope our results pave the way to robust exoplanet spectroscopy and accurate atmospheric retrievals with JWST data, ushering in the new era of high-precision remote sensing of exoplanetary atmospheres.

\section*{Acknowledgements}
We thank the anonymous referee for their valuable review and feedback. We thank NASA, ESA, CSA, STScI and everyone whose efforts have contributed to the JWST, and the exoplanet science community for the thriving current state of the field. This work is supported by research grants to N.M. from the MERAC Foundation, Switzerland, and the UK Science and Technology Facilities Council (STFC) Center for Doctoral Training (CDT) in Data intensive science at the University of Cambridge (STFC grant number ST/P006787/1). N.M. and M.H. acknowledge support from these sources towards the doctoral studies of M.H. M.H. thanks Subhajit Sarkar for initial discussions on the ERO dataset for WASP-96 b and Savvas Constantinou for discussion on pre-JWST transmission spectra of WASP-96 b and WASP-39 b using HST and ground-based telescopes.

\section*{Data availability}
This work is based on observations made with the NASA/ESA/CSA JWST. The publicly available data were obtained from the Mikulski Archive for Space Telescopes at the Space Telescope Science Institute, which is operated by the Association of Universities for Research in Astronomy, Inc., under NASA contract NAS 5-03127 for JWST. These observations are associated with programs \#1091, \#1366, \#1541, and \#2734. The authors acknowledge the JWST Transiting Exoplanet Community ERS team (PI: Batalha) and the ERO team (PI: Pontoppidan) for developing their observing programs with a zero-exclusive-access period. We provide the transmission spectra, the empirical covariance matrices, and the NIRISS SOSS throughput function derived in this work at \href{https://doi.org/10.5281/zenodo.7813271}{10.5281/zenodo.7813271}.

\bibliographystyle{mnras}
\bibliography{references, bibtex}

\appendix

\section{Effects of correlated noise} \label{app:effects_of_corr}

\begin{figure*}
	\includegraphics[width=2\columnwidth]{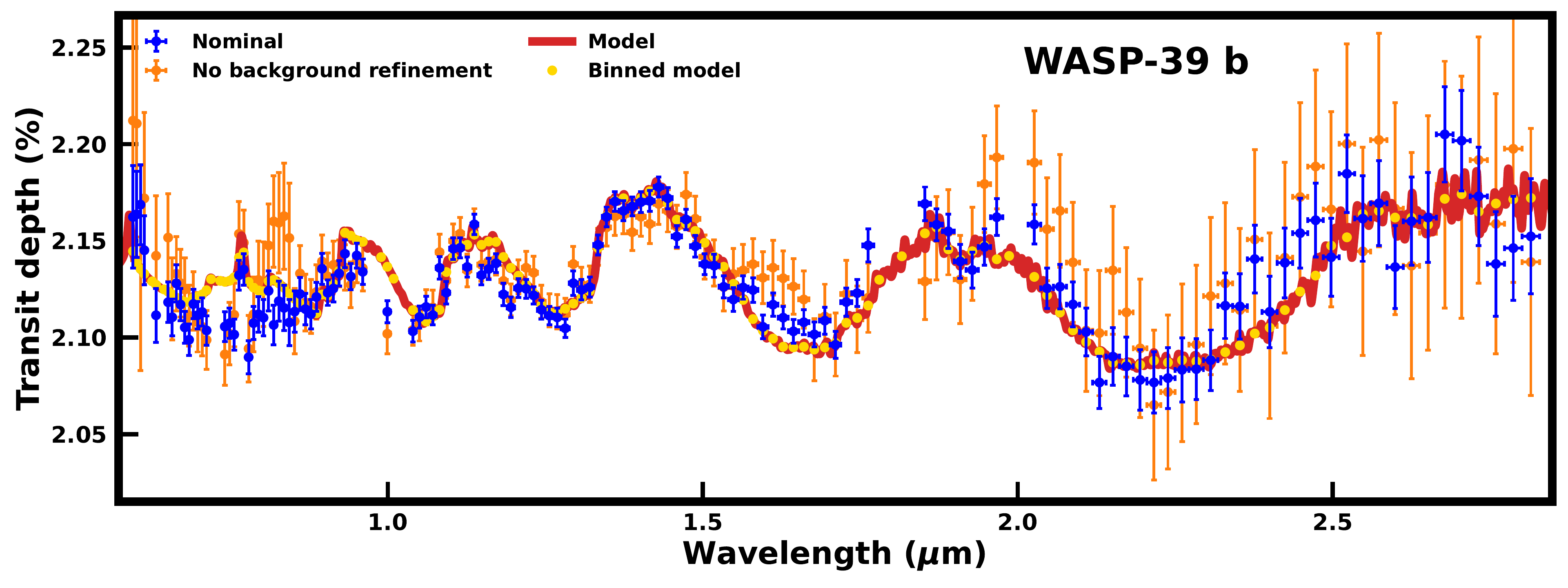}
    \caption{Effects of correlated noise on the transmission spectrum of WASP-39 b, binned at $R = 100$. The nominal spectrum, with the background level refined to reduce 1/f noise, is shown in blue, whereas a spectrum without refining the background, leading to stronger correlated noise, is shown in orange. The model spectrum, described in section \ref{sec:WASP_39}, is shown in red and binned to the same wavelength grid as the data in yellow. Apart from larger uncertainties, several spurious features are seen in the spectrum without background refinement that is not found in the nominal spectrum - features that could be mistaken to be of atmospheric origin.}
    \label{fig:wasp_39_spectrum_corr}
\end{figure*}

In figure \ref{fig:wasp_39_spectrum_corr}, we show the JWST/NIRISS SOSS spectrum of WASP-39 b with and without the background refinement used to mitigate 1/f noise, as described in section \ref{sec:extraction}. The latter case provides the worst-case scenario for the level of correlated noise in the spectrum and illustrates both an increase in white noise, seen as larger error bars, and long-range correlations, demonstrated by several spurious features that are not visible in the nominal spectrum.

\section{High-resolution spectra} \label{app:figures}

\begin{figure*}
    \begin{subfigure}{2\columnwidth}
    \includegraphics[width=\columnwidth]{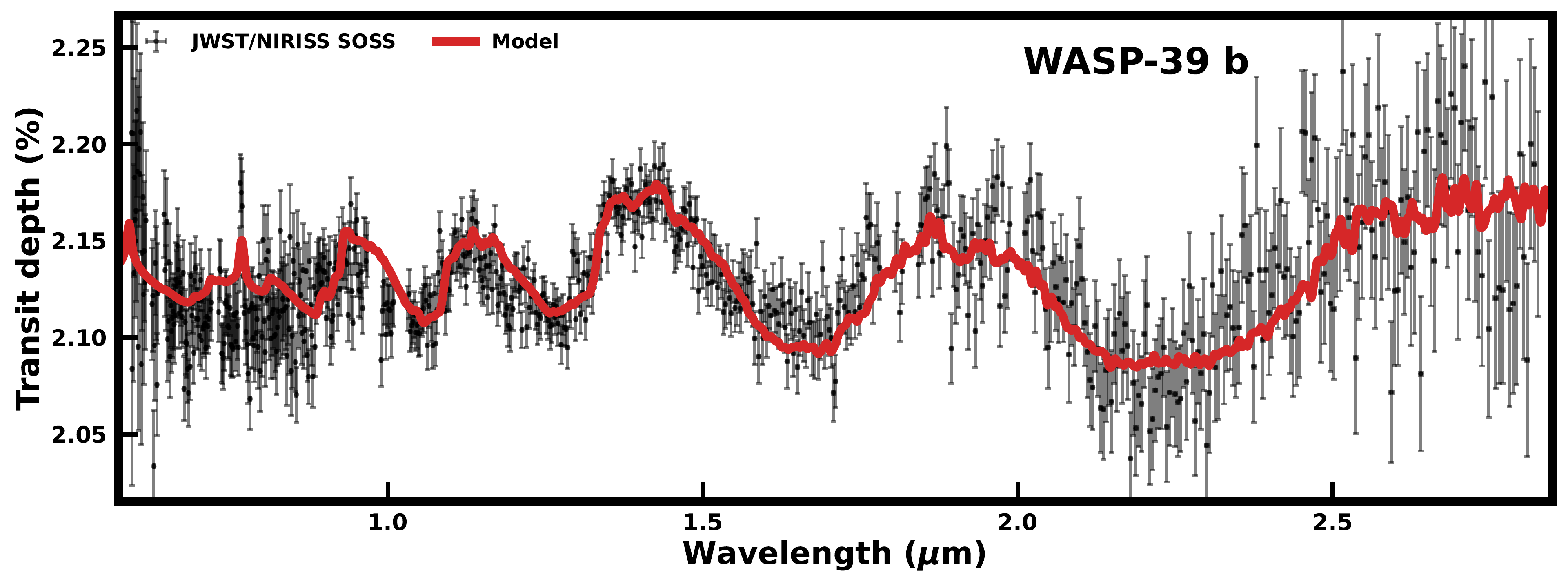}
    \end{subfigure}
    \begin{subfigure}{2\columnwidth}
    \includegraphics[width=\columnwidth]{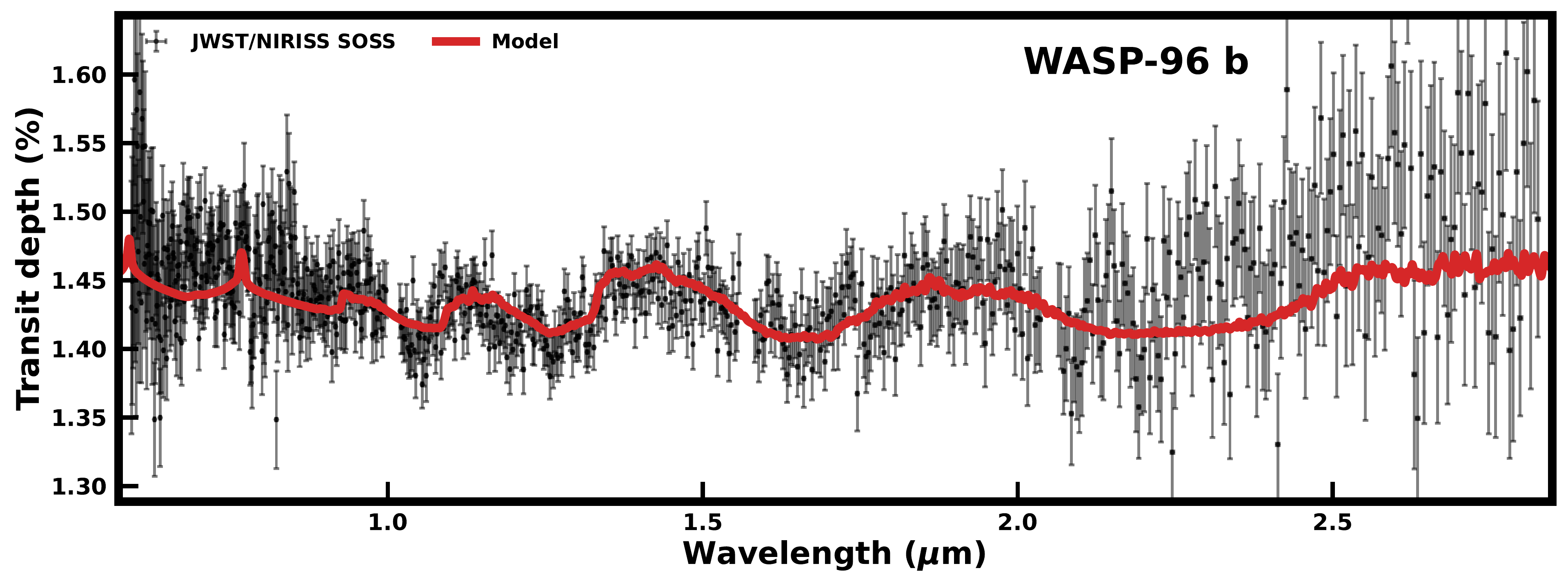}
\end{subfigure}
\caption{High-resolution JWST/NIRISS SOSS transmission spectra of WASP-39 b (top) and WASP-96 b (bottom), binned at $R = 500$, shown together with nominal models, as described in section \ref{sec:case_studies}}.
\label{fig:spectra_case_studies_500}
\end{figure*}

Figure \ref{fig:spectra_case_studies_500} shows the transmission spectra of WASP-39 b and WASP-96 b at $R = 500$. The most prominent spectral features for both planets are due to H$_2$O. In the case of WASP-39 b, the high resolution allows for the K feature at 0.77 $\mu$m to be resolved.

\section{Model comparison} \label{app:model_comp}

\begin{figure*}
	\includegraphics[width=2\columnwidth]{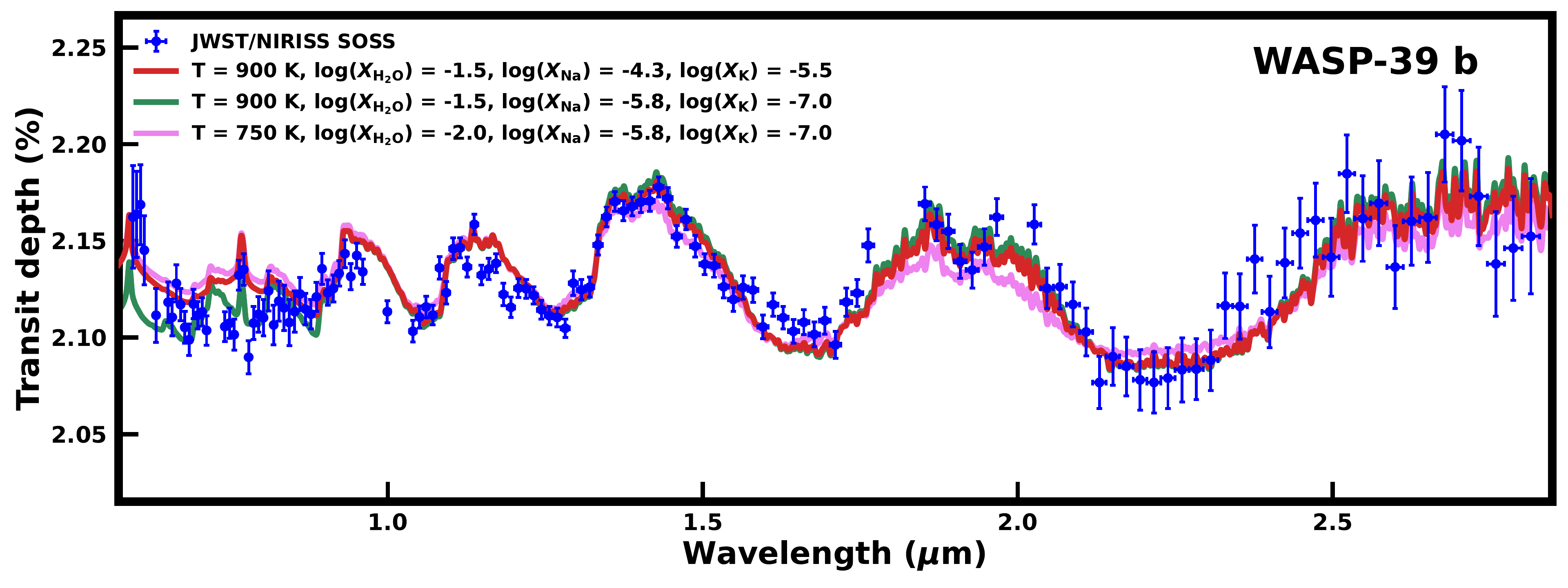}
    \caption{Comparison of forward models for WASP-39 b. We show three different models that each can explain the JWST/NIRISS SOSS data of WASP-39 b qualitatively, highlighting the level of degeneracies among the atmospheric parameters. The model in red is the nominal model, described in section \ref{sec:WASP_39}, while the two other models are variations with different temperatures and abundances as shown in the legend. The reference pressures of the models in red (nominal), green and pink are 0.35 mbar, 0.5 mbar and 1 mbar, respectively. We note that most of the differences between the models occur below 0.8 $\mu$m.}
    \label{fig:wasp_39_model_comp}
\end{figure*}

We note that different combinations of abundances, cloud coverage, and temperatures can be used to qualitatively explain the observations. This is  demonstrated in figure \ref{fig:wasp_39_model_comp}. Here, we show three model transmission spectra with different input parameters that are in similar qualitative agreement with the observed JWST/NIRISS SOSS spectrum of WASP-39 b. This underscores the need for robust atmospheric retrievals in the future for deriving accurate constraints on the atmospheric properties from such observations.

\section{Comparison with pre-JWST data} \label{app:spectra_pre_JWST}

\begin{figure*}
    \begin{subfigure}{2\columnwidth}
    \includegraphics[width=\columnwidth]{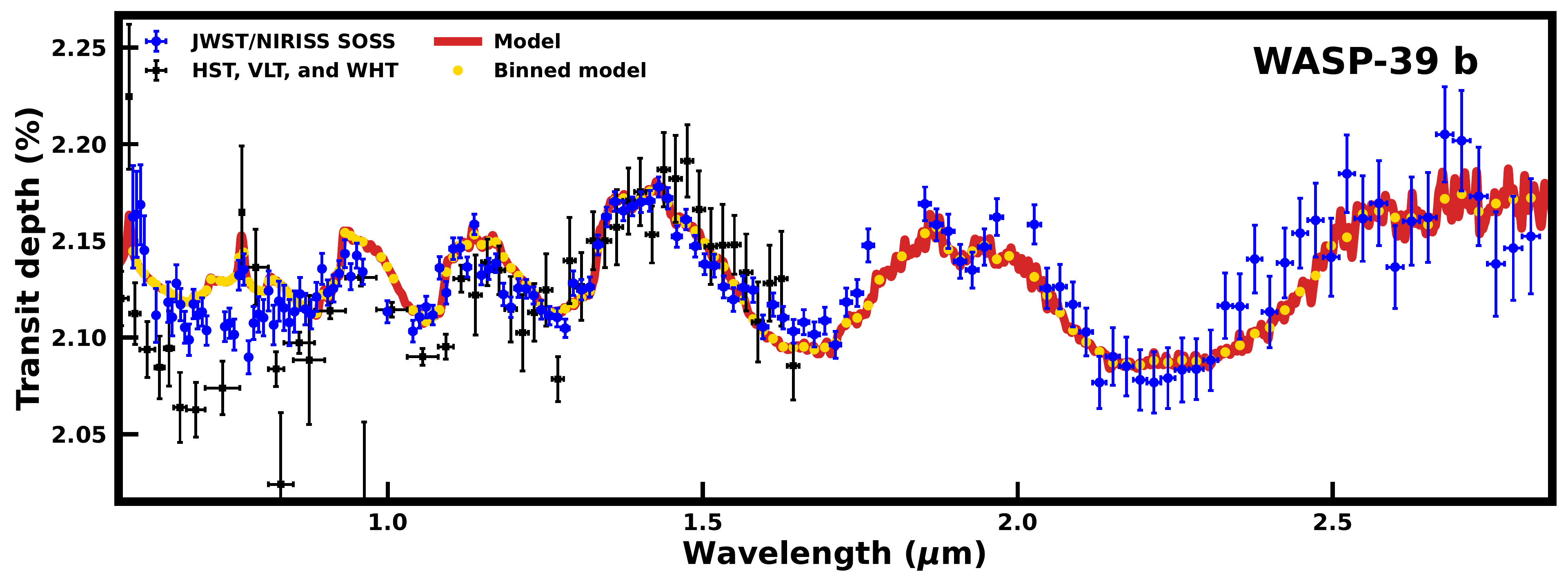}
    \end{subfigure}
    \begin{subfigure}{2\columnwidth}
    \includegraphics[width=\columnwidth]{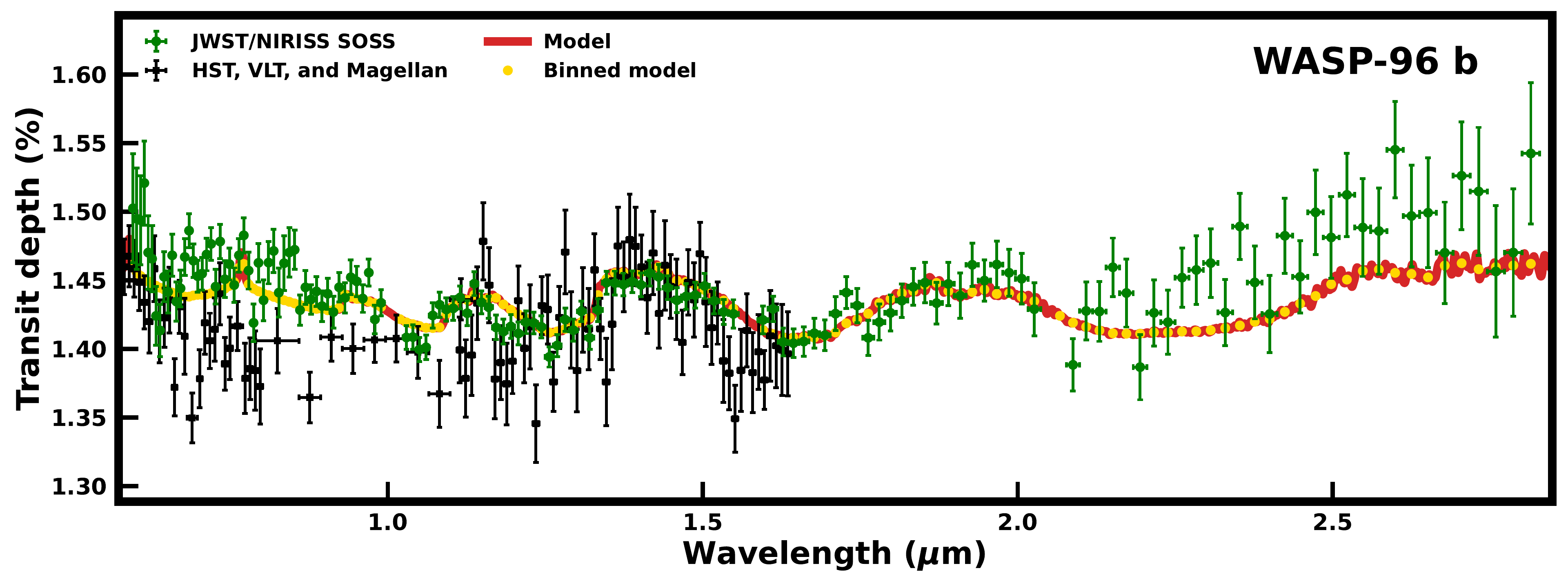}
\end{subfigure}
\caption{JWST/NIRISS SOSS transmission spectra of WASP-39 b (top) and WASP-96 b (bottom), binned at $R = 100$, shown together with nominal models, described in section \ref{sec:case_studies}, and pre-JWST observations for comparison. The HST, VLT, and Magellan observations of WASP-96 b are obtained from \protect\cite{nikolov_solar--supersolar_2022}, while the HST, VLT and WHT observations of WASP-39 b are obtained from \protect\cite{kirk_lrg-beasts_2019}. In the case of WASP-96 b, we note the difference between the observed spectra in the optical, as discussed in section \ref{sec:WASP_96}.}
\label{fig:spectra_pre_JWST}
\end{figure*}

In figure \ref{fig:spectra_pre_JWST}, we compare the JWST/NIRISS SOSS transmission spectra derived in this work to pre-JWST observations from HST and ground-based telescopes.

\label{lastpage}
\end{document}